\documentclass[fleqn,twoside]{article}
\usepackage{amsmath,amssymb,espcrc2,color,cite}
\usepackage[usenames,dvipsnames]{xcolor}
\usepackage[percent]{overpic}
\usepackage{graphicx,soul}
\usepackage[figuresright]{rotating}
\usepackage{ulem}

\newcommand{\url}[1]{{\tt #1}}

\newcommand{\gmt}{\ensuremath{(g-2)_\mu}}
\newcommand{\br}{{\rm BR}}
\newcommand{\bsg}{BR($b \to s \gamma$)}

\newcommand{\bsdmm}{\ensuremath{\br(B_{s, d} \to \mu^+\mu^-)}}

\newcommand{\ssi}{\ensuremath{\sigma^{\rm SI}_p}}

\newcommand{\ssd}{\ensuremath{\sigma^{\rm SD}_p}}

\newcommand{\MW}{\ensuremath{M_W}}

\newcommand{\Mh}{\ensuremath{M_h}}

\newcommand{\MA}{\ensuremath{M_A}}
\newcommand{\Min}{\ensuremath{M_{\rm in}}}
\newcommand{\MGUT}{\ensuremath{M_{\rm GUT}}}

\newcommand{\gl}{\ensuremath{{\tilde g}}}
\newcommand{\mgl}{\ensuremath{m_{\tilde g}}}
\newcommand{\msq}{\ensuremath{m_{\tilde q}}}

\newcommand{\stau}[1]{\ensuremath{\tilde \tau_{#1}}}
\newcommand{\stopone}{\ensuremath{\tilde t_{1}}}

\newcommand{\cha}[1]{\tilde \chi^\pm_{#1}}

\newcommand{\mcha}[1]{\ensuremath{m_{\tilde \chi^\pm_{#1}}}}
\newcommand{\neu}[1]{\ensuremath{\tilde \chi^0_{#1}}}
\newcommand{\mneu}[1]{\ensuremath{m_{\tilde \chi^0_{#1}}}}

\newcommand{\sq}{\tilde q}

\newcommand{\staue}{\tilde \tau_1}
\newcommand{\mstaue}{m_{\staue}}

\newcommand{\tb}{\ensuremath{\tan\beta}}

\newcommand{\tev}{\ensuremath{\,\, \mathrm{TeV}}}
\newcommand{\gev}{\ensuremath{\,\, \mathrm{GeV}}}
\newcommand{\mev}{\ensuremath{\,\, \mathrm{MeV}}}

\def\reffi#1{\mbox{Fig.~\ref{#1}}}

\definecolor{orange}{rgb}{1,0.5,0}

\definecolor{Gray}{named}{Gray}

\newcommand{\ETslash}{\ensuremath{/ \hspace{-.7em} E_T}}

\graphicspath{{plots/}}

\title{\vspace{-4.5cm}
\bf \LARGE Likelihood Analysis of the Sub-GUT MSSM in Light of LHC 13-TeV Data \\ \vspace{0.5em}}

\author{
{\bf J.C.~Costa}\address[Imperial]
   {High\,Energy\,Physics\,Group,\,Blackett\,Laboratory,\,Imperial\,College,\,Prince\,Consort\,Road,\,London\,SW7\,2AZ,\,UK},
{\bf E.~Bagnaschi}\address[DESY]
   {DESY, Notkestra{\ss}e 85, D--22607 Hamburg, Germany},
{\bf K.~Sakurai}\address[Warsaw]
{Institute of Theoretical Physics, Faculty of Physics, University of Warsaw, ul.~Pasteura 5, PL--02--093 Warsaw, Poland},
{\bf M.~Borsato}\address[USdC]{Instituto Galego de F{\' i}sica de Altas Enerx{\' i}as, Universidade de Santiago de Compostela, Spain},
{\bf O.~Buchmueller}{\addressmark[Imperial]
{\bf M.~Citron}\addressmark[Imperial],
{\bf A.~De~Roeck}\address[CERNEP]
   {Experimental Physics Department, CERN, CH--1211 Geneva 23, Switzerland; \\  Antwerp University, B--2610 Wilrijk, Belgium},
 {\bf M.J.~Dolan}\address[SLAC]
{ARC Centre of Excellence for Particle Physics at the Terascale, School of Physics, University of Melbourne, 3010, Australia},
{\bf J.R.~Ellis}\address[KCL]{Theoretical Particle Physics
  and Cosmology Group, Department of Physics, King's College London, London~WC2R~2LS, UK; \\
    National Institute of Chemical Physics and Biophysics, R{\" a}vala 10, 10143 Tallinn, Estonia; \\
Theoretical Physics Department, CERN, CH--1211 Geneva 23, Switzerland},
{\bf H.~Fl\"acher}\address[Bristol]
   {H.H.~Wills Physics Laboratory, University of Bristol, Tyndall Avenue, Bristol BS8 1TL, UK},
{\bf S.~Heinemeyer}\address[Madrid]
{Campus of International Excellence UAM+CSIC, Cantoblanco, E--28049 Madrid, Spain;\\
  Instituto de F\'{\i}sica Te{\'o}rica UAM-CSIC, C/ Nicolas Cabrera 13-15, E--28049 Madrid, Spain; \\
   Instituto de F\'{\i}sica de Cantabria (CSIC-UC), Avda. de Los Castros s/n,
    E--39005 Santander, Spain},
{\bf M.~Lucio}\addressmark[USdC],
{\bf D.~Mart\'inez~Santos}\addressmark[USdC],
{\bf K.A.~Olive}\address[Minnesota]
{William I.\ Fine Theoretical Physics Institute, School of Physics and
 Astronomy, University of Minnesota, Minneapolis, Minnesota 55455, USA},
{\bf A.~Richards}\addressmark[Imperial],
{\bf G.~Weiglein}\addressmark[DESY]}
}

\begin{document}
\begin{abstract}
\vspace{0.25cm}

We describe a likelihood analysis using {\tt MasterCode} of variants of the
MSSM in which the soft supersymmetry-breaking parameters are assumed
to have universal values at some scale \Min\ below the supersymmetric
grand unification scale \MGUT, as can occur in mirage mediation and other models.
In addition to \Min, such `sub-GUT' models have the 4 parameters of the CMSSM,
namely a common gaugino mass $m_{1/2}$, a common soft supersymmetry-breaking
scalar mass $m_0$, a common trilinear mixing parameter
{$A$} and the ratio of MSSM Higgs vevs $\tb$, {assuming that the Higgs mixing parameter $\mu > 0$}.
We {take into account} constraints on strongly- and electroweakly-interacting sparticles from {$\sim 36$/fb of LHC data at 13 TeV
and the LUX and 2017 PICO, XENON1T and PandaX-II searches for dark matter scattering}, in addition to the previous
{LHC and dark matter constraints} {as well as full sets of flavour and electroweak
constraints}. We find a preference for $\Min \sim 10^5$ {to $10^9 \gev$},
with $\Min \sim \MGUT$ disfavoured by $\Delta \chi^2 \sim 3$ due to the \bsdmm\ constraint. The lower limits
on strongly-interacting sparticles are largely determined by LHC searches, and similar to those in the CMSSM.
We find a preference for the LSP to be a {Bino or} Higgsino with $\mneu1 \sim 1 \tev$, with
{annihilation via heavy Higgs bosons $H/A$ and stop coannihilation, or chargino coannihilation,}
bringing the cold dark matter density into the cosmological range. We find that spin-independent
dark matter scattering is likely to be within reach of the planned LUX-Zeplin and XENONnT
experiments. We {probe the impact of the \gmt\ constraint}, finding similar results whether or not it is included.

\vspace{1cm}
\begin{center}
{\tt KCL-PH-TH/2017-45, CERN-PH-TH/2017-197, DESY 17-156, IFT-UAM/CSIC-17-089\\
{FTPI-MINN-17/19, UMN-TH-3703/17}}
\end{center}
%\vspace{2.0cm}
\end{abstract}
% Remove page number from the title page
\thispagestyle{empty}
\newpage

%\vspace{4.0cm}

\maketitle

%\tableofcontents

\newpage

%---------------------------------------------------------------------
\section{Introduction}
\label{sec:intro}
%---------------------------------------------------------------------
Models invoking the appearance of supersymmetry (SUSY) at the TeV scale
are being sorely tested by the negative results of high-sensitivity searches
for sparticles at the LHC~\cite{CMSWiki,ATLASWiki} and for the
scattering of dark matter particles~\cite{lux16,PICO,XENON1T,pandax}.
There have been many global analyses of the implications of these experiments
for specific SUSY models, mainly within the minimal supersymmetric
extension of the Standard Model (MSSM), in which the lightest supersymmetric
particle (LSP) is stable and a candidate for dark matter (DM). This may well be
the lightest neutralino, $\neu1$~\cite{EHNOS}, as we assume here.
Some of these studies have assumed universality of the soft SUSY-breaking
parameters at the GUT scale, e.g., in the constrained MSSM (the CMSSM)~\cite{mcold,mc9,Fittino,GAMBIT} and in
models with non-universal Higgs masses (the NUHM1,2)~\cite{mc9,mc10}. Other analyses have
taken a phenomenological approach, allowing free variation in the soft SUSY-breaking
parameters at the electroweak scale (the
pMSSM)~{\cite{pMSSM,mc11,gambit-pmssm7,pMSSM11}}.

A key issue in the understanding of the implications of the LHC searches for
SUSY is {the} exploration of regions of parameter space where compressed spectra
may reduce the sensitivity of searches for missing transverse energy, $\ETslash$.
These regions also have relevance to cosmology, since models with
sparticles that are nearly degenerate with the LSP allow for important
coannihilation processes that suppress the relic LSP number density, allowing heavier
values of $\mneu1$. The accompanying heavier SUSY spectra are also more
challenging for {the} LHC $\ETslash$ searches.

The CMSSM offers limited prospects for coannihilation, and examples that have
been studied in some detail include coannihilation with the lighter stau slepton,
$\stau1$~\cite{stauco,staucoann}, or the lighter stop squark, $\stopone$~\cite{stopcoann}. Other models offer the
possibilities of different coannihilation partners, such as the lighter chargino,
$\cha1$~{\cite{chacoann,mc11}}, some other slepton~\cite{pMSSM11} or squark flavour~\cite{mc-su5},
or the gluino~\cite{ELO,EELO}. In particular, the
pMSSM allows for all these possibilities, potentially also in combination~\cite{pMSSM11}.

In this paper we study the implications of LHC and DM searches for
an intermediate class of SUSY models, in which universality of the
soft SUSY-breaking parameters is imposed at some input scale \Min\ below
the GUT scale \MGUT\ but above the electroweak scale~\cite{sub-GUT,ELOS},
which we term `sub-GUT' models. Models in this
class are well motivated theoretically, since the soft SUSY-breaking
parameters in the visible sector may be induced by some dynamical mechanism
such as gluino condensation that kicks in below the GUT scale. Specific examples of
sub-GUT models include warped extra dimensions~\cite{wed} and mirage mediation~\cite{mirage}.

Mirage mediation can occur when two sources of supersymmetry breaking
play off each other, such as moduli mediation based, e.g., on moduli stabilization 
as in~\cite{KKLT} and anomaly mediation~\cite{anom}. The relative contributions of 
each source of supersymmetry breaking can be parametrized by the strength of the
moduli mediation, $\alpha$, and allows one to interpolate between nearly pure
moduli mediation (large $\alpha$) and nearly pure anomaly mediation ($\alpha \to 0$).
For example, gaugino masses, $M_i$, can be written as $M_i = M_s (\alpha + b_i g_i^2)$
where $M_s$ is related to the gravitino mass in anomaly mediation ($m_{3/2} = 16 \pi^2 M_s$), 
and $b_i, g_i$ are the beta functions and gauge couplings. This leads to a renormalization scale, 
$M_{in} = M_{GUT} e^{-8\pi^2/\alpha}$ at which 
gaugino masses and soft scalar masses take unified values, although there is no 
physical threshold at $M_{in}$ in this model. We are not concerned here with the 
detailed origin of $M_{in}$, simply postulating that there is a scale below the GUT scale
where the supersymmetry breaking masses are unified.

Sub-GUT models are of particular phenomenological interest, since the reduction
in the amount of renormalization-group (RG) running below \Min, compared to
that below \MGUT\ in the CMSSM and related models, leads naturally
to SUSY spectra that are more compressed~\cite{sub-GUT}. These may offer extended
possibilities for `hiding' SUSY via suppressed \ETslash\ signatures, as well as
offering enhanced possibilities for different coannihilation processes. Other
possible effects of the reduced RG running include a stronger lower limit on $\mneu1$
because of the smaller hierarchy with the gluino mass, a stronger lower limit
on the DM scattering cross section because of a smaller hierarchy
between \mneu1\ and the squark masses, and greater tension between LHC searches
and a possible SUSY explanation of the measurement of \gmt~\cite{newBNL,g-2}, because of the
smaller hierarchies between the gluino and squark
masses and the smuon and \neu1\ masses.

We use the {\tt MasterCode} framework~\cite{mcold,mc9,mc10,mc11,mc12,mc-su5,mc-amsb,pMSSM11,KdV,mcweb} to
study these issues in the sub-GUT generalization of the CMSSM, which has
5 free parameters, comprising \Min\ as well as a common gaugino mass $m_{1/2}$,
a common soft SUSY-breaking scalar mass $m_0$, a common trilinear mixing parameter
{$A$} and the ratio of MSSM Higgs vevs $\tb$, {assuming that the Higgs mixing parameter
$\mu > 0$, as may be suggested by \gmt}~\footnote{We have also made an exploratory study
for $\mu < 0$ with a limited sample, finding quite similar results within the statistical uncertainties.}. Our global analysis {takes into account} the
relevant CMS searches for strongly-and electroweakly-interacting sparticles
with the full 2016 sample of $\sim 36$/fb of data at 13
TeV~\cite{cms_0lep-mt2,cms_1lep-MJ,sus-16-039},
{and also considers the available results of searches for long-lived charged particles~\cite{LLsearches1,LLsearches2}~\footnote{{The
ATLAS SUSY searches with $\sim 36$/fb of data at 13~TeV~\cite{ATLASWiki} yield similar constraints.}}.}
We also include a complete set
of direct DM searches published in 2017, including the PICO limit on the
spin-dependent scattering cross section, \ssd~\cite{PICO}, as well as the first XENON1T limit~\cite{XENON1T}
and the most recent PandaX-II limit~\cite{pandax} on the spin-independent scattering cross section,
\ssi, as well as the previous LUX search~\cite{lux16}. {We also include full sets of relevant electroweak
and flavour constraints.}

We find in our global sub-GUT analysis a distinct preference for $\MW \ll \Min \ll \MGUT$,
with {values of $\Min \sim 10^5$ or $\sim 10^8$ to $10^9 \gev$} being preferred by $\Delta \chi^2 \sim 3$
compared to the CMSSM {(where $\Min = \MGUT$)}. This preference is driven
principally by {the ability of the sub-GUT MSSM to accommodate a value of \bsdmm\
smaller than in the Standard Model {(SM)}, as preferred by the current data~\cite{NatureBsmm,ATLASBsmm,1703.05747}}.
{As discussed later, this effect can be traced to the different RGE evolution of $A_t$
in the sub-GUT model}, which enables it have a different sign
from that in the CMSSM.
The lower limits on strongly-interacting sparticles
are similar to those in the CMSSM, being largely determined by LHC searches.
The favoured DM scenario is that the LSP is a {Bino or}
Higgsino with $\mneu1 \sim 1 \tev$, with the cold DM being brought into
the cosmological range by {annihilation via heavy Higgs bosons $H/A$ and stop coannihilation, or chargino coannihilation}.
{In contrast to the CMSSM and pMSSM11,
the possibility that $\mneu1 \ll 1 \tev$ is strongly disfavoured in the sub-GUT model, so the LHC constraints have
insignificant impact. The same is true of the LHC searches for long-lived charged particles.}

The likelihood functions for fits with and without the \gmt\ constraint are quite
similar, reflecting the anticipated difficulty in accounting for the \gmt\ anomaly in
the sub-GUT MSSM. Encouragingly, we find a preference for a range of \ssi\
just below the current upper limits, and within the prospective sensitivities
of the LUX-Zeplin (LZ)~\cite{LZ} and XENONnT~\cite{XENONnT} experiments.

The outline of this paper is as follows. In Section~2 we summarize the
experimental and astrophysical constraints we apply. Since we follow exactly our
treatments in~\cite{pMSSM11}, we refer the interested reader there for details.
Then, in Section~3 we summarize the {\tt MasterCode} framework and how we
apply it to the sub-GUT models. Our results are presented in Section~4.
Finally, Section~5 summarizes our conclusions and discusses future
perspectives for the sub-GUT MSSM.

\section{Experimental and Astrophysical Constraints}

\subsection{Electroweak {and Flavour Constraints}}
\label{sec:others}

Our treatments of these constraints are identical to those in~\cite{pMSSM11}, which
were based on Table~1 of~\cite{mc-su5} with the updates listed in Table~2 of~\cite{pMSSM11}.
Since we pay particular attention in this paper to the impact on the sub-GUT parameter space
of the \gmt\ constraint~\cite{newBNL}, we note that we assume
\begin{equation}
a_{\mu}^{\rm EXP} - a_{\mu}^{\rm SM} \; = \; (30.2 \pm 8.8 \pm 2.0_{\rm {MSSM}})\times10^{-10}
\label{gmt}
\end{equation}
to be the possible discrepancy with SM calculations~\cite{g-2} that may be explained by SUSY.
{As we shall see, the \bsdmm\ measurement~\cite{NatureBsmm,ATLASBsmm,1703.05747}
plays an important role in indicating a preferred region of the sub-GUT parameter space.}

\subsection{Higgs Constraints}
\label{sec:Higgs}

In the absence of published results on the Higgs boson based on Run~2 data, we use
in this global fit the published results from Run~1~\cite{ATLAS+CMSH}, as
incorporated in the {\tt HiggsSignals} code~\cite{HiggsSignals}.

Searches for heavy MSSM Higgs bosons are incorporated using the {\tt HiggsBounds} code~\cite{HiggsBounds},
which uses the results from Run 1 of the LHC. We also include the
ATLAS limit from
$\sim 36$/fb of data from the LHC at 13 TeV~\cite{HA13}.

\subsection{Dark Matter Constraints and\\ Mechanisms}
\label{sec:DM}

\noindent
{\it\bf Cosmological density}

Since $R$-parity is conserved in the MSSM, the LSP is a candidate
to provide the cold DM (CDM). We assume that the LSP is the lightest
neutralino $\neu1$~\cite{EHNOS}, and that it dominates the total CDM density.
For the latter we assume the Planck 2015 value: $\Omega_{\rm CDM} h^2 = 0.1186 \pm 0.0020_{\rm EXP} \pm
0.0024_{\rm TH}$~\cite{Planck15}.\\

\noindent
{\it Density mechanisms}

As in~\cite{pMSSM11},
we use the following set of measures related to particle masses to indicate when specific mechanisms are
important for bringing $\Omega_{\rm CDM} h^2$ into the Planck 2015 range, {which have been
validated by checks using {\tt Micromegas}~\cite{MicroMegas}.}\\

\noindent
$\bullet${\it Chargino coannihilation}

This may be important if the $\neu1$ is not much lighter than the lighter chargino, $\cha1$,
and we introduce the following coannihilation measure:
\begin{equation}
{\rm chargino} {\rm ~coann.~:} \qquad \left(\frac{\mcha1}{\mneu{1}} - 1 \right)  \,<\,  0.25 \, .
\label{Inoco}
\end{equation}
We shade green in the 2-dimensional plots in Section~4 the parts of the
68 and 95\% CL regions where (\ref{Inoco}) is satisfied. \\

\noindent
$\bullet${\it Rapid annihilation via direct-channel $H/A$ poles}

We find that LSP annihilation is enhanced significantly if the following condition is satisfied:
\begin{equation}
H/A {\rm ~funnel~:} \qquad
\left|\frac{\MA}{\mneu{1}} - 2 \right| \,<\,  0.1 \, ,
\label{Bfunnel}
\end{equation}
and shade in blue the parts of the 68 and 95\% CL regions of the two-dimensional plots
in Section~4 where (\ref{Bfunnel}) is satisfied.\\

\noindent
$\bullet${\it Stau coannihilation}

We introduce the following measure for stau coannihilation:
\begin{equation}
{{\tilde \tau}} {\rm ~coann.~:} \qquad \left(\frac{m_{\tilde \tau_1}}{\mneu{1}} - 1 \right)  \,<\,  0.15 \, ,
\label{Ellco}
\end{equation}
{and shade in pink the corresponding area of the 68 and 95\% CL regions of the two-dimensional sub-GUT parameter planes.
We do not find regions where coannihilation with other charged slepton species, or with sneutrinos, is important.}\\

\noindent
$\bullet${\it Stop coannihilation}

{We introduce the following measure for stop coannihilation:
\begin{equation}
{{\tilde t_1}} {\rm ~coann.~:} \qquad \left(\frac{m_{\tilde t_1}}{\mneu{1}} - 1 \right)  \,<\,  0.15 \, ,
\label{stopco}
\end{equation}
and shade in yellow the corresponding area of the 68 and 95\% CL regions of the two-dimensional sub-GUT parameter planes.
We do not find regions where coannihilation with other squark species, or with gluinos, is important.}\\

\noindent
$\bullet${\it Focus-point region}

{The sub-GUT parameter space has a focus-point region where the DM annihilation rate is enhanced because the LSP
$\neu1$ has an enhanced Higgsino component as a result of near-degeneracy in the neutralino
mass matrix. We introduce the following measure to characterize this possibility:
\begin{equation}
{\rm focus~point~:} \qquad \left( \frac{\mu}{\mneu1} \right) - 1  \,<\, 0.3 \, ,
\label{focuspoint}
\end{equation}
and shade in cyan the corresponding area of the 68 and 95\% CL regions of the two-dimensional sub-GUT parameter planes.}\\

\noindent
$\bullet${\it Hybrid regions}

{In addition to regions where one of the above DM mechanisms is dominant, there are also various `hybrid' regions where
more than one mechanism is important. These are indicated in the two-dimensional planes below
by shadings in mixtures of the `primary' colours above, which are shown in the corresponding figure legends.
For example, there are prominent regions where both chargino coannihilation and direct-channel
$H/A$ poles are important, whose shading is darker than the blue of regions where $H/A$ poles are dominant.}\\

\noindent
{\it\bf Direct DM searches}

We apply the constraints from direct searches for weakly-interacting dark
matter particles via both spin-independent and -dependent scattering on nuclei. In addition to the
2016 LUX constraint on \ssi~\cite{lux16}, we use the 2017 XENON1T~\cite{XENON1T} and
PandaX-II~\cite{pandax} constraints on the spin-independent DM scattering,
which we combine in a joint two-dimensional likelihood function in the $(\mneu1, \ssi)$ plane.
We estimate the spin-independent nuclear scattering matrix element assuming
$\sigma_0 = 36 \pm 7 \mev$ and $\Sigma_{\pi N} = 50 \pm 7 \mev$ as in~\cite{EOSavage,SSARD}~\footnote{We
note {that a recent analysis using covariant baryon chiral perturbation theory yields a very similar
central value of $\Sigma_{\pi N}$~\cite{NewSigmaTermEstimate}. However, we emphasize that there
are still considerable uncertainties in the estimates
of $\sigma_0$ and $\Sigma_{\pi N}$ and hence the $\langle N | {\bar s} s | N \rangle$ matrix element
that is important for \ssi~\cite{larger}.}},
and the spin-dependent nuclear scattering matrix element assuming
$\Delta u = + 0.84 \pm 0.03, \Delta d = - 0.43 \pm 0.03$ and $\Delta s = - 0.09 \pm 0.03$~\cite{EOSavage,SSARD}.
We implement the recent PICO~\cite{PICO} constraint on the spin-dependent dark
matter scattering cross-section on protons, \ssd.\\

\noindent
{\it\bf Indirect astrophysical searches for DM} \\
 {As discussed in~\cite{pMSSM11}, there are considerable uncertainties in the use of IceCube data~\cite{IceCube}
 to constrain \ssd\ and, as we discuss below, the global fit yields a prediction that lies well below the current PICO~\cite{PICO}
 constraint on \ssd\ and the current IceCube sensitivity, so we do not include the IceCube data
in our global fit}.\\

\subsection{13 TeV LHC Constraints}
\label{sec:LHC}

\noindent
{\it Searches for gluinos and squarks}

We {implement} the CMS simplified model searches with $\sim 36$/fb of data at 13~TeV for
events with jets and $\ETslash$ but no leptons~\cite{cms_0lep-mt2}
and for events with jets, $\ETslash$ and a single lepton~\cite{cms_1lep-MJ}, using
the {\tt Fastlim} approach~\cite{Fastlim}. We use~\cite{cms_0lep-mt2} to constrain $\gl \gl \to [ q {\bar q} \neu1 ]^2$ and
$[ b {\bar b} \neu1 ]^2$, and ${\tilde q} {\tilde {\bar q}} \to [ q \neu1 ] [{\bar q} \neu1 ]$, and
use~\cite{cms_1lep-MJ} to constrain ${\tilde g} {\tilde g} \rightarrow [ t {\bar t} \neu1 ]^2$.
Details are given in~\cite{pMSSM11}.\\

\noindent
{\it Stop and sbottom searches}

We {also implement} the CMS simplified model searches with $\sim 36$/fb of data at 13~TeV
in the jets + 0 \cite{cms_0lep-mt2} and 1 \cite{cms_1lep-MJ} lepton final states
to constrain ${\tilde t_1} {\tilde {\bar t}}_1 \to [ t \neu1 ] [{\bar t} \neu1 ]$, $[ c \neu1 ] [{\bar c} \neu1 ]$
in the compressed-spectrum region, $[ b W^{+} \neu1 ] [{\bar b} W^{-} \neu1 ]$ via $\cha1$
intermediate states and
${\tilde b_1} {\tilde {\bar b}}_1 \to [ b \neu1 ] [{\bar b} \neu1 ]$, again using {\tt Fastlim}
as described in detail in~\cite{pMSSM11}.\\

\noindent
{\it Searches for electroweak inos}

We {also consider} the CMS searches for electroweak inos
in multilepton final states with $\sim 36$/fb of data at 13~TeV~\cite{sus-16-039},
constraining $\cha1 \neu2 \to [W \neu1] [Z \neu1], 3 \ell^\pm + 2 \neu1$
via ${\tilde \ell}^\pm/\tilde \nu$ intermediate states, and $3 \tau^\pm + 2 \neu1$
via ${\tilde \tau}^\pm$ intermediate states using {\tt Fastlim}~\cite{Fastlim} as described in~\cite{pMSSM11}.
These analyses can also be used to constrain the production of
electroweak inos in the decays of coloured sparticles, since
these searches do not impose conditions on the number of jets.
{However, as we discuss below, in the sub-GUT model the above-mentioned searches for
strongly-interacting sparticles impose such strong limits on the $\mneu1$
and $\mcha1$ that the searches for electroweak inos do not have significant impact
on the preferred parameter regions.}\\

\noindent
{\it Searches for long-lived or stable charged particles}

{We also consider {\it a posteriori} the search for long-lived charged particles published in~\cite{LLsearches1},
which are sensitive to lifetimes $\gtrsim$~ns, and the search for massive charged particles
that escape from the detector without decaying~\cite{LLsearches2}. However, these also
do not have significant impact on the preferred parameter regions, as we discuss in detail below,
{and are not included in our global fit}.}\\

\section{Analysis Framework}
\label{sec:framework}

\subsection{Model Parameters}
\label{sec:sub-GUT}

As mentioned above, the five-dimensional sub-GUT MSSM parameter space
we consider in this paper comprises a gaugino mass parameter $m_{1/2}$,
a soft SUSY-breaking scalar mass parameter $m_0$ and a trilinear soft
SUSY-breaking parameter $A_0$ that are assumed to be universal at some
input mass scale \Min, and the ratio of MSSM Higgs vevs, $\tb$.
Table~\ref{tab:ranges} displays the ranges of these parameters sampled in our analysis,
as well as their divisions into segments, which define boxes in the five-dimensional
parameter space.

%%%%%%%%%%%%%%%%%%%%%% T A B L E %%%%%%%%%%%%%%%%%%%%%%%%%%%%%%%%%%%%%%%%%
\begin{table}[htb!]
\begin{center}
%\hspace{-0.5cm}
\resizebox{0.475\textwidth}{!}{
\begin{tabular}{c c c}
Parameter   &  \; \, Range      & \# of  \\
            &             & segments   \\
\hline \hline
	\Min      &  $(10^{3}, 10^{16} ) \gev$  & 6 \\
$m_{1/2}$       &  $(0, 6) \tev$  & 2 \\
$m_0$       &  $(0, 6) \tev$  & 2 \\
$A_0$        &  $(- 15,  10) \tev$  & 2 \\
\tb         &  ( 1  , 60)      & 2 \\
\hline
\multicolumn{2}{c}{Total \# of boxes}& 96 \\
\hline
\end{tabular}}
\caption{\it The ranges of the sub-GUT MSSM parameters sampled, together with the numbers of
segments into which they are divided, together with the total number of sample boxes shown in the last row.
This sample is for positive values of the Higgs mixing parameter, $\mu$. As already noted, a smaller sample
for $\mu < 0$ gives similar results.
Note that our sign convention for $A$ is opposite to that used in {\tt SoftSusy}~\cite{Allanach:2001kg}.}
\label{tab:ranges}
\end{center}
\end{table}
%%%%%%%%%%%%%%%%%%%%%% T A B L E %%%%%%%%%%%%%%%%%%%%%%%%%%%%%%%%%%%%%%%%

\subsection{Sampling Procedure}
\label{sec:sampling}

We sample the boxes in the five-dimensional sub-GUT MSSM parameter space
using the {\tt MultiNest} package~\cite{multinest}, choosing for each box a prior such that
80\% of the sample has a flat distribution within the nominal box, and 20\% of the sample is in
normally-distributed tails extending outside the box. This eliminates features associated with the
boundaries of the 96 boxes, by providing a smooth overlap between them. {In total, our
sample includes $\sim 112$ million points with $\Delta \chi^2 < 100$.}

\subsection{The {\tt MasterCode}}
\label{sec:MasterCode}

The {\tt MasterCode} framework~\cite{mcold,mc9,mc10,mc11,mc12,mc-su5,mc-amsb,pMSSM11,KdV,mcweb},
interfaces and combines consistently various private and public codes
using the SUSY Les Houches Accord (SLHA)~\cite{SLHA}. This analysis
uses the following codes: {\tt SoftSusy~3.7.2}~\cite{Allanach:2001kg}
for the MSSM spectrum, {\tt FeynWZ}~\cite{Svenetal} for the electroweak precision observables,
{\tt SuFla}~\cite{SuFla} and {\tt SuperIso}~\cite{SuperIso} for flavour observables,
{\tt FeynHiggs~2.12.1-beta}~\cite{FeynHiggs} for \gmt\ and calculating
Higgs properties, {\tt HiggsSignals~1.4.0}~\cite{HiggsSignals} and {\tt HiggsBounds~4.3.1}~\cite{HiggsBounds}
for experimental constraints on the Higgs sector,
{\tt Micromegas~3.2}~\cite{MicroMegas} for the DM
relic density, {\tt SSARD}~\cite{SSARD} for the spin-independent and -dependent elastic scattering cross-sections
\ssi\ and \ssd, {\tt SDECAY~1.3b}~\cite{Sdecay} for sparticle branching ratios and (as already mentioned)
{\tt Fastlim}~\cite{Fastlim} to recast LHC 13~TeV constraints on events with $\ETslash$.

\section{Results}

\subsection{{Results for $\mathbf{\Min, m_0}$ and $\mathbf{m_{1/2}}$}}

%%%%%%%%%%%%%%%%%%%%%%%%% F I G U R E %%%%%%%%%%%%%%%%%%%%%%%%%%%%%%%%%%%%%%%%
\begin{figure*}[]
\centering
\vspace{-3cm}
\includegraphics[width=0.475\textwidth]{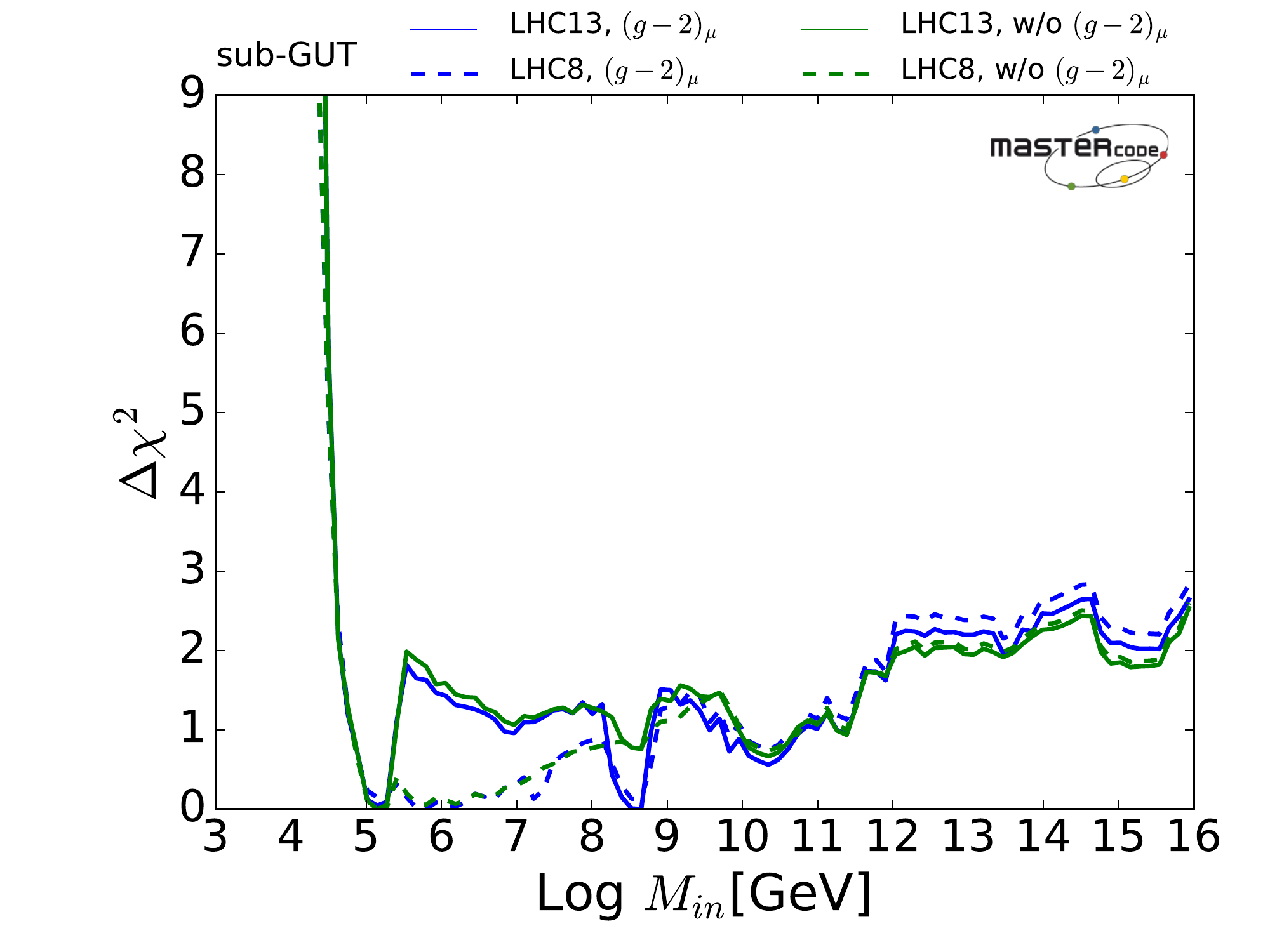}
\includegraphics[width=0.475\textwidth]{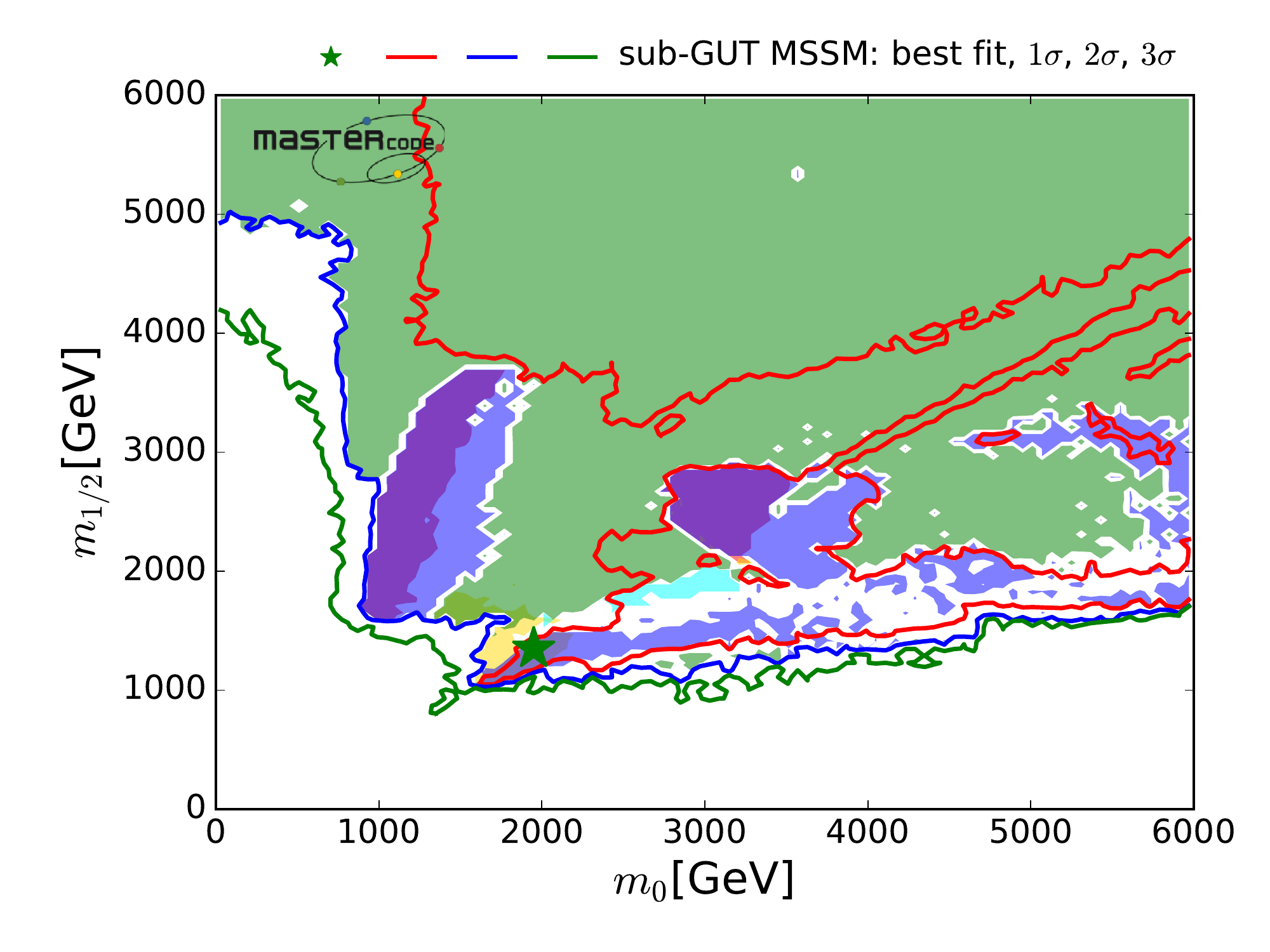} \\
\vspace{-0.3cm}
\includegraphics[width=0.475\textwidth]{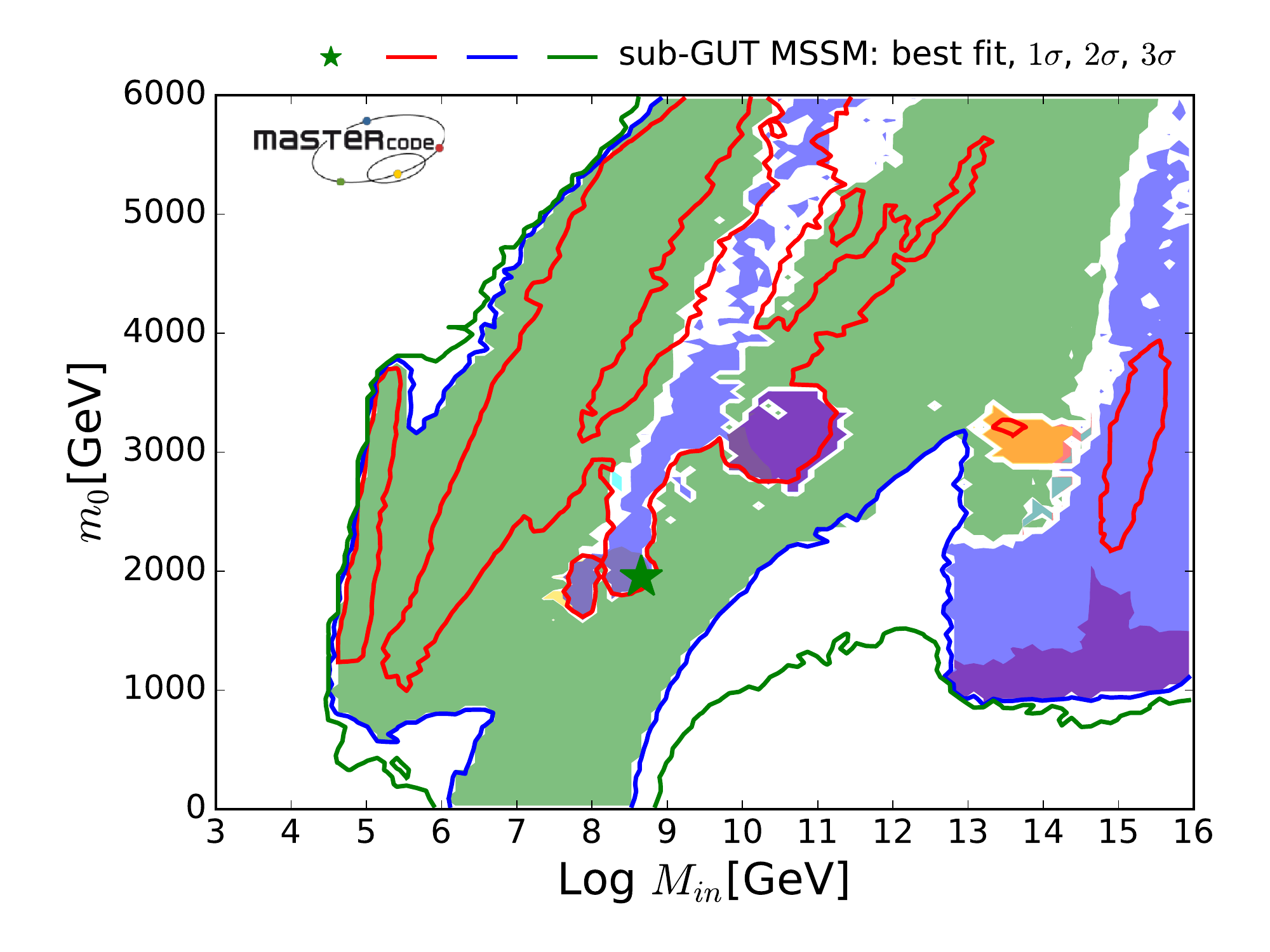}
\includegraphics[width=0.475\textwidth]{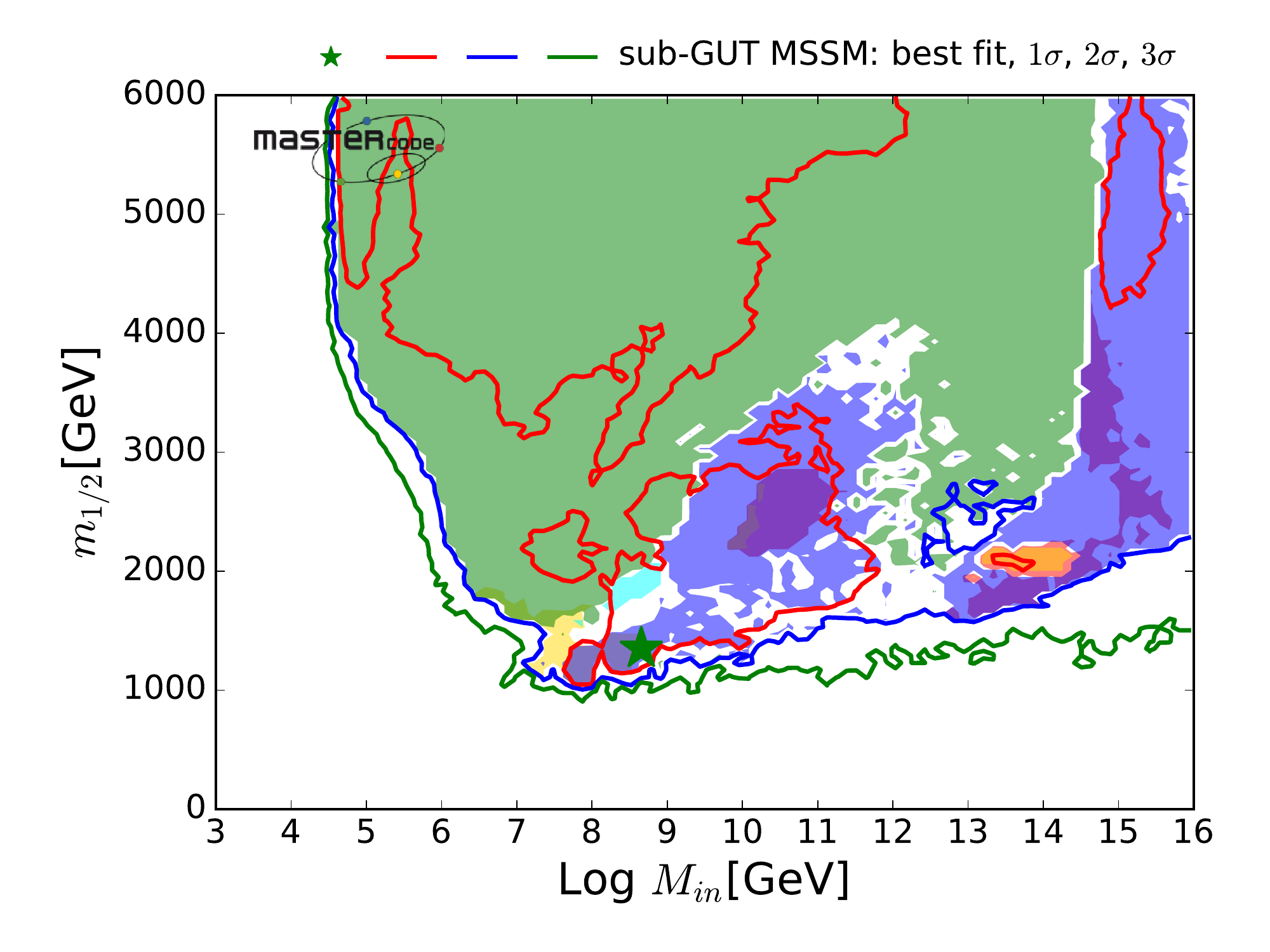} \\
\vspace{-0.3cm}
\includegraphics[width=0.475\textwidth]{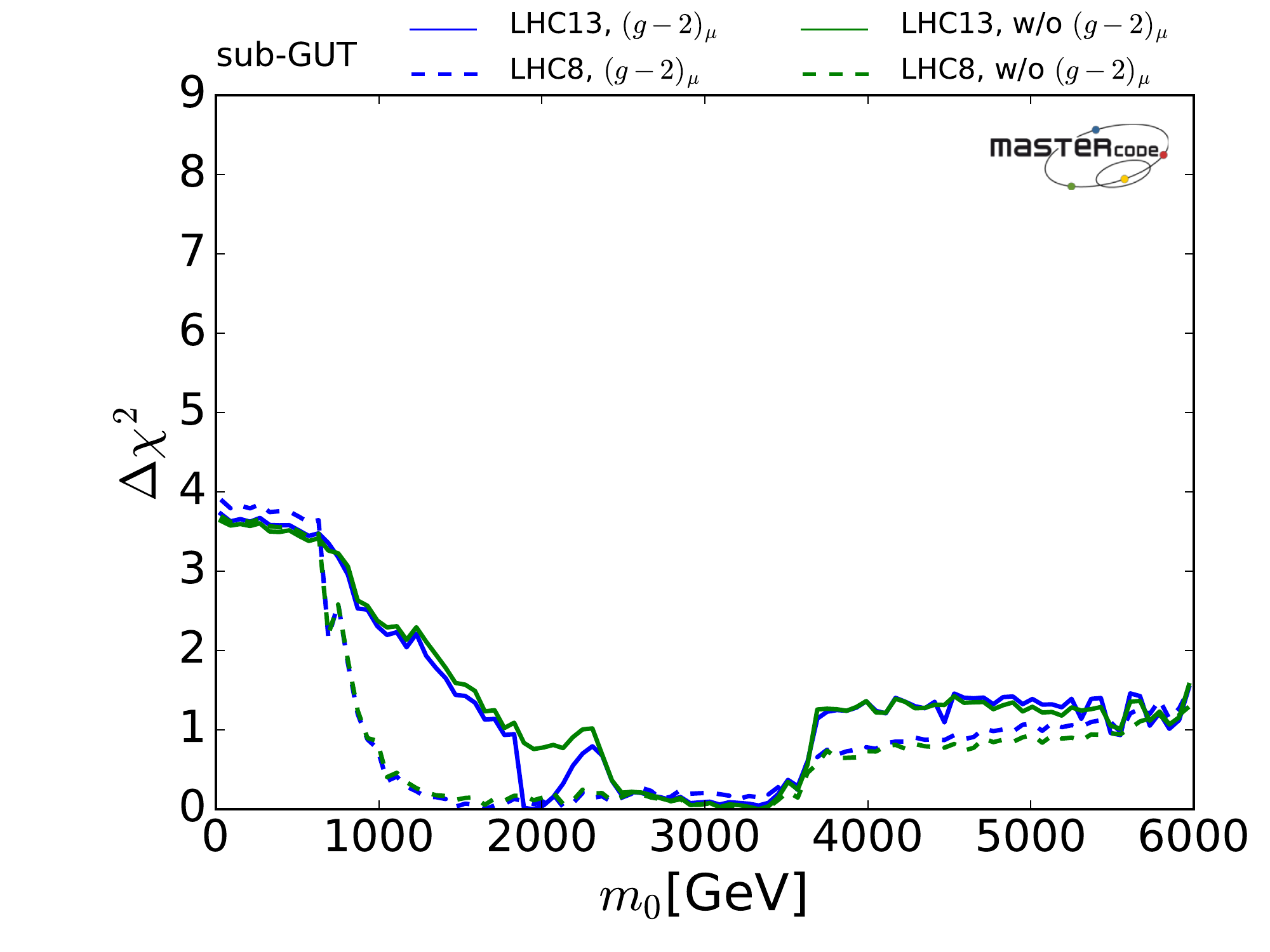}
\includegraphics[width=0.475\textwidth]{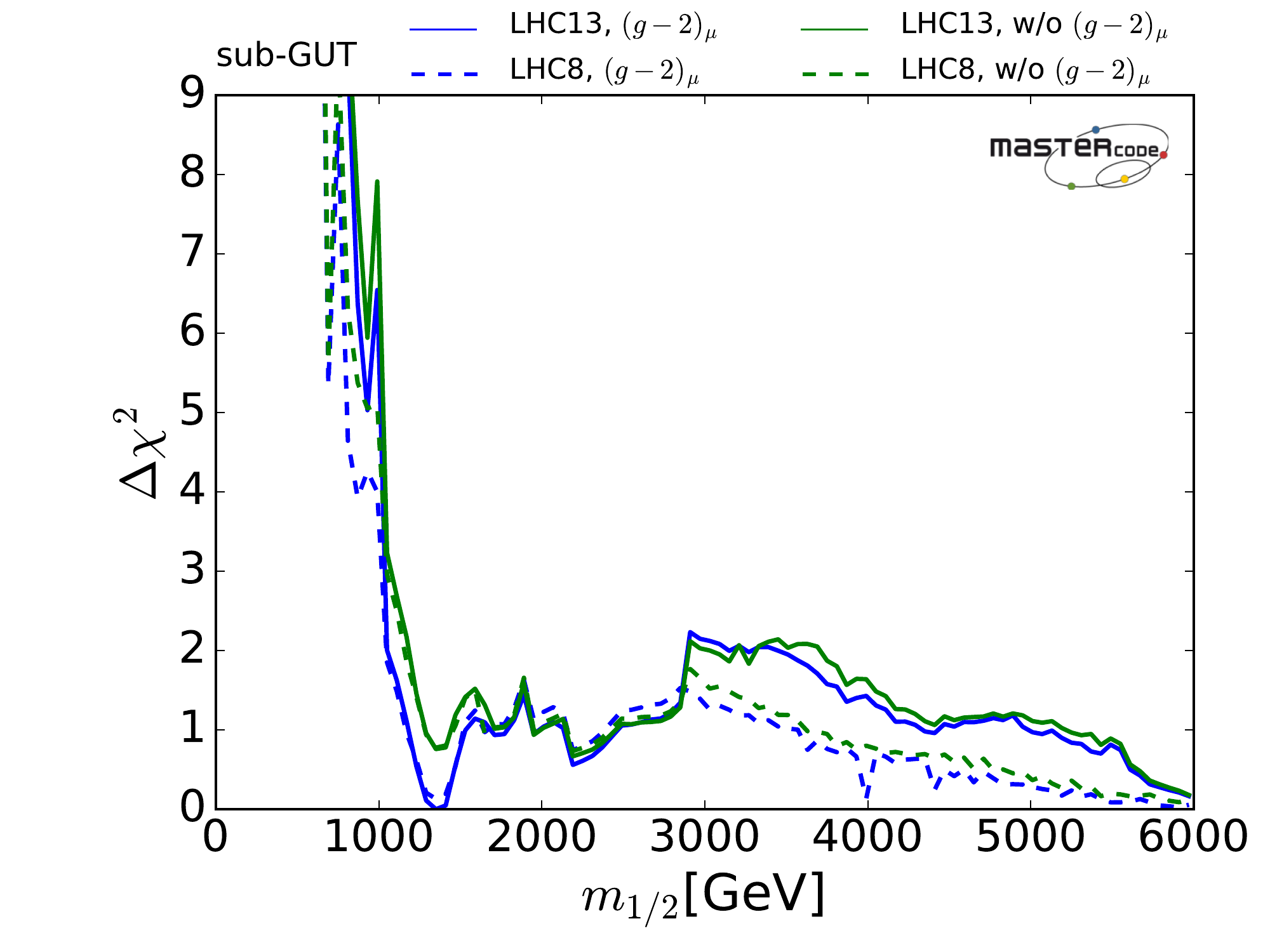} \\
\vspace{-0.2cm}
\includegraphics[width=0.9\textwidth]{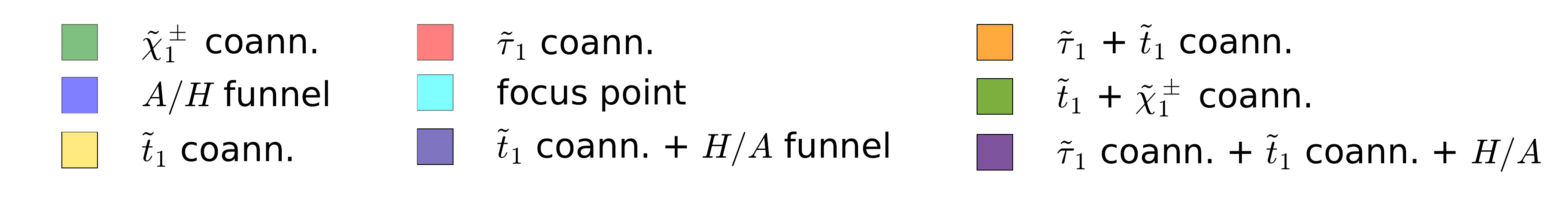} \\
\vspace{-0.7cm}
\caption{\it Profile likelihood functions in the sub-GUT MSSM.
Top left: One-dimensional profile likelihood function for \Min.
Top right: Two-dimensional projection of the likelihood function in the $(m_0, m_{1/2})$ plane.
Middle left: Two-dimensional projection of the likelihood function in the $(\Min, m_{0})$ plane.
Middle right: Two-dimensional projection of the likelihood function in the $(\Min, m_{1/2})$ plane.
Bottom left: One-dimensional profile likelihood function for $m_0$.
Bottom right: One-dimensional profile likelihood function for $m_{1/2}$.
Here and in subsequent one-dimensional plots, the {solid lines include the constraints from
$\sim 36$/fb of LHC data at 13~TeV and the dashed lines drop them,
and the {blue} lines include \gmt, whereas the {green} lines
drop these constraints}. Here and in subsequent two-dimensional plots, the red (blue) (green)
contours are boundaries of the 1-, 2- and 3-$\sigma$ regions, and the shadings correspond to the
DM mechanisms indicated in the legend.}
\label{fig:Minm0m12}
\end{figure*}
%%%%%%%%%%%%%%%%%%%%%%%%% F I G U R E %%%%%%%%%%%%%%%%%%%%%%%%%%%%%%%%%%%%%%%%

{The top left panel of Fig.~\ref{fig:Minm0m12} displays the one-dimensional profile $\chi^2$
likelihood function for $\Min$, as obtained under various assumptions~\footnote{{This and
subsequent figures were made using {\tt Matplotlib}~\cite{matplotlib}, unless otherwise noted}.}. In this and subsequent
one-dimensional plots, the solid lines represent the results of a fit including results from $\sim 36$/fb
of data from the LHC at 13 TeV (LHC13), whereas the dashed lines omit these results, and the {blue} lines include
\gmt, whereas the {green} lines are obtained when this constraint is dropped.

We observe in the top left panel of Fig.~\ref{fig:Minm0m12} a preference for {$\Min \simeq 4.2 \times 10^8 \gev$
when the LHC 13-TeV data and \gmt\ are both included (solid blue line), falling to $\simeq 5.9 \times 10^5 \gev$ when
the 13-TeV data are dropped (dashed blue line). There is little difference between the global $\chi^2$ values at these two minima, but
values of $\Min < 10^5 \gev$ are strongly disfavoured.} {The rise in $\Delta \chi^2$ when $\Min$ increases to $\sim 10^6 \gev$
and the LHC 13-TeV data are included (solid lines) is largely due to the contribution of \bsdmm. At lower $\Min$, the $H \to \tau^+ \tau^-$ constraint allows
a larger value of $\tb$, which leads (together with an increase in  the magnitude of $A$) to greater negative interference in the supersymmetric
contribution to \bsdmm, as preferred by the data.}

{For both fits including the LHC 13-TeV data {(solid lines)}, the $\Delta \chi^2$ function $\sim 1$
for most of the range $\Min \in (10^5,10^{11}) \gev$, apart from localized dips, whereas
$\Delta \chi^2$ rises to $\gtrsim 2$ for $\Min \gtrsim 10^{12} \gev$.
As already mentioned and discussed in more detail later, the reduction in the global $\chi^2$ function
for {$\Min \lesssim 10^{12} \gev$} arises because for these values of $\Min$ the sub-GUT model can accommodate better
the measurement of \bsdmm, whose central experimental value is somewhat lower than in the SM.}

{When the \gmt\ constraint is dropped, as shown by the green lines in {top left panel of} Fig.~\ref{fig:Minm0m12},
there is a minimum of $\chi^2$ around $\Min \simeq 1.6 \times 10^5 \gev$, whether the LHC 13-TeV constraint
is included, or not. The values of the other input parameters at the best-fit points with and without these data are
also very similar, as are the values of $\Delta \chi^2$. On the other hand, the values of $\Delta \chi^2$ for $\Min \in (10^5, 10^{8}) \gev$
are generally smaller when the LHC 13-TeV constraints are dropped, {the principal effect being due to the
$H/A \to \tau^+ \tau^-$ constraint.}

In contrast, when $\Min \gtrsim 10^{9} \gev$
the $\Delta \chi^2$ function {in the top left panel of Fig.~\ref{fig:Minm0m12}} is
quite similar whether the LHC 13-TeV and \gmt\ constraints
are included or not, though $\Delta \chi^2 \gtrsim 0.5$ {lower} when the \gmt\ constraint is dropped,
as seen by comparing the green and blue lines. {This is because the tension between
\gmt\ and LHC data is increased when $M_3/M_1$ is reduced, as occurs because of the smaller
RGE running when $\Min < \MGUT$. Conversely, lower $\Min$ is relatively more favoured when \gmt\
is dropped, leading to this increase in $\Delta \chi^2$ at high $\Min$ though the total $\chi^2$ is reduced.}

{We list in Table~\ref{tab:inputs} the parameters of the best-fit points when we drop
one or both of the \gmt\ and LHC13 constraints, as well as the values of the global $\chi^2$ function
at the best-fit points. We see that the best-fit points without \gmt\
are very similar with and without the LHC 13-TeV constraint. On the other hand, the best-fit points
with \gmt\ have quite different values of the other input parameters, as well as larger values of $\Min$,
particularly when the LHC 13-TeV data are included.}

The top right panel of Fig.~\ref{fig:Minm0m12} displays the $(m_0, m_{1/2})$
plane when the \gmt\ and LHC13 constraints are applied. Here and in subsequent planes,
the {green star indicates the best-fit point, whose input parameters are
listed in Table~\ref{tab:inputs}: {it lies in a hybrid {stop} coannihilation and
rapid $H/A$ annihilation region.}

{This parameter plane and others} in Fig.~\ref{fig:Minm0m12} and subsequent figures
also display the 68\% CL (1-$\sigma$), 95\% CL (2-$\sigma$)
and 99.7\% (3-$\sigma$) contours in the fit including both \gmt\ and the LHC13 data
as red, blue and green lines}, respectively. We note,
here and subsequently, that the green 3-$\sigma$ contours are generally close to the
blue 2-$\sigma$ contours, indicating a relatively rapid increase in $\chi^2$,
{and that} the $\chi^2$ function is relatively flat for $m_0, m_{1/2} \gtrsim 1 \tev$. The regions
inside the 95\% CL contours are colour-coded according to the dominant DM mechanisms,
as shown in the legend beneath Fig.~\ref{fig:Minm0m12}~\footnote{{In regions left uncoloured
none of the DM mechanism dominance criteria are satisfied.}}. Similar results for this and other planes
are obtained when either or both of the \gmt\ and LHC13 constraints are dropped.

{We see that chargino coannihilation is important in the upper part of the $(m_0, m_{1/2})$ plane
shown in the top right panel of Fig.~\ref{fig:Minm0m12},
but rapid annihilation via the $H/A$ bosons becomes important for lower $m_{1/2}$,
often hybridized with other mechanisms including stop and stau coannihilation.
We also note smaller regions with $m_{1/2} \sim 1.5$ to $3 \tev$ where stop
coannihilation and focus-point mechanisms are dominant.}

\begin{table*}[htb!]
\centering
	\begin{tabular}{c c c c c c c }
		& $m_0$ [GeV] & $m_{1/2}$ [GeV] & $A_0$ [GeV] & $\tan\beta$ & $M_{in}$ [GeV] & $\chi^2$ \\
	\hline \hline
	\multicolumn{7}{c}{With \gmt} \\
	\hline
		With 13-TeV & 1940 & 1370 & - 6860 & 36 & $4.1 \times 10^8$ & 99.56 \\
		Without 13-TeV & 1620 & 6100 & - 8670 & 45 & $5.7 \times 10^5$ & 99.38 \\
	\hline
	\multicolumn{7}{c}{Without \gmt} \\
	\hline
		With 13-TeV & 3550 & 6560 & - 14400 & 45 & $1.6 \times 10^5$ & 88.73 \\
		Without 13-TeV & 3340 & 6390 & - 14260 & 45 & $1.6 \times 10^5$ & 88.67 \\
	\hline
	\end{tabular}
		\caption{\it {Values of the sub-GUT input parameters at the best-fit points
		with and without \gmt\ and the LHC 13-TeV data.}}
		\label{tab:inputs}
\end{table*}

{The middle left panel of Fig.~\ref{fig:Minm0m12} shows the {corresponding} $(\Min, m_0)$ plane,
where we see a significant positive correlation between the variables that is particularly noticeable in the 68\% CL
region. In most of this and the 95\% CL region with $\Min \lesssim 10^{13} \gev$ the relic LSP density is controlled
by chargino coannihilation, though with patches where rapid annihilation via the $A/H$
bosons is important, partly in hybrid combinations.
In contrast, the $(\Min, m_{1/2})$ plane shown in the middle right panel of Fig.~\ref{fig:Minm0m12}
does not exhibit a strong correlation between the variables. We see again the importance
of chargino coannihilation, with the $A/H$ mechanism becoming more important for lower $m_{1/2}$
and larger $\Min$, and for all values of $m_{1/2}$ for $\Min \gtrsim 10^{14} \gev$.}

{Also visible in the middle row of planes are small regions
with $\Min \sim 10^{13}$ to $10^{14} \gev$ where stau coannihilation is dominant,
partly hybridized with stop coannihilation.
The reduction in the global $\chi^2$
function for $\Min \lesssim 10^{12} \gev$ visible in the top left panel of Fig.~\ref{fig:Minm0m12}
is associated with the 68\% CL regions in this range of $\Min$ visible in the two middle planes
of Fig.~\ref{fig:Minm0m12}.}

The one-dimensional profile likelihood functions for $m_0$ and $m_{1/2}$
are shown in the bottom panels of Fig.~\ref{fig:Minm0m12}. We note once again the
similarities between the results with/without \gmt\ {(blue/green lines)}
and the LHC13 constraints {(solid/dashed lines)}.
The flattening of the $\chi^2$ function for $m_0$ at small values reflects the
extension to $m_0 = 0$ of the 95\% CL region in the top right panel of
Fig.~\ref{fig:Minm0m12}. On the other hand,  the $\chi^2$ function for $m_{1/2}$
rises rapidly at small values, {reflecting the close spacing of the 95 and 99.7\%
CL contours for $m_{1/2} \sim 1 \tev$ seen in the same plane.} The impact of the LHC13 constraints is visible in
the differences between the solid and dashed curves at small {$m_{0}$, in particular. The
\gmt\ constraint has less impact, as shown by the smaller differences between the green
and blue curves. We see that the $\chi^2$ function for $m_0$ rises by $\gtrsim 1$ at large mass values,
whereas that for $m_{1/2}$ falls monotonically at large values.
The $\chi^2$ function for $m_{1/2}$ exhibits a local maximum at $m_{1/2}\sim 3 \tev$,
which corresponds to the separation between the two 68\% CL regions in
the top right plane of Fig.~\ref{fig:Minm0m12}. These are dominated by chargino
coannihilation (larger $m_{1/2}$, green shading) and by rapid annihilation
via $A/H$ bosons (smaller $m_{1/2}$, blue shading) and other mechanisms, respectively.}

\subsection{Squarks and gluinos}

%%%%%%%%%%%%%%%%%%%%%%%%% F I G U R E %%%%%%%%%%%%%%%%%%%%%%%%%%%%%%%%%%%%%%%%
\begin{figure*}[h]
\centering
\includegraphics[width=0.475\textwidth]{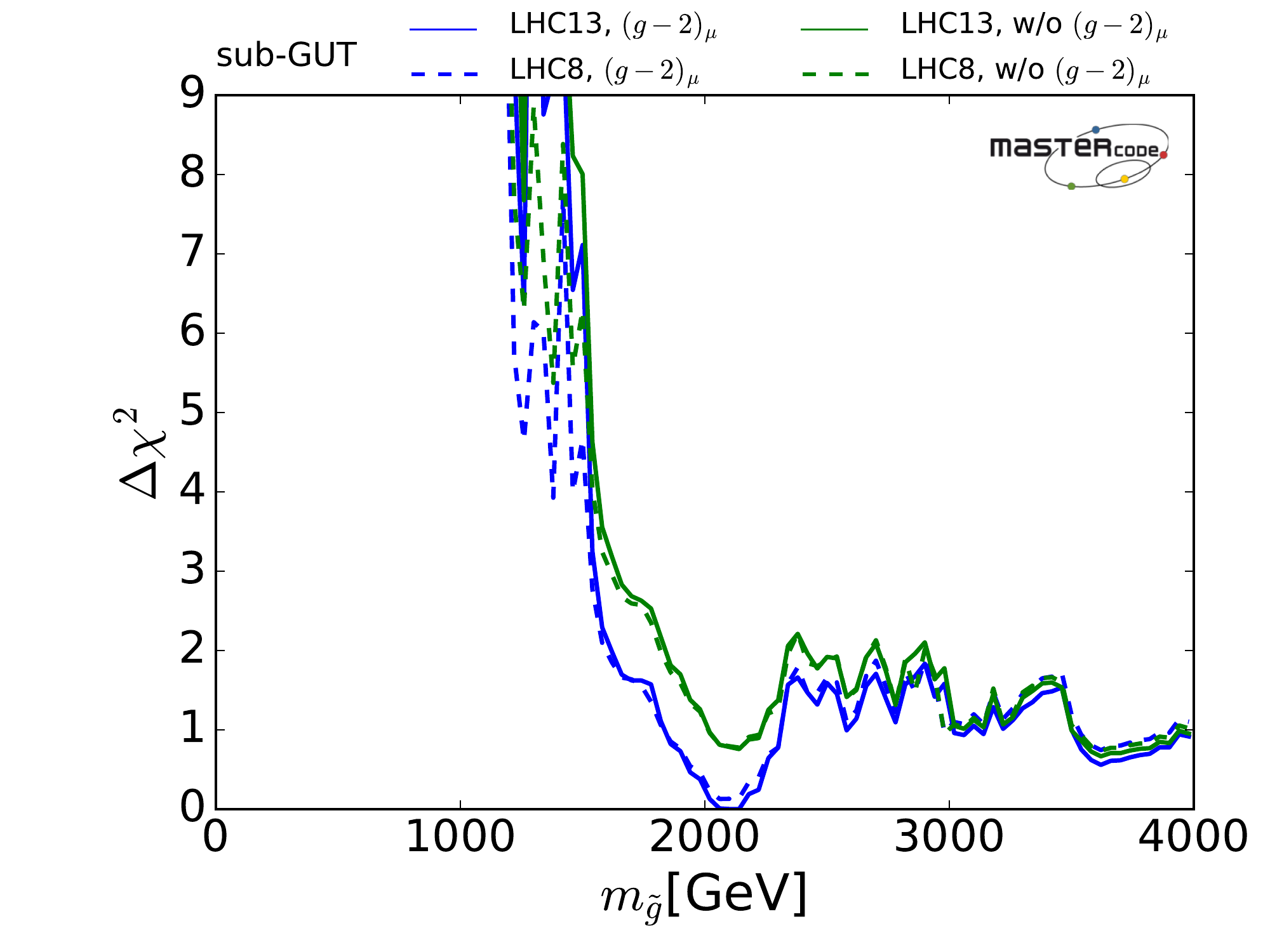}
\includegraphics[width=0.475\textwidth]{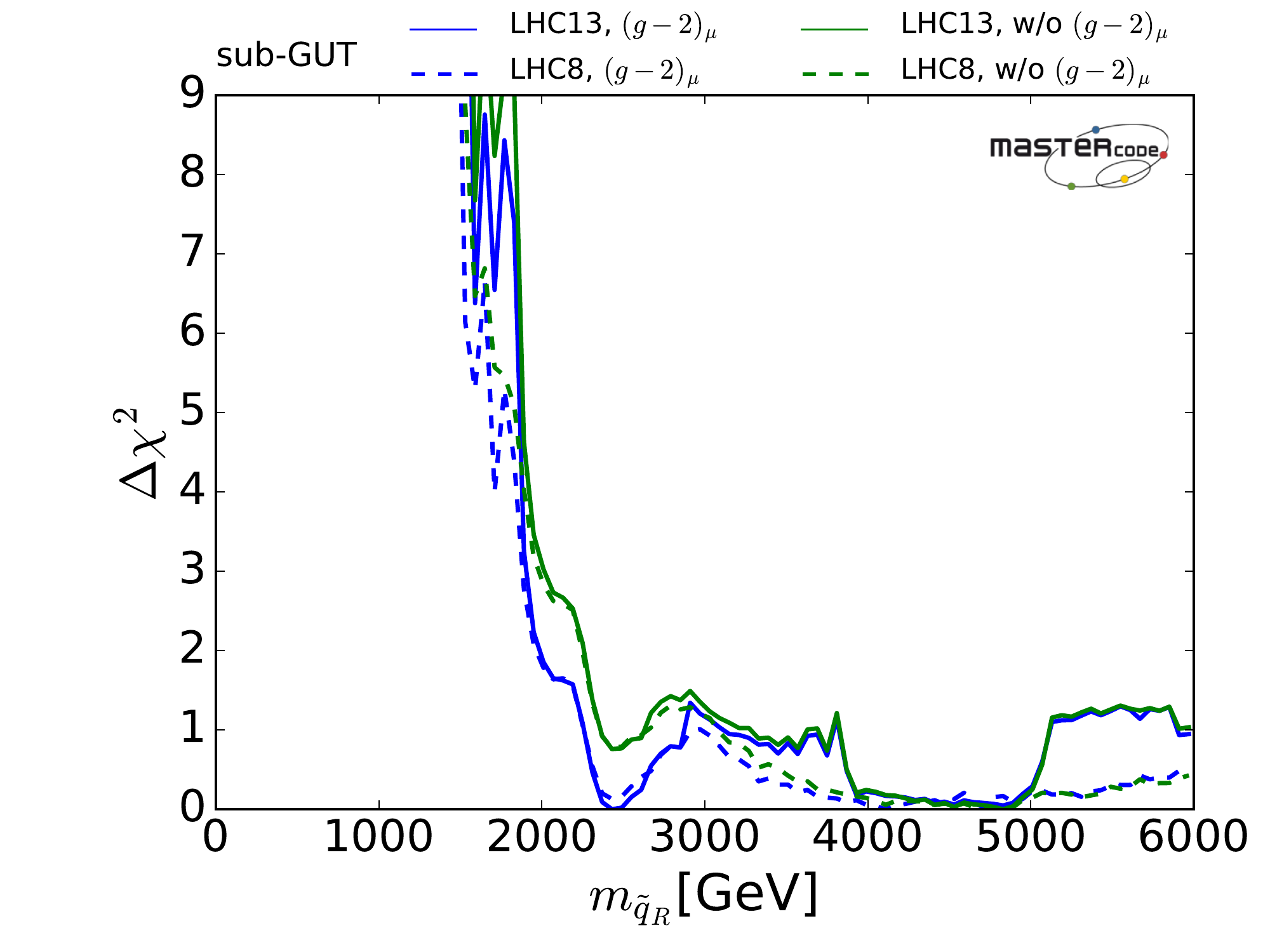} \\
\includegraphics[width=0.475\textwidth]{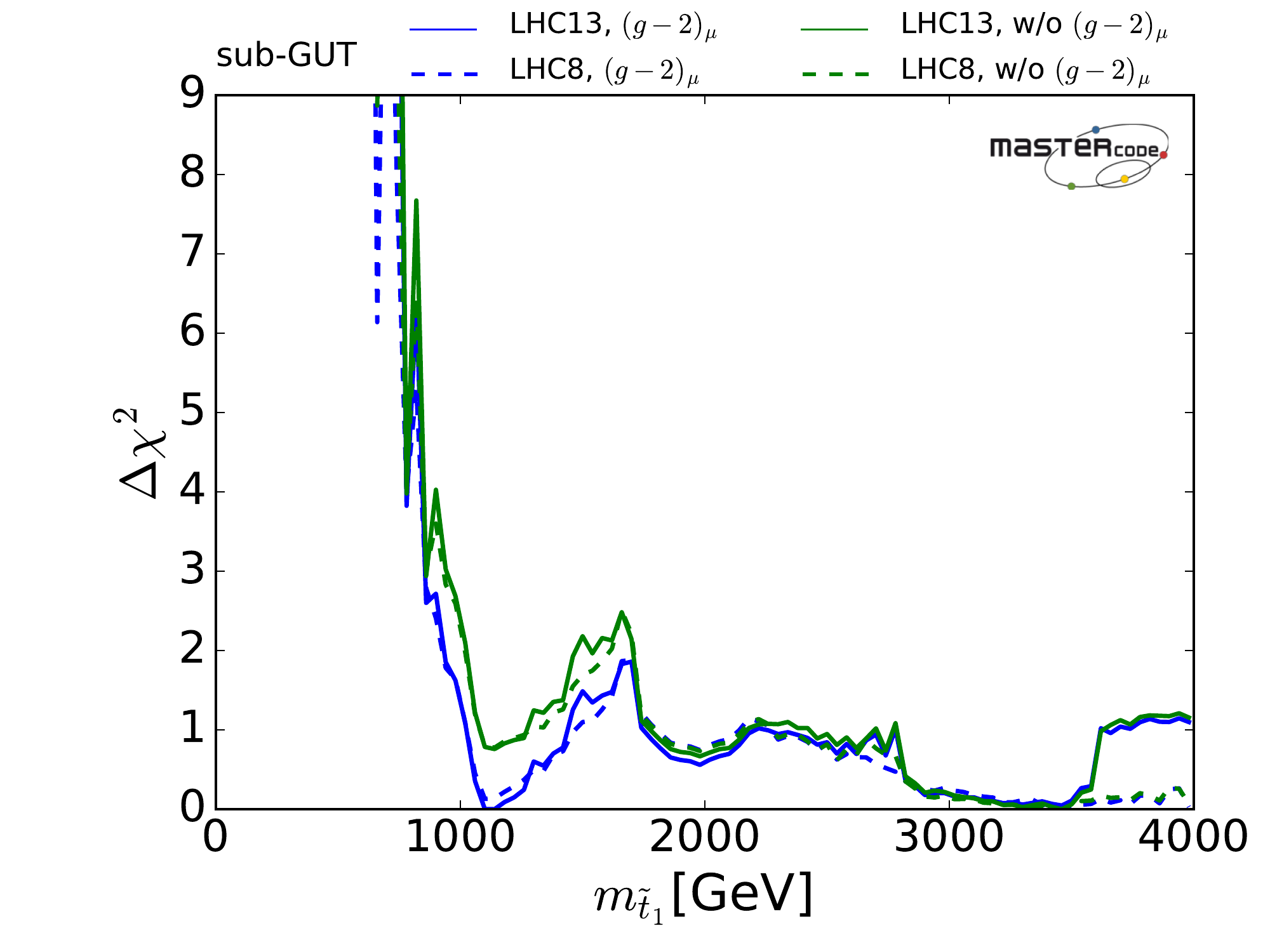}
\includegraphics[width=0.475\textwidth]{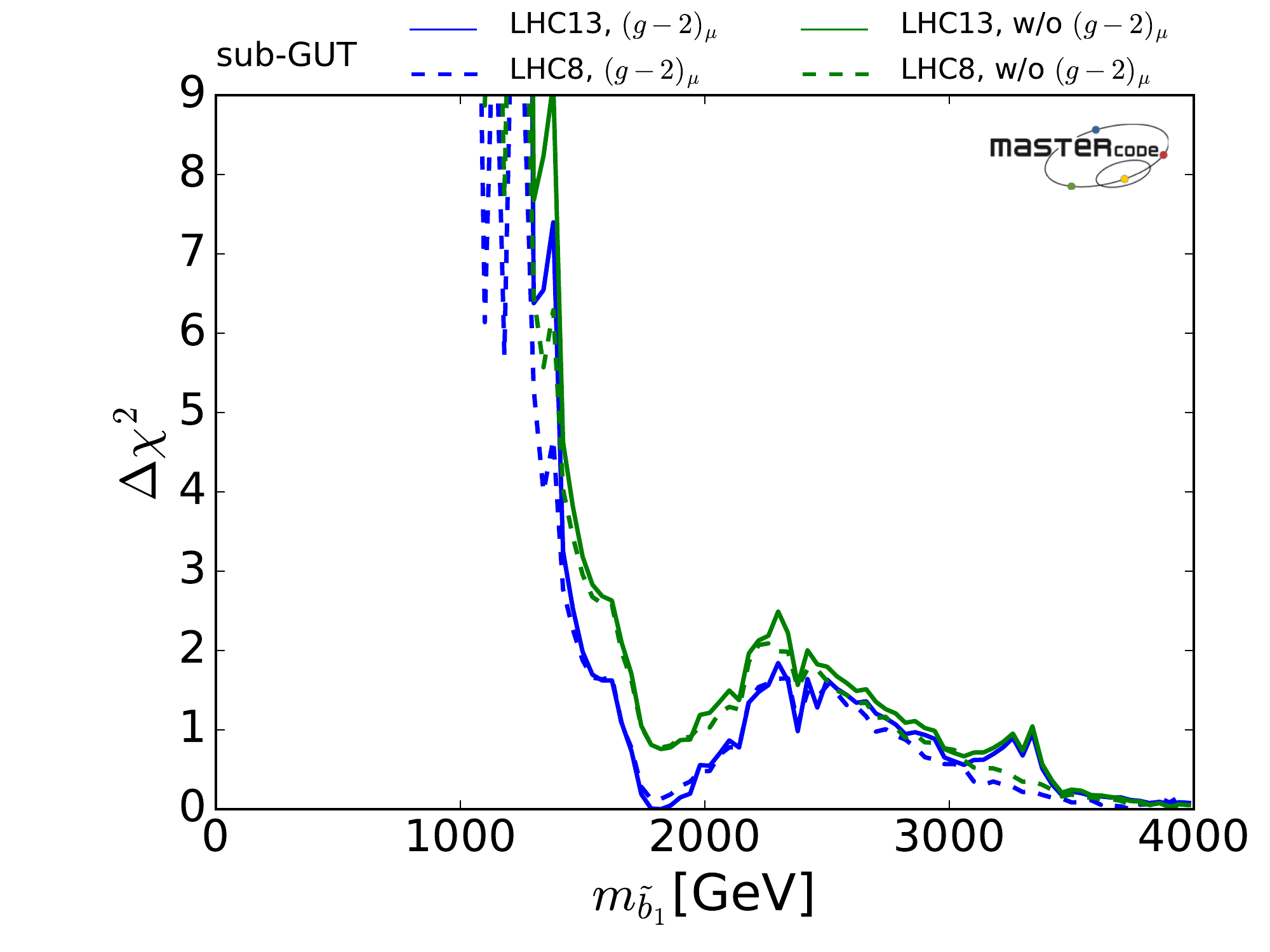} \\
\caption{\it One-dimensional profile likelihood functions for $m_{\tilde g}$ (upper left panel),
$m_{\tilde q_R}$ (upper right panel), $m_{\tilde t_1}$ (lower left panel)
  and $m_{\tilde b_1}$ (lower right panel).
}
\label{fig:gluinosquarkmasses}
\end{figure*}
%%%%%%%%%%%%%%%%%%%%%%%%% F I G U R E %%%%%%%%%%%%%%%%%%%%%%%%%%%%%%%%%%%%%%%%

{The various panels of Fig.~\ref{fig:gluinosquarkmasses} show the limited impact of
the LHC 13-TeV constraints on the possible masses of strongly-interacting sparticles {in the sub-GUT model,
comparing the solid and dashed curves}.
The upper left panel shows that the 95\% CL lower limit on {$\mgl \sim 1.5 \tev$, whether the
LHC 13-TeV data and the \gmt\ constraint are included or not. However, the best-fit
value of $\mgl$ increases from $\sim 2 \tev$ to a very large value when {\gmt\ is dropped,
although the $\Delta \chi^2$ price for $\mgl \sim 2 \tev$ is $\sim 1$.}
The upper right panel shows similar features
in the profile likelihood function for $m_{\sq_R}$ (that for $m_{\tilde q_L}$ is similar),
with a 95\% CL lower limit of $\sim 2 \tev$, which
is again quite independent of the inclusion of \gmt\ {and the 13-TeV data}.
The lower panels of Fig.~\ref{fig:gluinosquarkmasses}
show the corresponding profile likelihood functions for $m_{\tilde t_1}$ (left panel) and
$m_{\tilde b_1}$ (right panel). We see that these could both be considerably lighter than the gluino
and the first- and second-generation squarks, with 95\% CL lower limits $m_{\tilde t_1} \sim 900 \gev$
and $m_{\tilde b_1} \sim 1.5 \tev$, respectively.}

\subsection{The lightest neutralino and lighter chargino}

%%%%%%%%%%%%%%%%%%%%%%%%% F I G U R E %%%%%%%%%%%%%%%%%%%%%%%%%%%%%%%%%%%%%%%%
\begin{figure*}[]
\vspace{-1.5cm}
\centering
\includegraphics[width=0.475\textwidth]{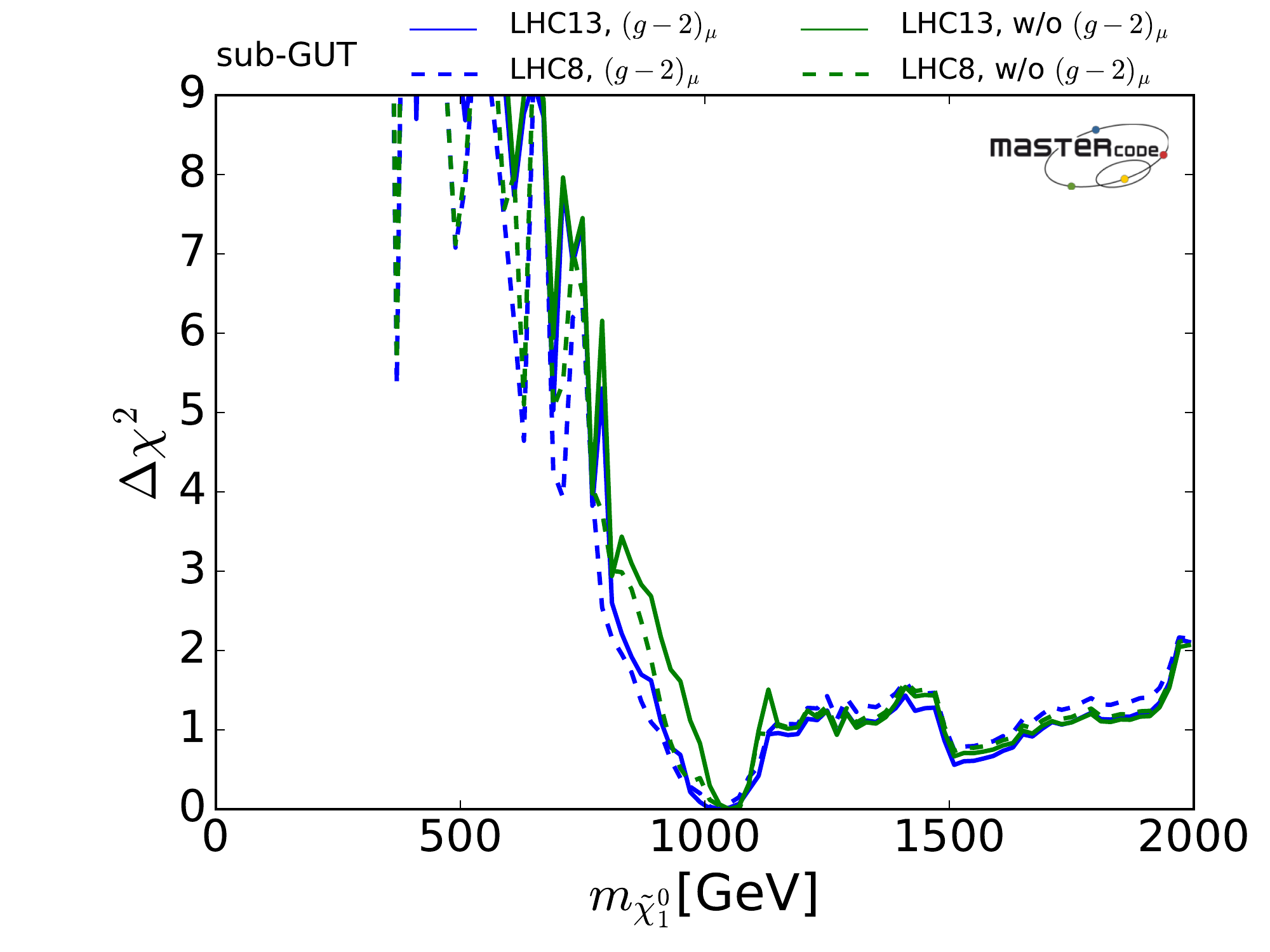}
\includegraphics[width=0.475\textwidth]{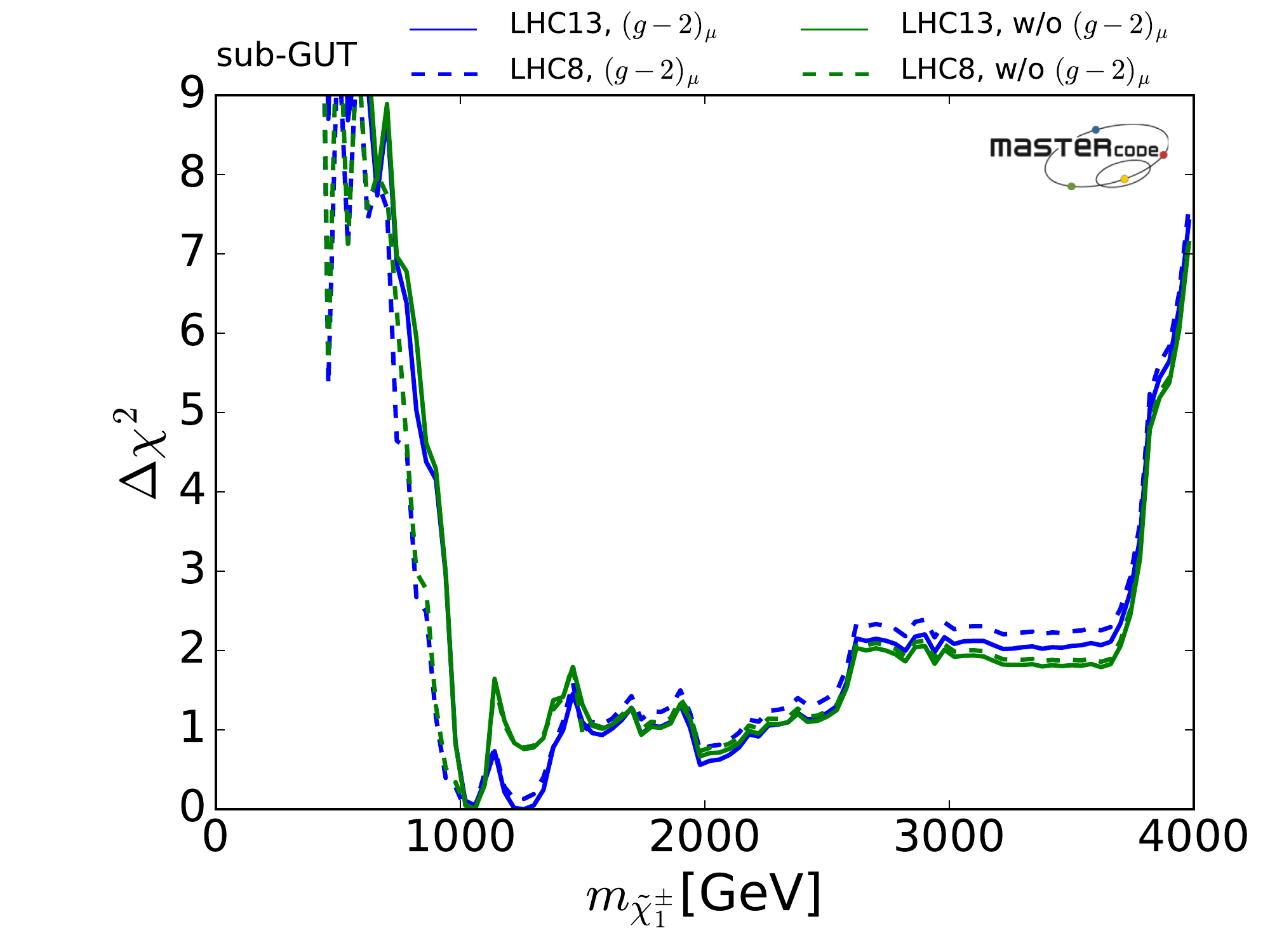} \\
\includegraphics[width=0.475\textwidth]{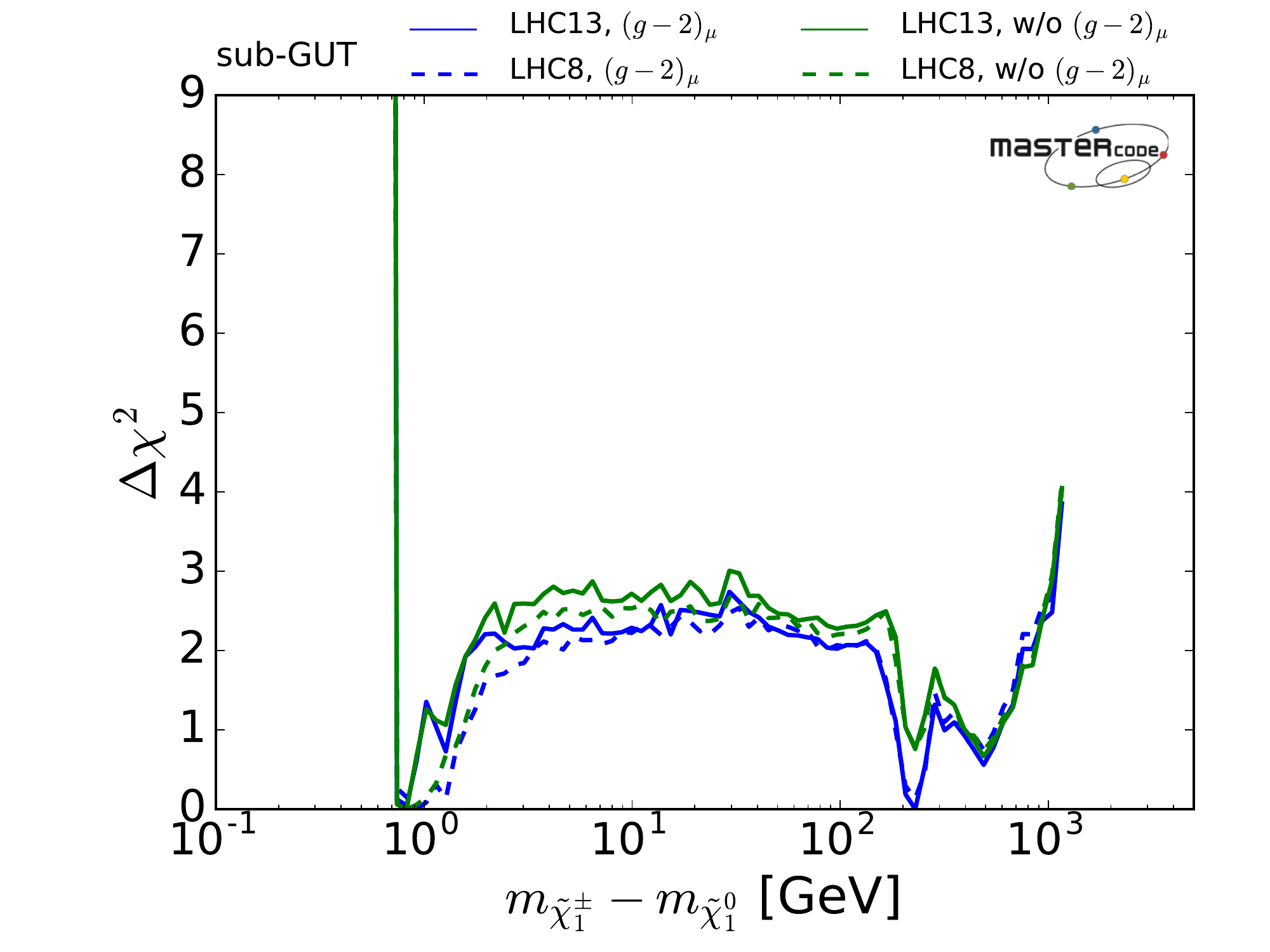}
\includegraphics[width=0.475\textwidth]{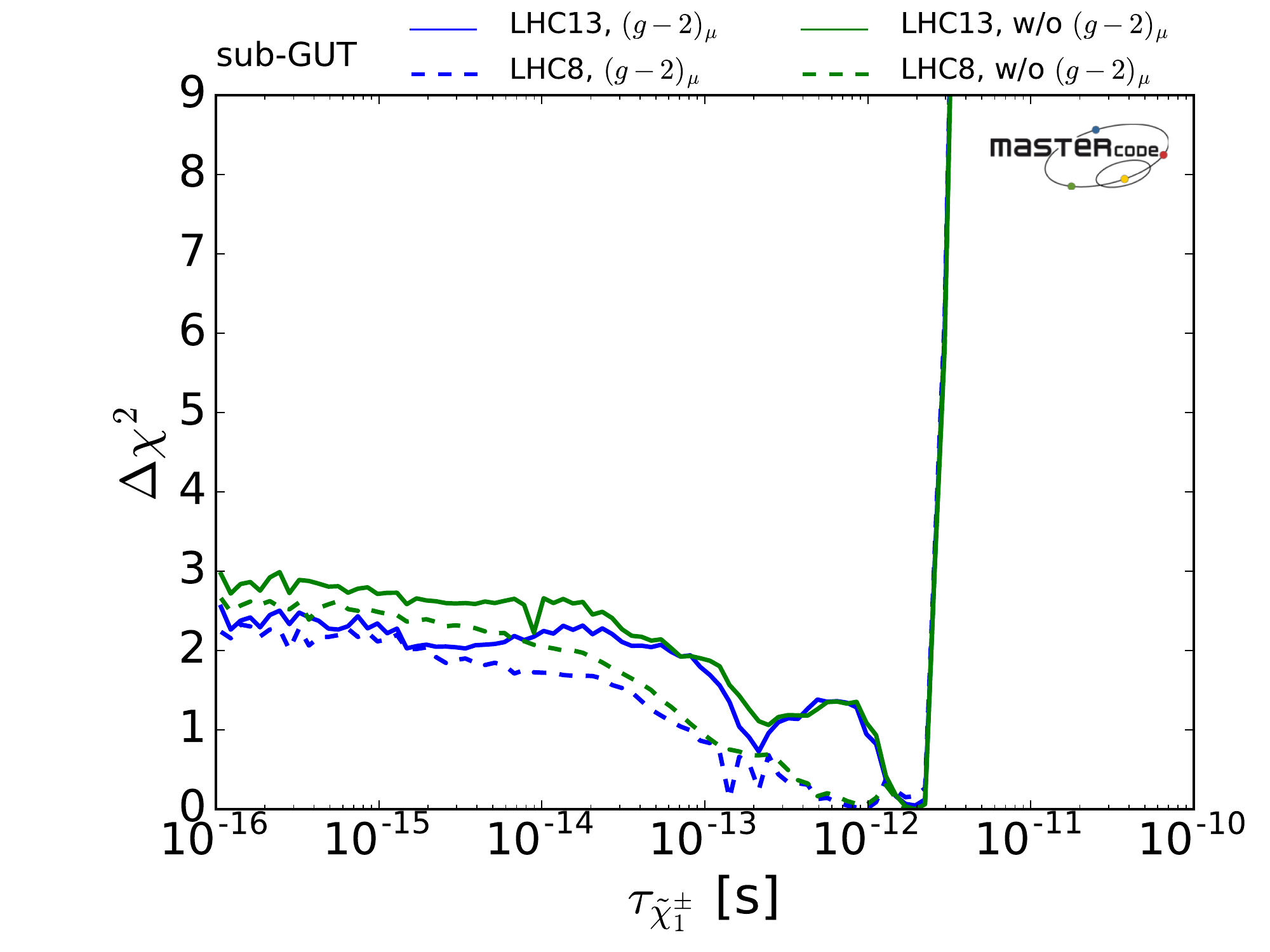} \\
\includegraphics[width=0.475\textwidth]{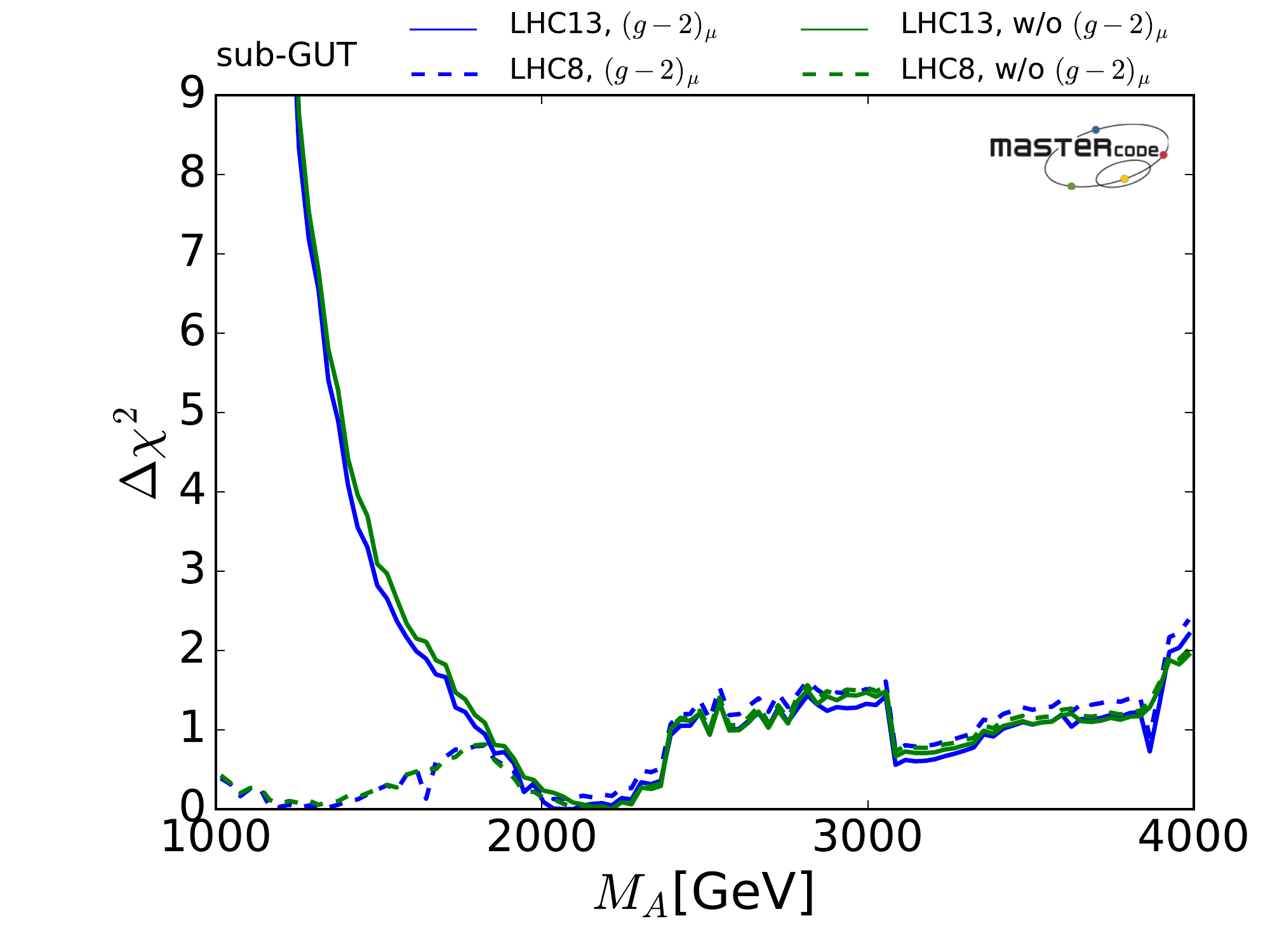}
\includegraphics[width=0.475\textwidth]{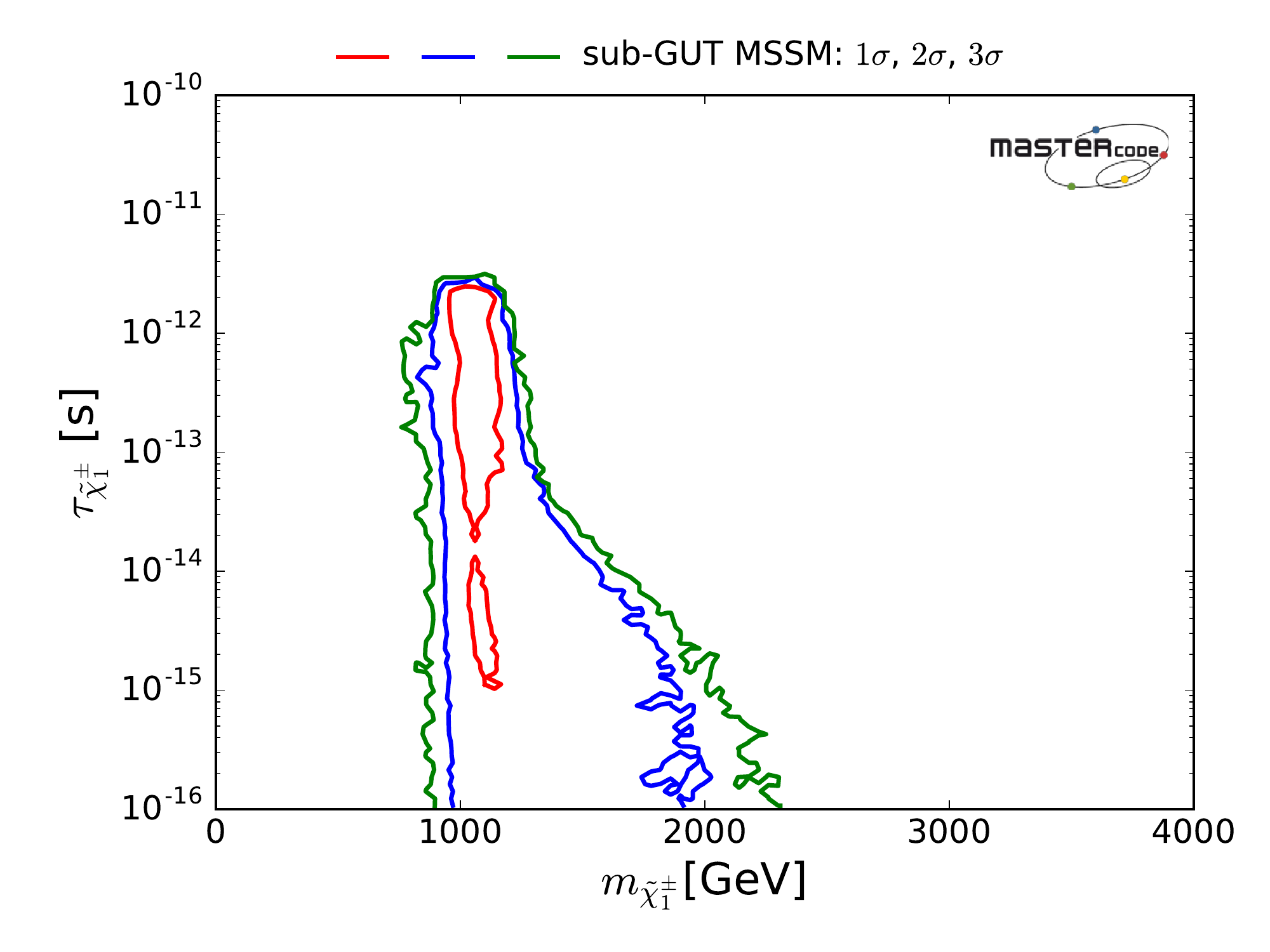} \\
\caption{\it One-dimensional profile likelihood functions for $\mneu1$ (top left panel)
and $\mcha1$ (top right panel), $\mcha1 - \mneu1$ (middle left panel)
the $\cha1$ lifetime (middle right panel) and $\MA$ (bottom left panel).
The bottom right panel shows the regions of the $(\mcha1, \tau_{\cha1})$ plane
 with $\tau_{\cha1} \ge 10^{-15}$~s that are allowed
in the fit including the \gmt\ and LHC 13-TeV constraints at the 68 (95) (99.7)\% CL in 2 dimensions,
i.e., $\Delta \chi^2 < 2.30 (5.99) (11.83)$, enclosed by the red (blue) (green) contour.}
\label{fig:neutralinocharginomass}
\end{figure*}
%%%%%%%%%%%%%%%%%%%%%%%%% F I G U R E %%%%%%%%%%%%%%%%%%%%%%%%%%%%%%%%%%%%%%%%

{The top left panel of Fig.~\ref{fig:neutralinocharginomass} shows the profile
likelihood function for $\mneu1$, and the top right panel shows that for $\mcha1$.
We see that in all the cases considered (with and without the \gmt\ and LHC13 constraints),
{the value of $\Delta \chi^2$ calculated using the LHC constraints on strongly-interacting
sparticles is {larger than} $4$ for $\mneu1 \lesssim 750 \gev$ and $\mcha1 \lesssim 800 \gev$.
Therefore, the LHC electroweakino searches~\cite{sus-16-039} have no impact on the 95\% CL regions
in our 2-dimensional projections of the sub-GUT parameter space, and we do not include
the results of~\cite{sus-16-039} in our global fit.}

{We now examine the profile likelihood functions for the fractions of Bino, Wino and Higgsino
in the $\neu1$ composition:
\begin{equation}
\neu1 \; = \; N_{11} {\tilde B} + N_{12} {\tilde W^3} + N_{13} {\tilde H_u} + N_{14} {\tilde H_d} \, ,
\label{composition}
\end{equation}
which are shown in \reffi{fig:1Dcompo}. As usual, results from an analysis including the 13-TeV data
{are shown as solid lines and without them as dashed lines, with \gmt\ as blue lines and without it
as green} lines. The top left panel shows that in the LHC 13-TeV case with \gmt\ an almost pure $\tilde B$
composition of the $\neu1$ is preferred, $N_{11} \to 1$, though the possibility that this component is almost absent
is only very slightly disfavoured. Conversely, before the LHC 13-TeV data there
was a very mild preference for $N_{11} \to 0$, and this is still the case if \gmt\ is dropped.
The upper right panel shows that a small $\tilde W^3$ component in the $\neu1$ is
strongly preferred in all cases.
Finally, the lower panel confirms that small $\tilde H_{u,d}$ components are preferred when
the LHC 13-TeV and \gmt\ constraints are applied, but large $\tilde H_{u,d}$ components are preferred
otherwise.}

%%%%%%%%%%%%%%%%%%%%%%%%% F I G U R E %%%%%%%%%%%%%%%%%%%%%%%%%%%%%%%%%%%%%%%%
\begin{figure*}[h]
\centering
\includegraphics[width=0.475\textwidth]{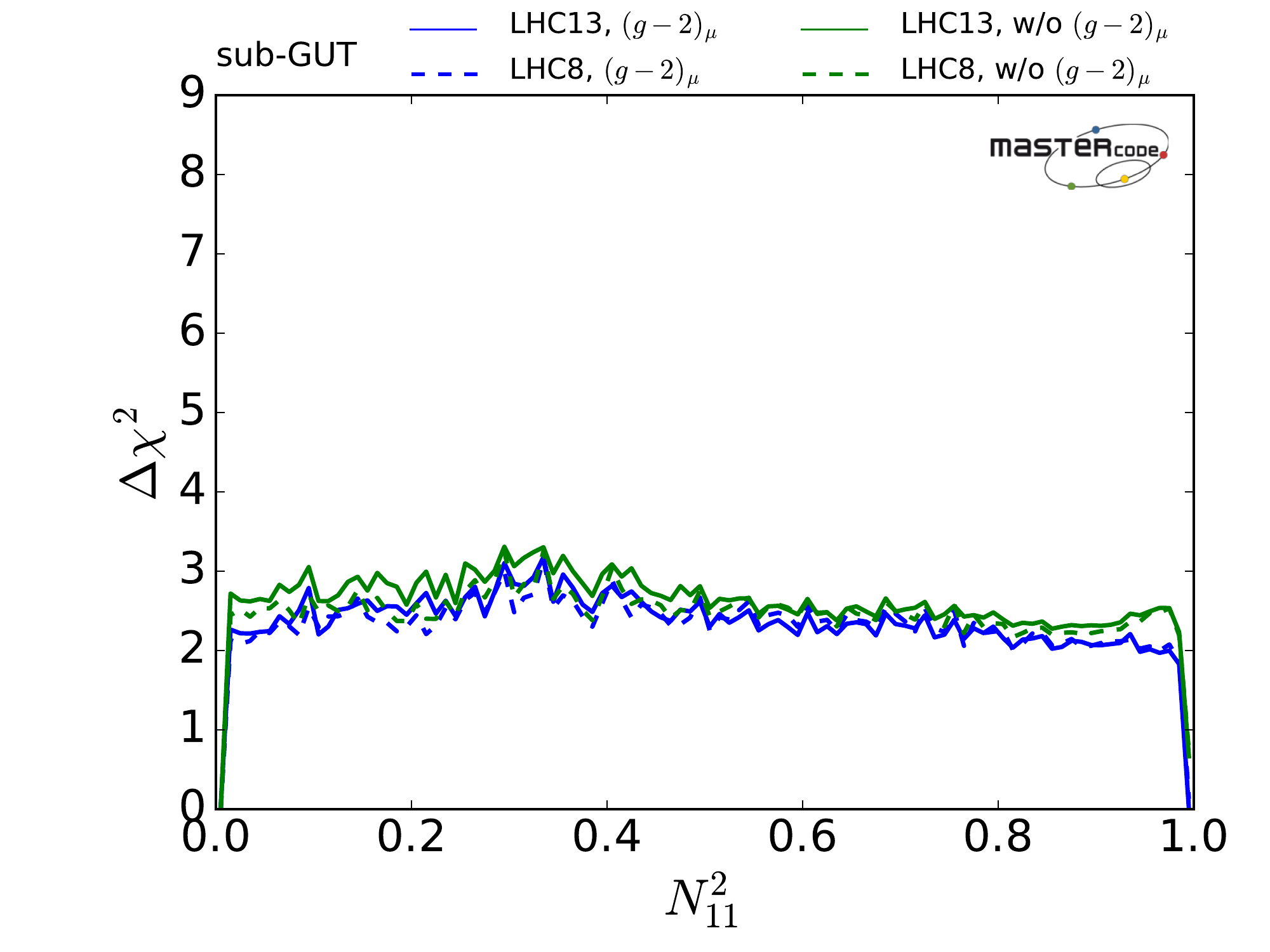}
\includegraphics[width=0.475\textwidth]{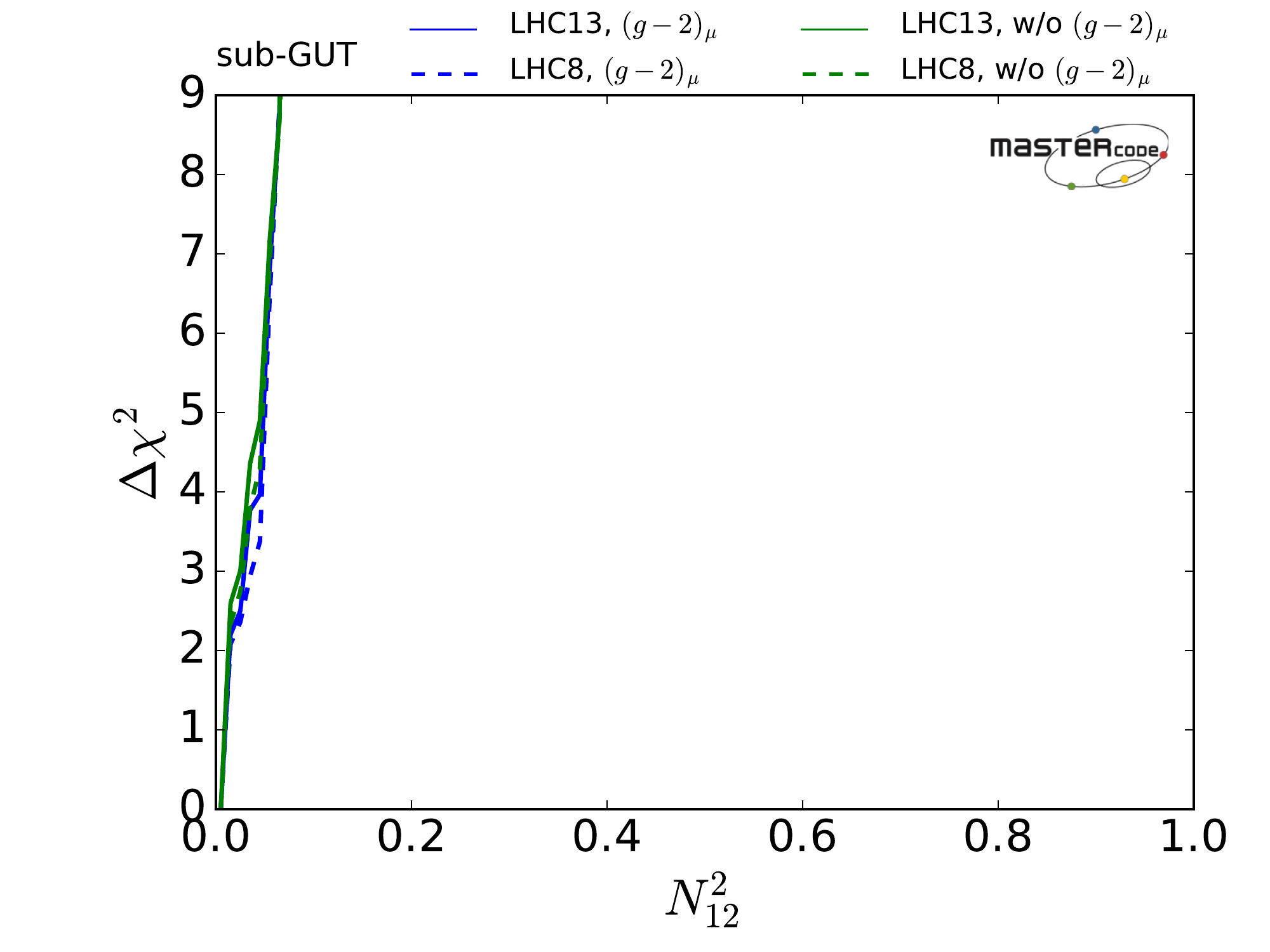} \\
%\vspace{2cm}
\centering
\includegraphics[width=0.475\textwidth]{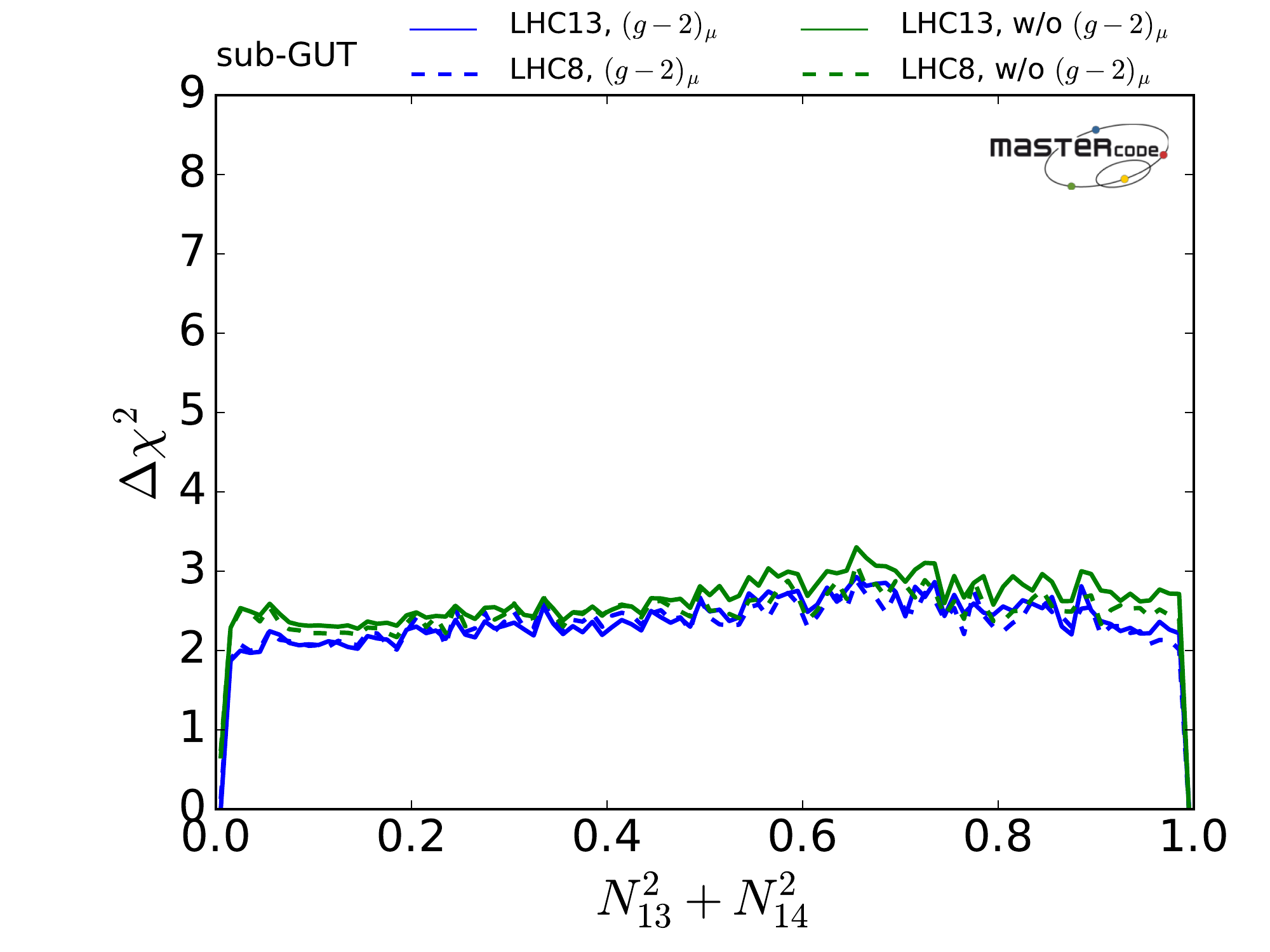}
\caption{\it {Plots of the one-dimensional profile likelihood  for the $\tilde B$ fraction in the LSP $\neu1$
(upper left), for the $\tilde W^3$ fraction (upper right)
and for the $\tilde H_{u,d}$ fraction (lower panel).}}
\label{fig:1Dcompo}
\end{figure*}
%%%%%%%%%%%%%%%%%%%%%%%%% F I G U R E %%%%%%%%%%%%%%%%%%%%%%%%%%%%%%%%%%%%%%%%

{The $\neu1$ compositions favoured at the 1-, 2- and 3-$\sigma$ levels (blue, yellow and red)
are displayed in Fig.~\ref{fig:triangles} for fits including LHC 13-TeV data with (without) the
\gmt\ constraint in the left (right) panel. We see that these regions are quite similar in the two panels,
and correspond to small Wino admixtures. On the other hand, the Bino fraction
$N_{11}^2$ and the Higgsino fraction $N_{13}^2 + N_{14}^2$ are relatively unconstrained at the 95\% CL.
The best-fit points are indicated by green stars,
and the left panel shows again that in the fit with \gmt\ the LSP is an almost pure Bino, whereas
an almost pure Higsino composition is favoured in the fit without \gmt, as also seen
in Table~\ref{tab:lspcomposition}. These two extremes have very similar $\chi^2$ values
in each of the fits displayed.}

%%%%%%%%%%%%%%%%%%%%%%%%% F I G U R E %%%%%%%%%%%%%%%%%%%%%%%%%%%%%%%%%%%%%%%%
\begin{figure*}[h]
\centering
\includegraphics[width=0.475\textwidth]{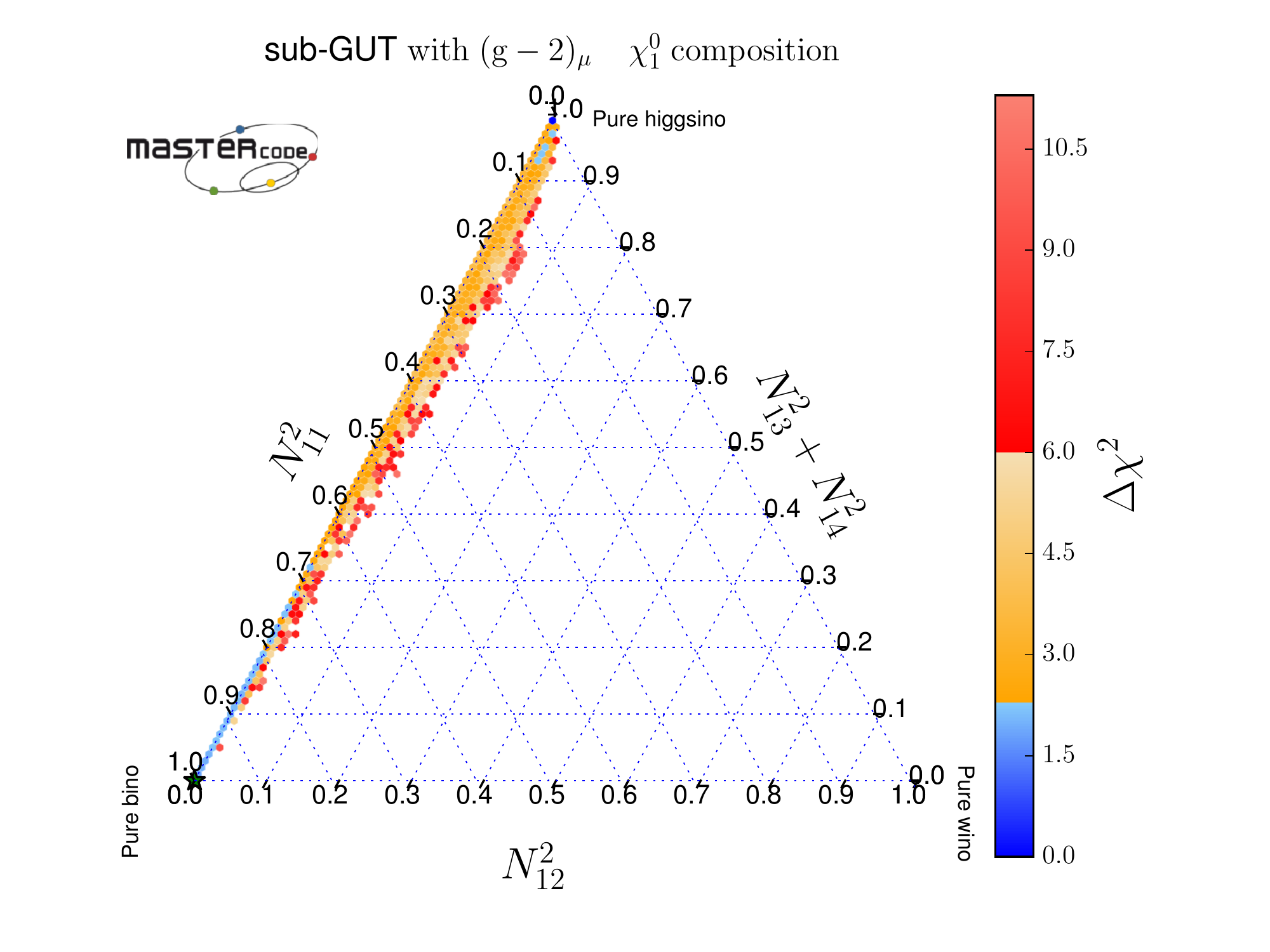}
\includegraphics[width=0.475\textwidth]{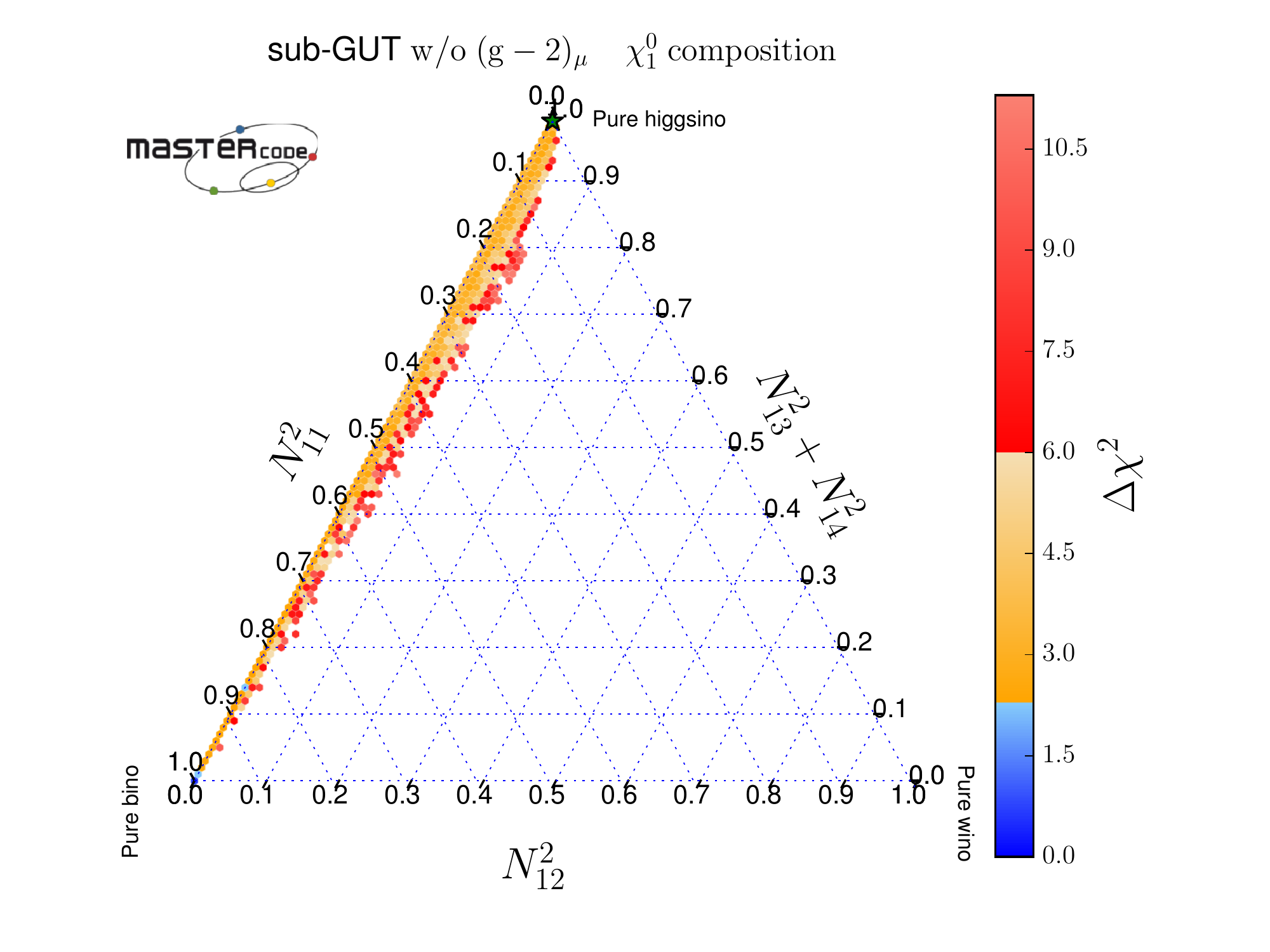}\\
\caption{\it {Triangular presentations of the $\neu1$ composition in the fit
with LHC 13-TeV including (dropping) the \gmt\ constraint in the left (right) panel.
The 1-, 2- and 3-$\sigma$ regions in the plots are coloured blue, yellow and red, and the best-fit points are
indicated by green stars.} }
\label{fig:triangles}
\end{figure*}
%%%%%%%%%%%%%%%%%%%%%%%%% F I G U R E %%%%%%%%%%%%%%%%%%%%%%%%%%%%%%%%%%%%%%%%

\begin{table}[htb!]
\centering
        \begin{tabular}{c c c c c }
		& $\tilde B$ & $\tilde W_3$ & $\tilde H_u$ & $\tilde H_d$ \\
        \hline \hline
        \multicolumn{5}{c}{With \gmt} \\
        \hline
                With 13-TeV & 0.999 & -0.010 & 0.041 & -0.025 \\
                Without 13-TeV & 0.007 & -0.011 & 0.707 & -0.707 \\
        \hline
        \multicolumn{5}{c}{Without \gmt} \\
        \hline
                With 13-TeV & 0.006 & -0.010 & 0.707 & -0.707 \\
                Without 13-TeV & 0.007 & -0.011 & 0.707 & -0.707 \\
        \hline
        \end{tabular}
                \caption{\it {Composition of the \neu1 LSP at the best-fit points
                with and without \gmt\ and the LHC 13-TeV data.}}
                \label{tab:lspcomposition}
\end{table}

The global $\chi^2$ function is minimized for $\mneu1 \simeq 1.0 \tev$, which is typical of scenarios
with a Higgsino-like LSP whose density is brought into the Planck 2015 range
by coannihilation with a nearly-degenerate Higgsino-like chargino $\cha1$.
Indeed, we see in the top right panel of Fig.~\ref{fig:neutralinocharginomass} that
$\chi^2$ is minimized when also $\mcha1 \simeq \mneu1 \simeq 1.0 \tev$.
{Table ~\ref{tab:lspcomposition} displays the LSP composition of the sub-GUT model
at the best-fit points with and without \gmt and the LHC 13-TeV data. We see again that the \neu1 LSP is
mainly a Higgsino with almost equal $\tilde H_u$ and $\tilde H_d$ components, except in the fit
with both LHC 13-TeV data and \gmt\ included, in which case it is an almost pure Bino.}

Looking at the middle left panel of Fig.~\ref{fig:neutralinocharginomass}, we see that the best-fit
point has a chargino-LSP mass difference that {may be ${\cal O}(1) \gev$
or $\sim 200$ to $300 \gev$, with similar $\chi^2$} in all the cases considered, namely
with and without the \gmt\ and LHC13 constraints.  As seen in the middle right panel
of Fig.~\ref{fig:neutralinocharginomass}, in the {more degenerate case the preferred
chargino lifetime $\tau_{\cha1} \sim 10^{-12}$~s.}
{The current LHC searches for long-lived charged particles~\cite{LLsearches1} therefore do not impact
this chargino coannihilation region, and are also not included in our global fit.}

{The top right panel of Fig.~\ref{fig:neutralinocharginomass} displays
an almost-degenerate local minimum of $\chi^2$
with $\mcha1 \sim 1.3 \tev$, corresponding to a second, local minimum of
$\chi^2$ where $\mcha1 - \mneu1 \sim 200$ to $300 \gev$, as seen in the middle left panel.
In this region the relic density is brought into the Planck 2015 range by rapid
annihilation through $A/H$ bosons, as can be inferred from the bottom left
panel of Fig.~\ref{fig:neutralinocharginomass}, where we see that at this secondary minimum
$\MA \simeq 2 \tev \simeq 2 \mneu1$. The $\cha1$ lifetime in this region is too short
to appear in the middle and bottom right panels of Fig.~\ref{fig:neutralinocharginomass},
and too short to have a separated vertex signature at the LHC.}

Finally, the bottom right panel of Fig.~\ref{fig:neutralinocharginomass}
shows the regions of the $(\mcha1, \tau_{\cha1})$ plane {with $\tau_{\cha1} \in
(10^{-16}, 10^{-10})$~s that are allowed
in the fit including the \gmt\ and LHC 13-TeV constraints at the 68 (95) (99.7) \% CL in 2 dimensions,
i.e., $\Delta \chi^2 < 2.30 (5.99) (11.83) $. Since the chargino would decay into a very soft track and a neutralino,
detecting a separated vertex in the region around the best-fit point would be very challenging.}

\subsection{Sleptons}

%%%%%%%%%%%%%%%%%%%%%%%%% F I G U R E %%%%%%%%%%%%%%%%%%%%%%%%%%%%%%%%%%%%%%%%
\begin{figure*}[]
\centering
\includegraphics[width=0.475\textwidth]{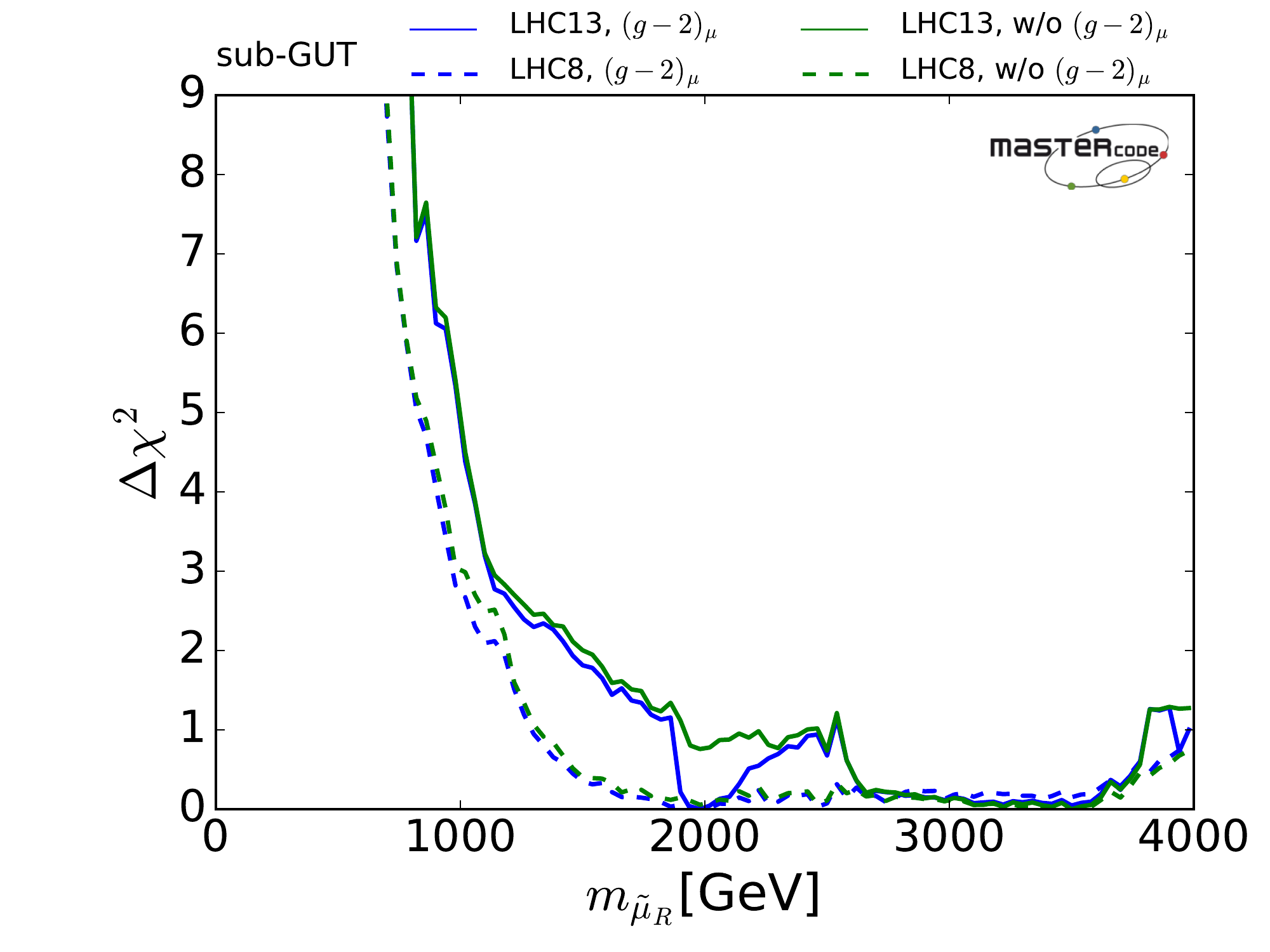}
\includegraphics[width=0.475\textwidth]{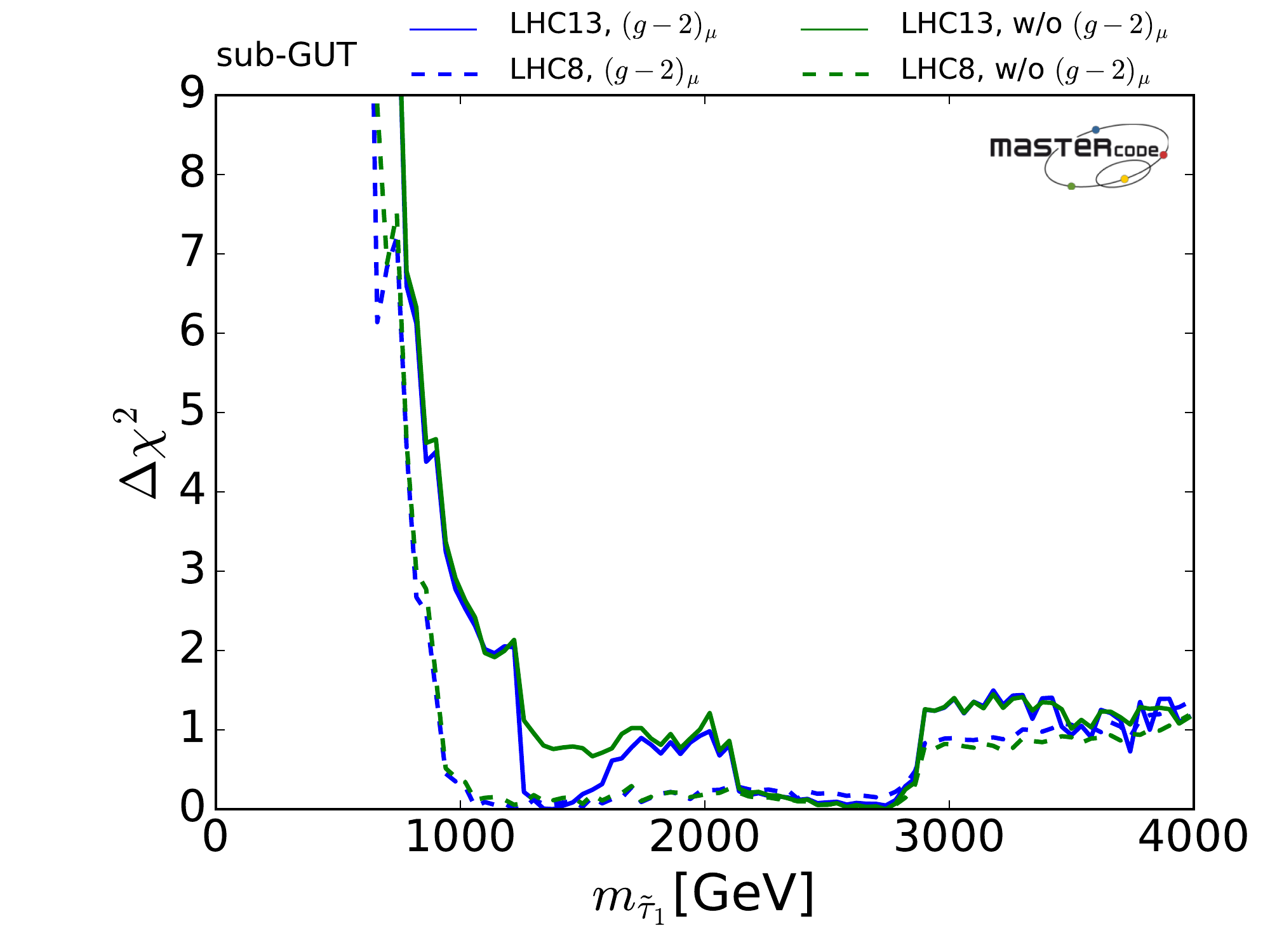} \\
\includegraphics[width=0.475\textwidth]{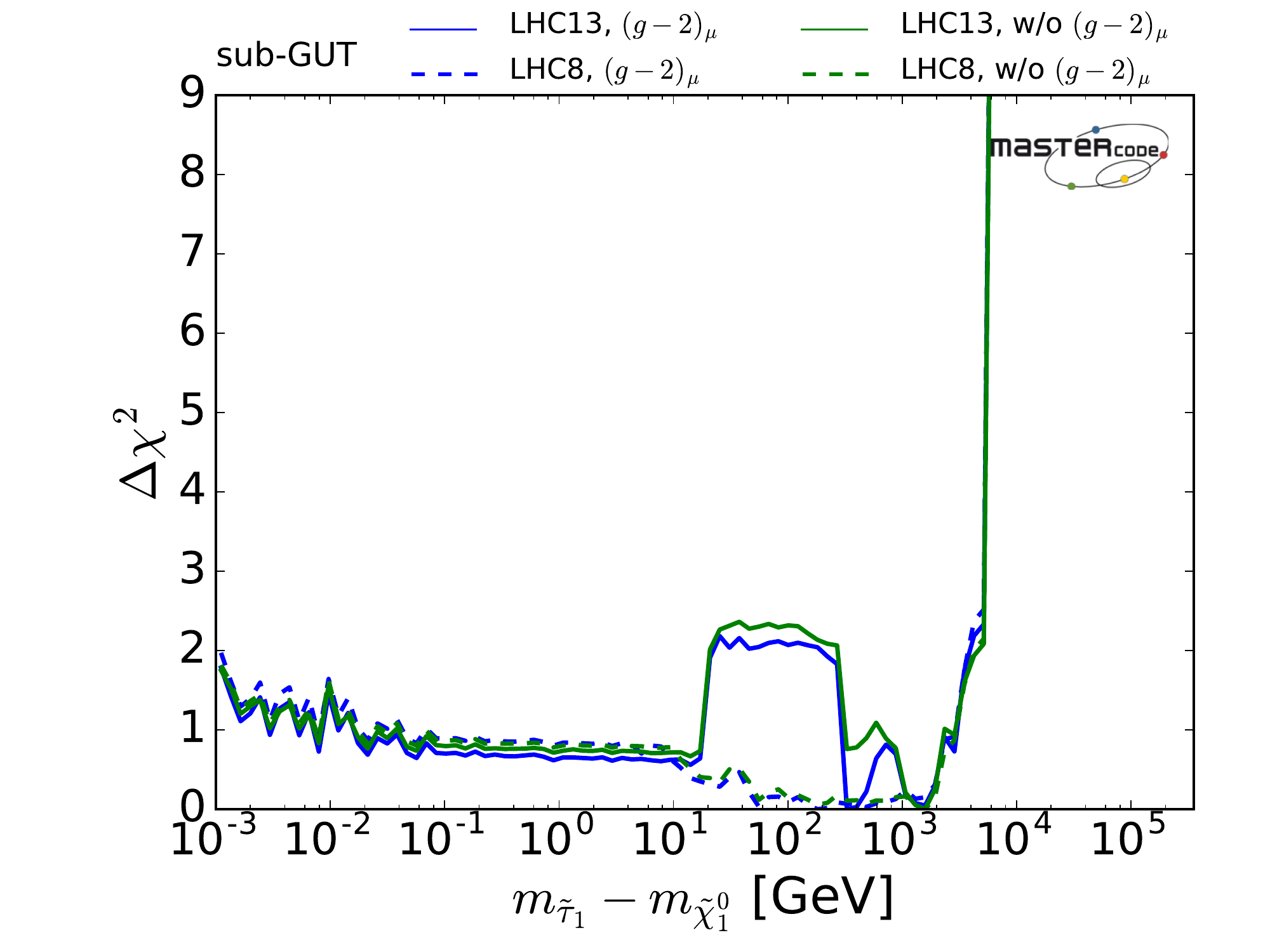}
\includegraphics[width=0.475\textwidth]{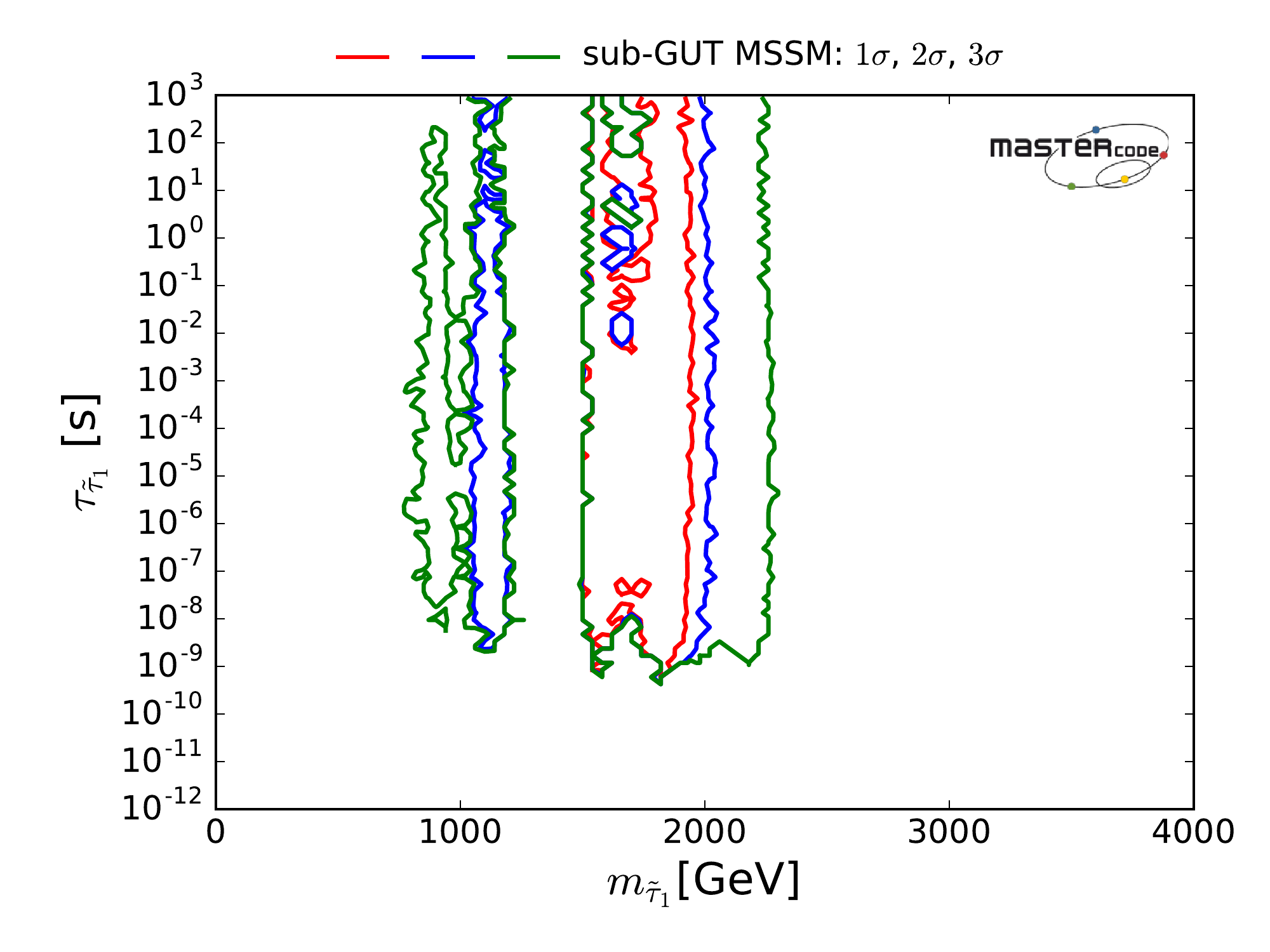} \\
\caption{\it One-dimensional profile likelihood functions for $m_{\tilde \mu_R}$ (upper left panel),
$m_{\tilde \tau_1}$ (upper right panel) and $m_{\tilde \tau_1} - \mneu1$ (lower left panel).
The lower right panel shows the $(m_{\tilde \tau_1}, \tau_{\tilde \tau_1})$ plane, colour-coded as
indicated in the right-hand legend. The 68 (95) (99.7)\% CL regions in 2 dimensions,
i.e., $\Delta \chi^2 < 2.30 (5.99) (11.83)$, are enclosed by the red (blue) (green) contours.}
\label{fig:staumassdifferencelifetimesmuonmass}
\end{figure*}
%%%%%%%%%%%%%%%%%%%%%%%%% F I G U R E %%%%%%%%%%%%%%%%%%%%%%%%%%%%%%%%%%%%%%%%

The upper left panel of Fig.~\ref{fig:staumassdifferencelifetimesmuonmass} shows the
profile likelihood function for $m_{\tilde \mu_R}$ (that for $m_{\tilde e_R}$ is
indistinguishable, the ${\tilde \mu_L}$ and ${\tilde e_L}$ are slightly heavier).
{We see that in the sub-GUT model small values of $m_{\tilde \mu_R}$ were already
disfavoured by earlier LHC data (dashed lines), and that this tendency has been reinforced by the LHC 13-TeV data
(compare the solid lines). The same is true whether
the \gmt\ constraint is included or dropped (compare the blue and green curves)}.

 The upper right panel Fig.~\ref{fig:staumassdifferencelifetimesmuonmass} shows the
corresponding profile likelihood function for $m_{\tilde \tau_1}$, which shares many
similar features. However, we note that the $\chi^2$ function for $m_{\tilde \tau_1}$ is
generally lower than that for {$m_{\tilde \mu_R} \in (1, 2) \tev$, though the 95\%
lower limits on $m_{\tilde \tau_1}$ and $m_{\tilde \mu_R}$ are quite similar,
and both are $\simeq 1 \tev$ when the LHC} 13-TeV constraints
are included in the fit.

The lower left panel of Fig.~\ref{fig:staumassdifferencelifetimesmuonmass} shows
that {very small values of $m_{\tilde \tau_1} - \mneu1$ in
the stau coannihilation region are allowed at the $\Delta \chi^2 \sim 1$ level in all the fits
with the \gmt\ constraint, rising to $\Delta \chi^2 \gtrsim 2$ for $m_{\tilde \tau_1} - \mneu1 \gtrsim 20 \gev$
when the LHC 13-TeV data are included.}

{The lower right panel of Fig.~\ref{fig:staumassdifferencelifetimesmuonmass} shows the
$(m_{\tilde \tau_1}, \tau_{\tilde \tau_1})$ plane, where we see that
$\tau_{\tilde \tau_1} \in (10^{-7}, 10^3)$~s is allowed at the 68\% CL,
for {$1600 \gev \lesssim m_{\tilde \tau_1} \lesssim 2000 \gev$ and at the 95\% CL also for $m_{\tilde \tau_1} \sim 1100 \gev$.
This region of parameter space is close to the tip of the stau coannihilation strip.
Lower ${\tilde \tau_1}$ masses are strongly disfavoured by the LHC constraints, {particularly at 13 TeV}, as seen in
the upper right panel of Fig.~\ref{fig:staumassdifferencelifetimesmuonmass}.
The heavier ${\tilde \tau_1}$ masses with lower $\Delta \chi^2$ seen there
do not lie in the stau coannihilation strip, and have larger $\mstaue - \mneu1$
and hence smaller lifetimes that are not shown in the lower right panel of
Fig.~\ref{fig:staumassdifferencelifetimesmuonmass}.
Because of the lower limit on $\mstaue$ seen in this panel,}
neither the LHC search for long-lived charged particles~\cite{LLsearches1}
nor the LHC search for (meta-)stable massive charged particles that exit the
detector~\cite{LLsearches2} {are relevant for} our global fit.

In view of this, and the fact that the search for long-lived particles~\cite{LLsearches1}
is also insensitive in the chargino coannihilation region, as discussed above, the results
of~\cite{LLsearches1,LLsearches2} are not included in the calculation of the global likelihood function.}

\subsection{$\mathbf{\gmt}$}

%%%%%%%%%%%%%%%%%%%%%%%%% F I G U R E %%%%%%%%%%%%%%%%%%%%%%%%%%%%%%%%%%%%%%%%
\begin{figure*}[]
\centering
\includegraphics[width=0.475\textwidth]{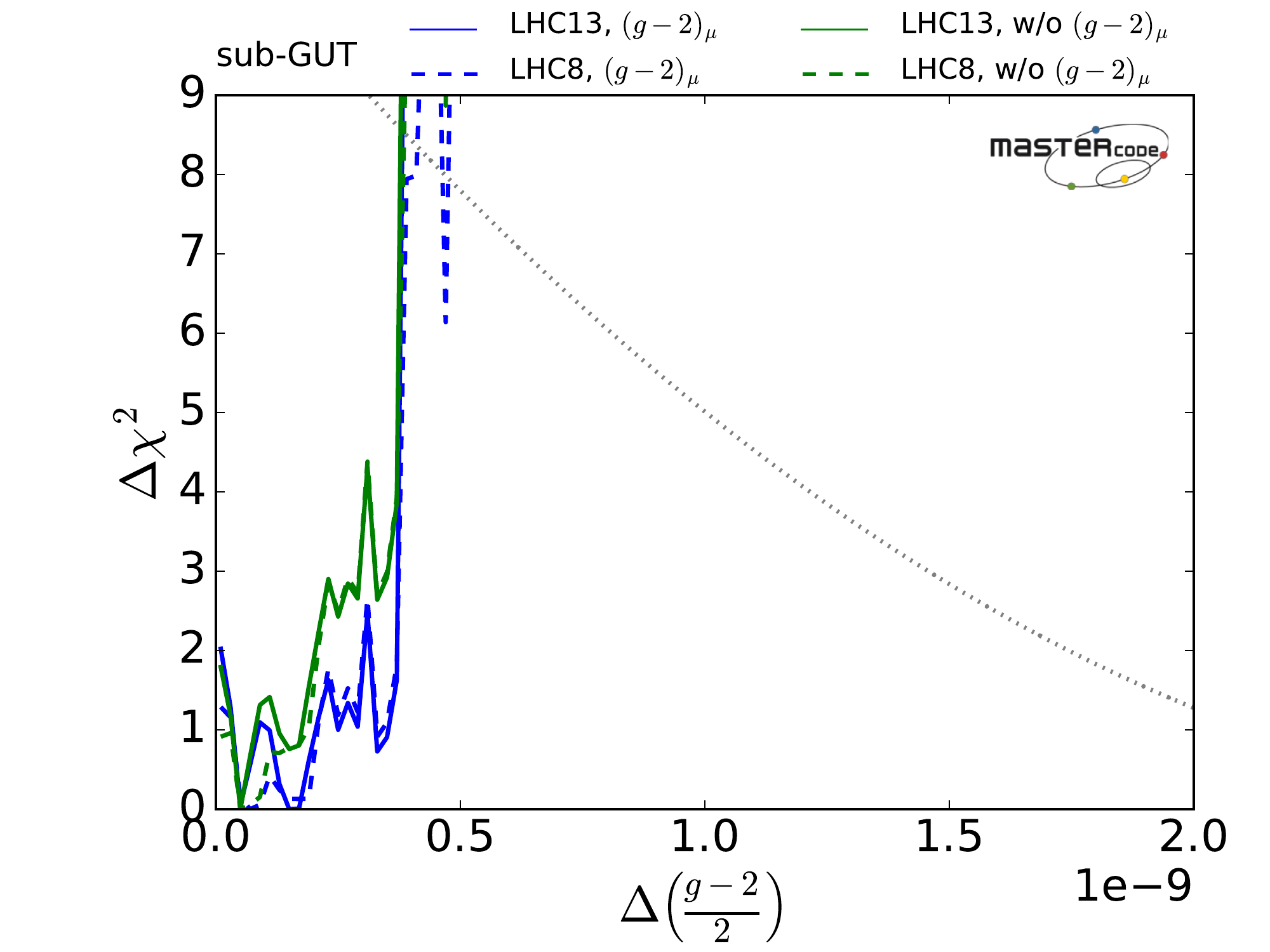}
\includegraphics[width=0.475\textwidth]{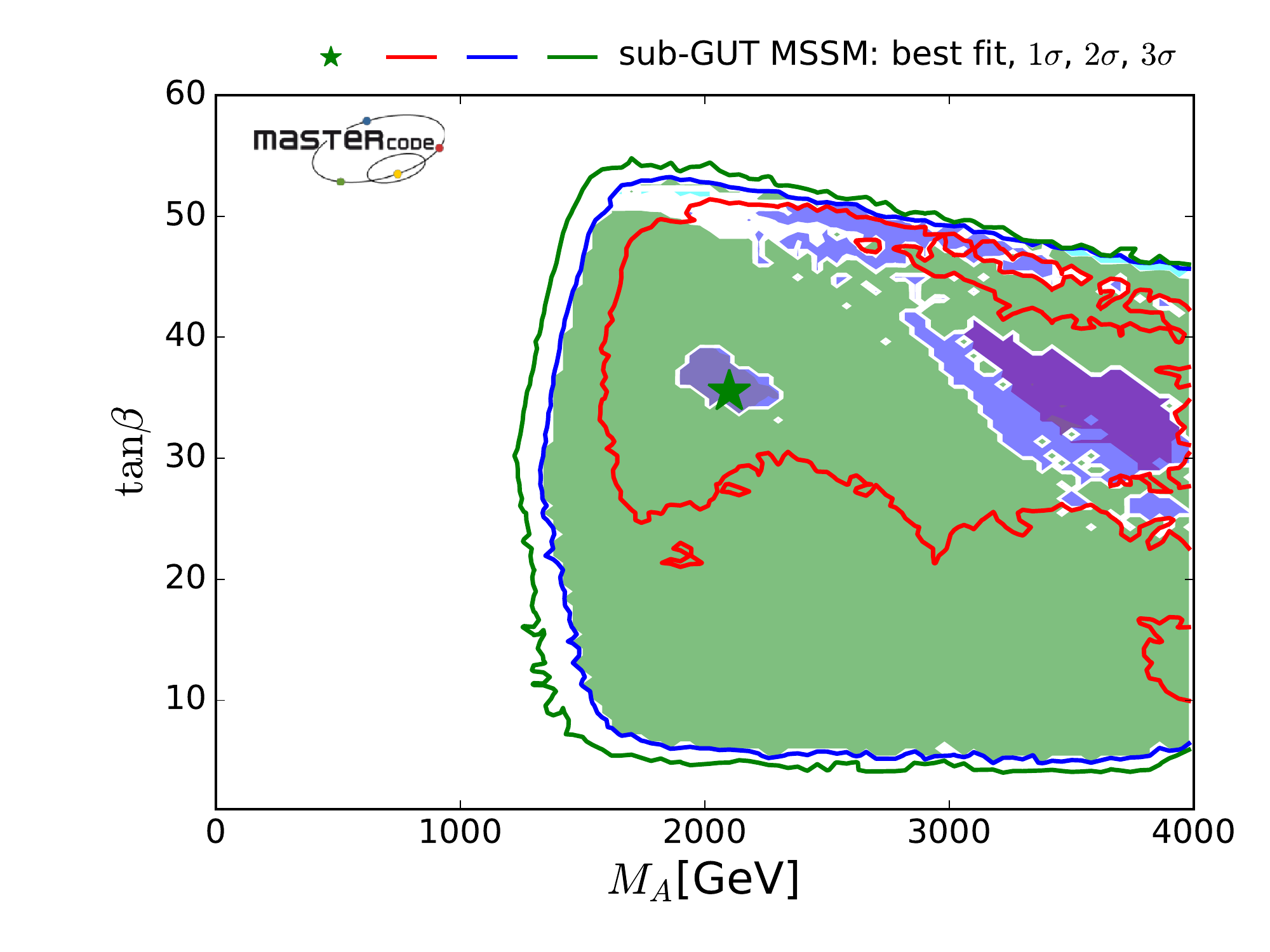} \\
\includegraphics[width=0.9\textwidth]{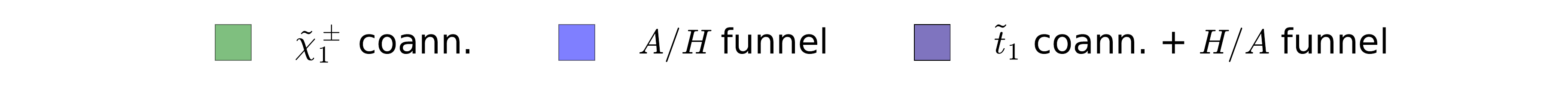} \\
\caption{\it Left panel: One-dimensional profile likelihood function for \gmt, where the dotted line
shows the $\chi^2$ contribution due to the \gmt\ constraint alone.
Right panel: Two-dimensional projection of the likelihood function in the $(\MA, \tb)$ plane.}
\label{fig:g-2tbMA}
\end{figure*}
%%%%%%%%%%%%%%%%%%%%%%%%% F I G U R E %%%%%%%%%%%%%%%%%%%%%%%%%%%%%%%%%%%%%%%%

{We see in the left panel of Fig.~\ref{fig:g-2tbMA} that only a small contribution to \gmt\ is
possible in sub-GUT models, the profile likelihood functions with and without the LHC 13-TeV
data and \gmt\ being all quite similar. {This is because in the sub-GUT model with low \Min\
the LHC searches for strongly-interacting sparticles constrain the $\tilde \mu$ mass more
strongly than in the GUT-scale CMSSM.} The dotted line shows the $\Delta \chi^2$ contribution due to
our implementation of the \gmt\ constraint alone. We see that in all cases it contributes
$\Delta \chi^2 \gtrsim 9$ to the global fit.}

\subsection{The $(\MA, \tb)$ Plane}

{The right panel of Fig.~\ref{fig:g-2tbMA} shows the $(\MA, \tb)$ plane {when the LHC 13-TeV data
and the \gmt\ constraint are included in the fit.}
We see that {$\MA \gtrsim 1.3 \tev$ at the 95\% CL and that, whereas} $\tb \sim 5$ is allowed
at the 95\% CL. Larger values $\tb \gtrsim 30$ are favoured at the 68\% CL, and the best-fit
point has $\tb \simeq 36$. {(This increases to $\tb \sim 45$ if either the LHC 13-TeV and/or
\gmt\ constraint is dropped.)} As in the previous two-dimensional projections of the sub-GUT
parameter space, the 99.7\% (3-$\sigma$) CL contour lies close to that for the 95\% CL.}

\subsection{$\mathbf{B}$ Decay Observables}

%%%%%%%%%%%%%%%%%%%%%%%%% F I G U R E %%%%%%%%%%%%%%%%%%%%%%%%%%%%%%%%%%%%%%%%
\begin{figure*}[]
\begin{flushright}
\includegraphics[width=0.5\textwidth]{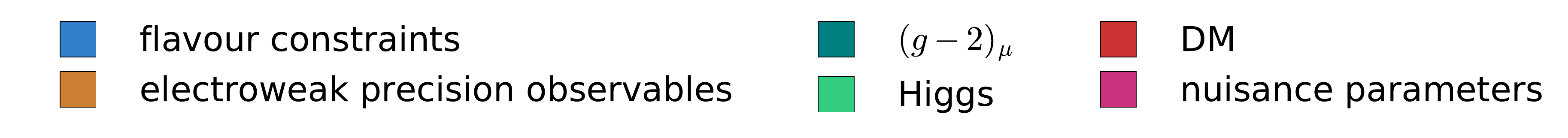} \\
\end{flushright}
\vspace{-0.52cm}
\centering
\includegraphics[width=0.475\textwidth]{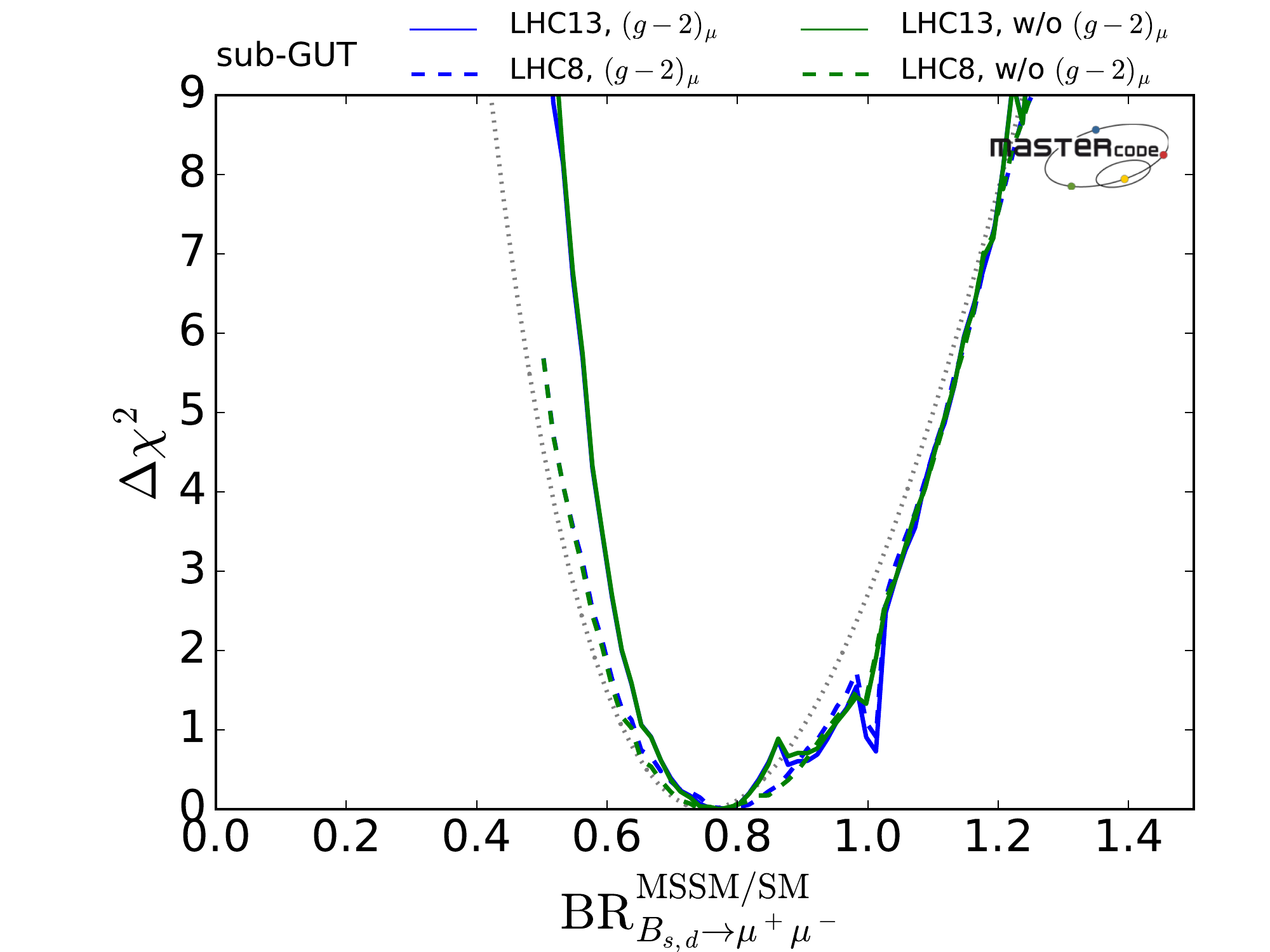}
\includegraphics[width=0.475\textwidth]{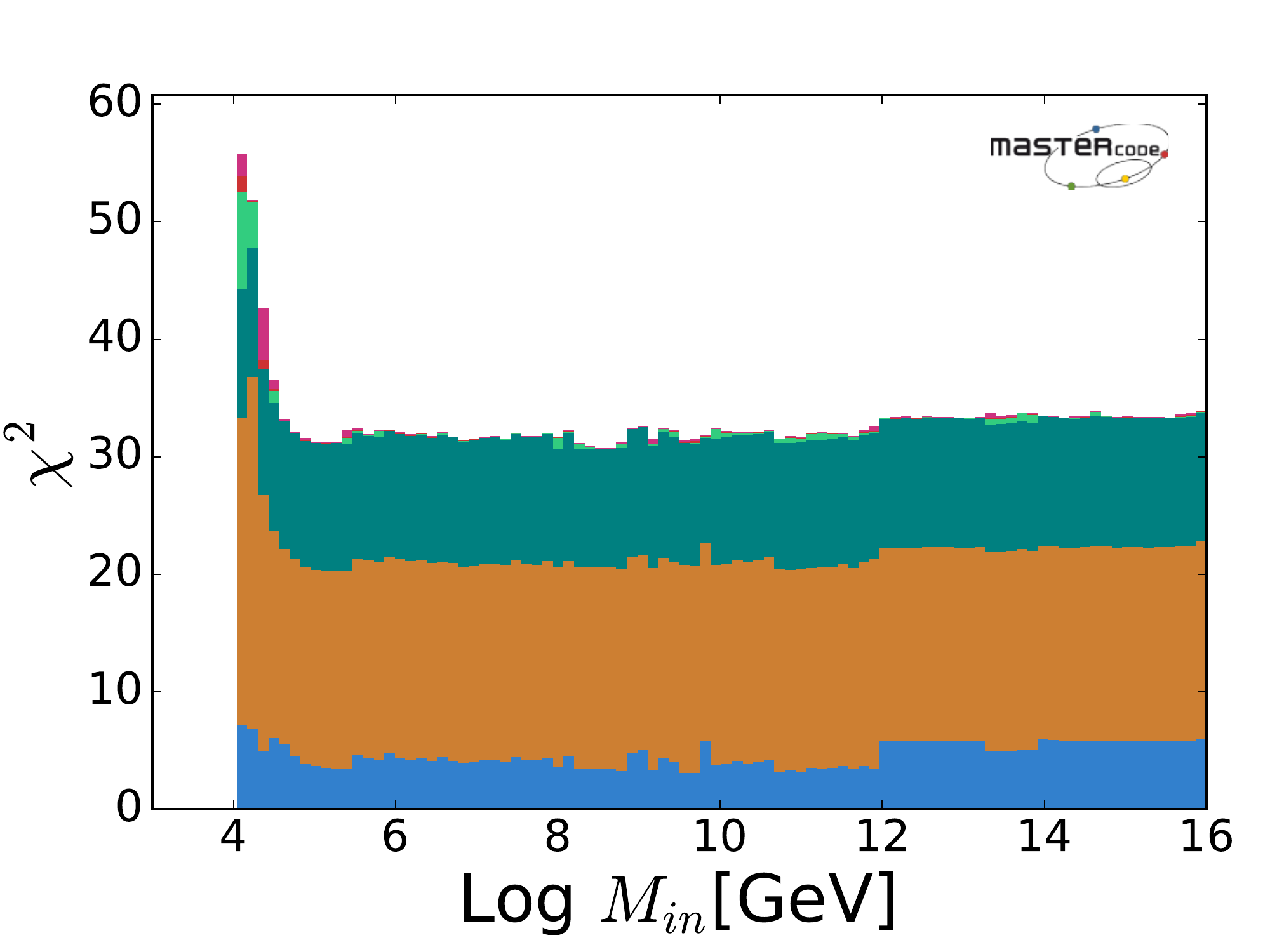} \\
\vspace{-0.6cm}
\caption{\it Left panel: One-dimensional profile likelihood function for \bsdmm, where the dotted line
shows the $\chi^2$ contribution due to the \bsdmm\ constraint alone.
	Right panel: Breakdown of the contributions to the global $\chi^2$ as functions of $\Min$.
	 {The shadings correspond to the different classes of observables, as indicated in the legend.}
	}
\label{fig:Bsdmm}
\end{figure*}
%%%%%%%%%%%%%%%%%%%%%%%%% F I G U R E %%%%%%%%%%%%%%%%%%%%%%%%%%%%%%%%%%%%%%%%

We see in the left panel of Fig.~\ref{fig:Bsdmm} that values of \bsdmm\ smaller
than that in the SM are favoured. The sub-GUT models {with $\mu > 0$ that we have studied} can accommodate comfortably
the preference seen in the data (dotted line) for such a small value of \bsdmm~\footnote{This is also the case
for the smaller sub-GUT sample with $\mu < 0$ that we have studied.},
which is not the case in models such as the CMSSM that impose universal
boundary conditions on the soft supersymmetry-breaking parameters at the GUT scale, {if $\mu > 0$.}
The right panel of Fig.~\ref{fig:Bsdmm} shows how the contributions of the flavour (blue shading) and other observables to the
global likelihood function depend on \Min\ for values between $10^4$ and $10^{16} \gev$.
{This variation in the flavour contribution (which is dominated by \bsdmm)  is largely
responsible for the sub-GUT preference for $\Min < \MGUT$ seen in the top left
panel of Fig.~\ref{fig:Minm0m12}. Values of $\Min \in (10^5, 10^{12}) \gev$ can accommodate very well the
experimental value of \bsdmm.}

{This preference is made possible by the different RGE running
in the sub-GUT model, which can change the sign of the product $A_t \mu$ that controls
the relative signs of the SM and SUSY contributions to the $B_{s,d} \to \mu^+ \mu^-$ decay amplitudes,
permitting negative interference that reduces \bsdmm.}
{As already discussed, the reduction in \bsdmm\ and the global $\chi^2$
function for $10^8 \gev \lesssim \Min \lesssim 10^{12} \gev$
is associated with the blue 68\% CL regions with
{$\Min \lesssim 10^{12} \gev$ seen in the middle panels} of Fig.~\ref{fig:Minm0m12}.
On the other hand, we see in Fig.~\ref{fig:bsg} that sub-GUT models favour values of \bsg\
that are close to the SM value.}

The contributions to the global $\chi^2$ function of other classes of observables as
functions of $\Min$ are also exhibited in the right panel of Fig.~\ref{fig:Bsdmm}. In addition
to the aforementioned reduction in the flavour contribution when $\Min \lesssim 10^{12} \gev$
(blue shading), {there is a coincident (but smaller)} increase in the contribution of the electroweak
precision observables (orange shading) {related to tension in the electroweak symmetry-breaking conditions}. The other
contributions to the global $\chi^2$ function, namely the nuisance parameters (red shading), Higgs mass
(light green), \gmt\ (teal) and DM (red), vary smoothly for $\Min \sim 10^{12} \gev$.

%%%%%%%%%%%%%%%%%%%%%%%%% F I G U R E %%%%%%%%%%%%%%%%%%%%%%%%%%%%%%%%%%%%%%%%
\begin{figure*}[]
\centering
\includegraphics[width=0.5\textwidth]{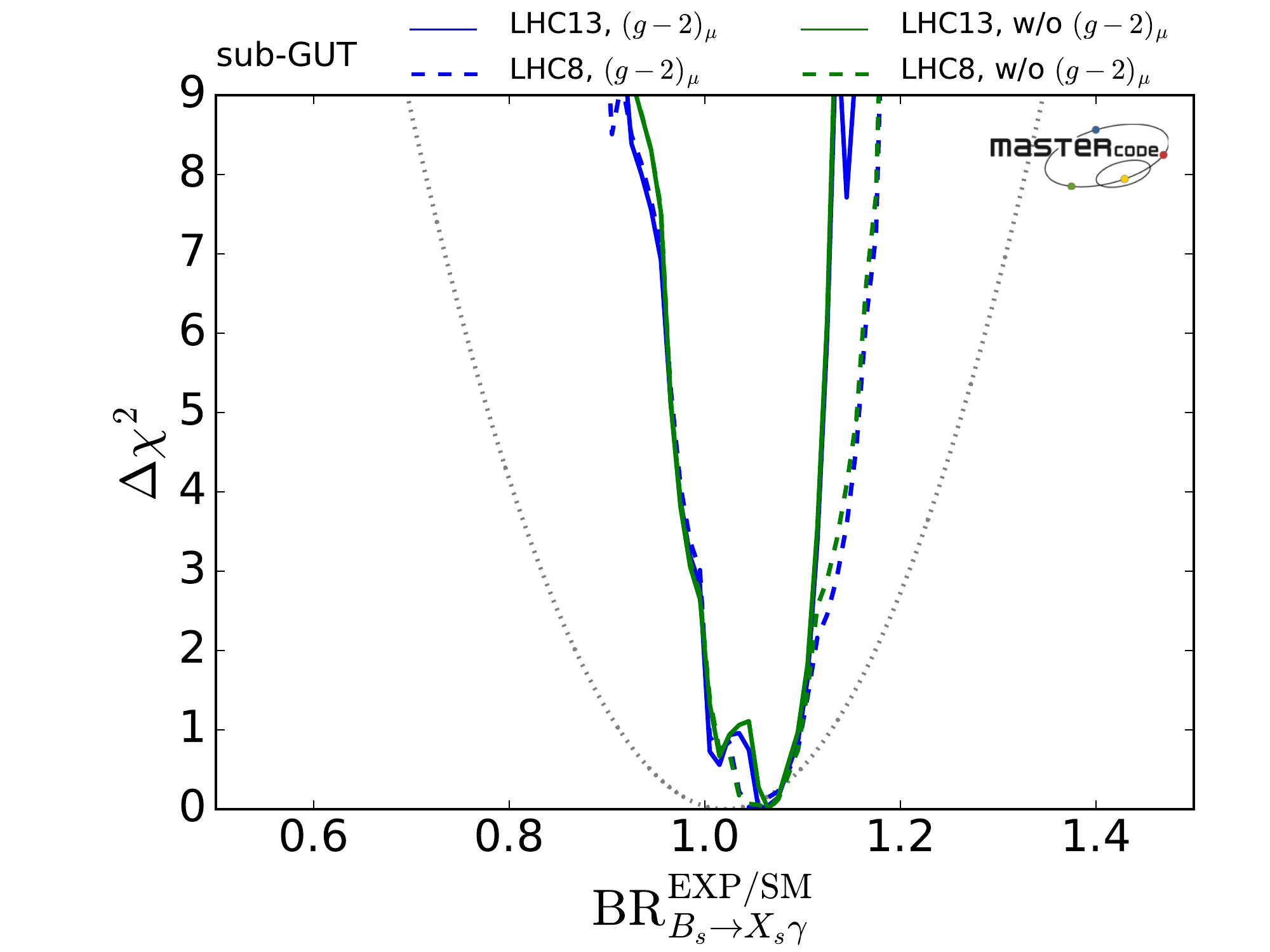}
\caption{\it One-dimensional profile likelihood function for \bsg, {showing the experimental constraint as a dotted line.}
}
\label{fig:bsg}
\end{figure*}
%%%%%%%%%%%%%%%%%%%%%%%%% F I G U R E %%%%%%%%%%%%%%%%%%%%%%%%%%%%%%%%%%%%%%%%

\subsection{Higgs Mass}

%%%%%%%%%%%%%%%%%%%%%%%%% F I G U R E %%%%%%%%%%%%%%%%%%%%%%%%%%%%%%%%%%%%%%%%
\begin{figure*}[]
\centering
\includegraphics[width=0.5\textwidth]{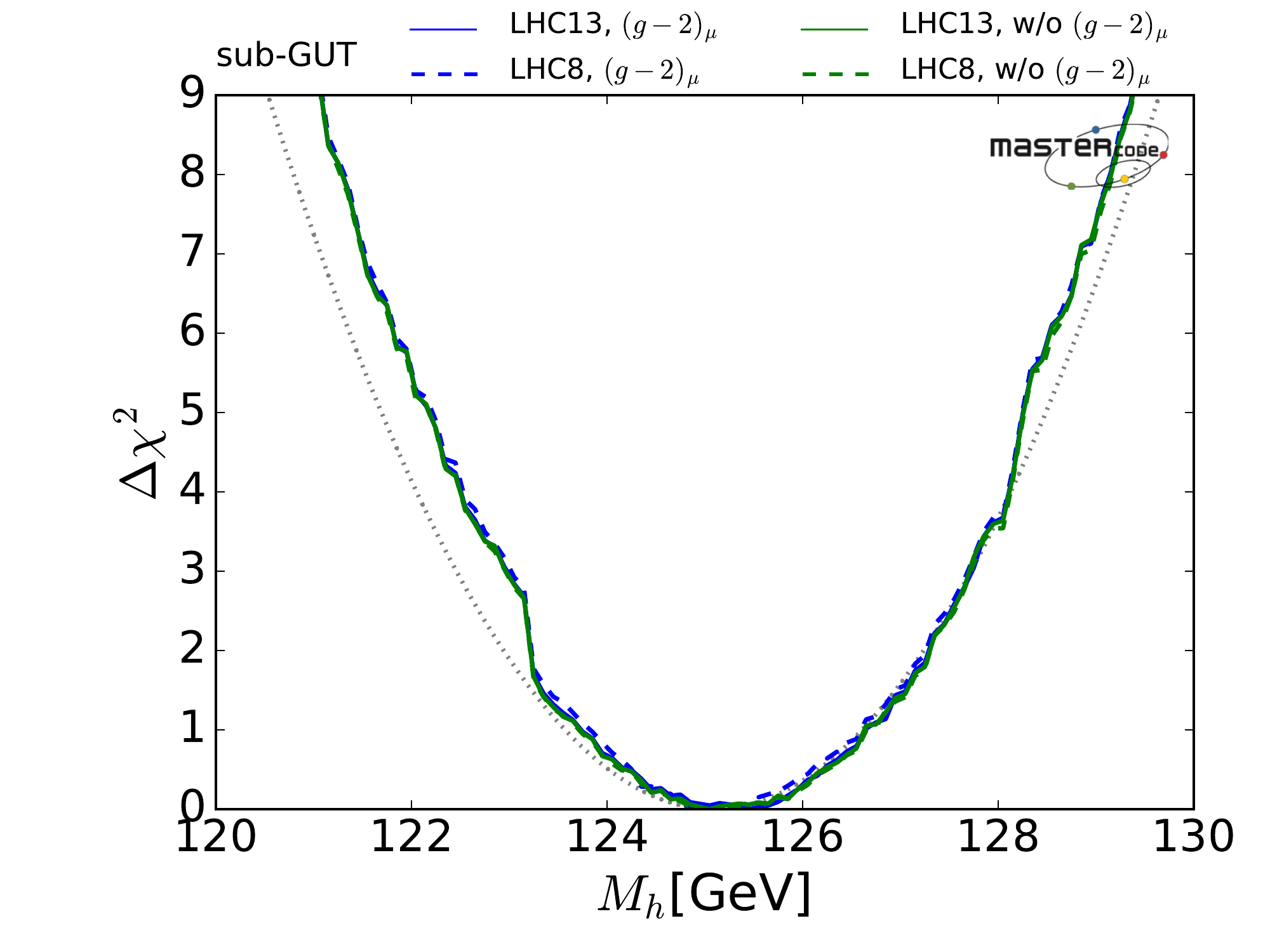}
\caption{\it {One-dimensional profile likelihood function for $\Mh$, where the dotted line
shows the $\chi^2$ contribution due to the \gmt\ constraint alone.}}
\label{fig:Mh}
\end{figure*}
%%%%%%%%%%%%%%%%%%%%%%%%% F I G U R E %%%%%%%%%%%%%%%%%%%%%%%%%%%%%%%%%%%%%%%%

{We see in Fig.~\ref{fig:Mh} that the profile likelihood function
for $\Mh$ lies within the contribution of the direct experimental constraint convoluted with the
uncertainty in the {\tt FeynHiggs} calculation of $\Mh$ (dotted line). We
infer that there is no tension between the direct experimental measurement of $\Mh$ and
the other observables included in our global fit. We have also calculated (not shown) the branching ratios for
Higgs decays into $\gamma \gamma$, $Z Z^*$ and $gg$ {(used as a proxy for $gg \to h$ production)}, finding that they
are expected to be very similar to their values in the SM, with 2-$\sigma$ ranges that lie
well within the current experimental uncertainties.}

\subsection{Searches for Dark Matter Scattering}

%%%%%%%%%%%%%%%%%%%%%%%%% F I G U R E %%%%%%%%%%%%%%%%%%%%%%%%%%%%%%%%%%%%%%%%
\begin{figure*}[h]
\centering
\includegraphics[width=0.495\textwidth]{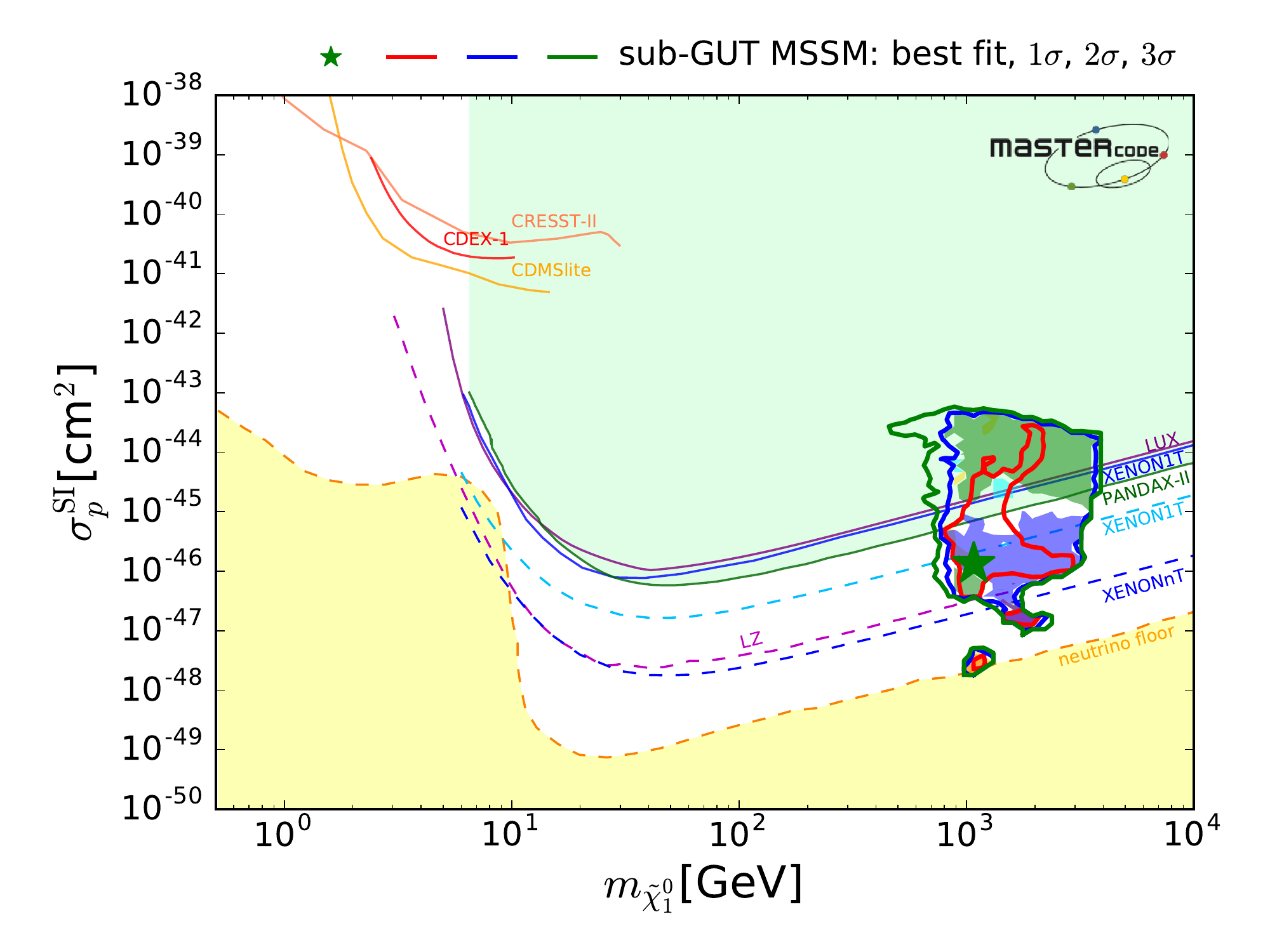}
\includegraphics[width=0.495\textwidth]{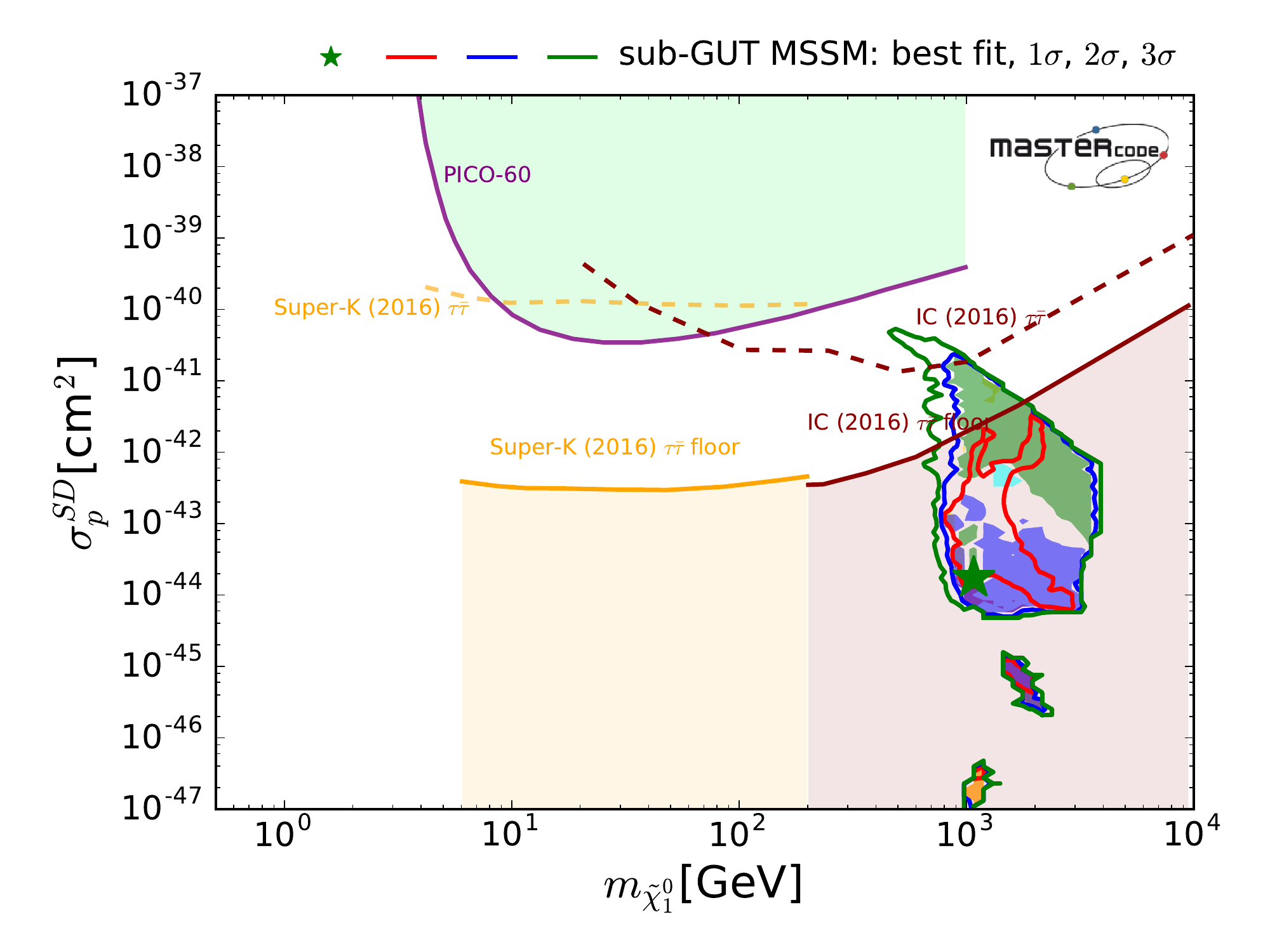} \\
\includegraphics[width=0.9\textwidth]{subGUT_dm_legend.pdf} \\
\vspace{-0.5cm}
\caption{\it Left panel: Two-dimensional profile likelihood function for the
nominal value of \ssi\ calculated using the {\tt SSARD} code~\protect\cite{SSARD} in the $(\mneu1, \ssi)$ plane,
displaying also the upper limits established by the LUX~\protect\cite{lux16},
XENON1T~\protect\cite{XENON1T} and PandaX-II Collaborations~\protect\cite{pandax}
shown as solid black, blue and green contours, respectively.
The projected future 90\% CL sensitivities of the
LUX-Zeplin (LZ)~\protect\cite{Mount:2017qzi} and XENON1T/nT~\protect\cite{Aprile:2015uzo} experiments are shown as dashed magenta and blue lines, respectively,
and the neutrino background `floor'~\protect\cite{Snowmass} is shown as a dashed orange line with yellow shading below.
Right panel: Two-dimensional profile likelihood function for the
nominal value of \ssd\ calculated using the {\tt SSARD} code~\protect\cite{SSARD} in the $(\mneu1, \ssd)$ plane,
showing also the upper limit established by the PICO Collaboration~\protect\cite{PICO}.{We also show the indirect limits
from the Icecube~\protect\cite{IceCube} and Super-Kamiokande~\protect\cite{SK} experiments, assuming $\neu1 \neu1 \to \tau^+ \tau^-$
dominates, as well as the `floor' for \ssd\ calculated in~\protect\cite{Ng}.}
}
\label{fig:ssissd}
\end{figure*}
%%%%%%%%%%%%%%%%%%%%%%%%% F I G U R E %%%%%%%%%%%%%%%%%%%%%%%%%%%%%%%%%%%%%%%%

{The left panel of Fig.~\ref{fig:ssissd} shows the nominal predictions for the
spin-independent DM scattering cross-section \ssi\ obtained using
the {\tt SSARD} code~\cite{SSARD}. We caution that there are considerable uncertainties in the calculation
of \ssi, which are taken into account in our global fit. Thus points with nominal values of \ssi\
above the experimental limit may nevertheless lie within the 95\% CL range for the global fit.
We see that sub-GUT models favour a range of \ssi\ close to the present limit
from the LUX, XENON1T and PandaX-II experiments~\footnote{{We also show, for completeness, the
CRESST-II~\cite{CRESST-II}, CDMSlite~\cite{CDMSlite} and CDEX~\cite{CDEX} constraints on \ssi,
which do not impact range of \mneu1
found in our analysis.}}. Moreover, at the 95\% CL, the nominal sub-GUT
predictions for \ssi\ are within the projected reaches of the LZ and XENON1/nT experiments.
However, they are subject to the considerable uncertainty in the \ssi\ matrix element, and might even fall below the
neutrino `floor' shown as a dashed orange line in~\cite{Snowmass}.}

{We see in the right panel of Fig.~\ref{fig:ssissd} that the sub-GUT predictions
for the spin-dependent DM scattering cross-section \ssd\ lie somewhat
below the present upper limit from the PICO direct DM search experiment.
Spin-dependent DM scattering is also probed by indirect searches for neutrinos
produced by the annihilations of neutralinos trapped inside the Sun after scattering on protons
in its interior. If the neutralinos annihilate into $\tau^+ \tau^-$,
the IceCube experiment sets the strongest such indirect limit~\cite{IceCube},
{and we also show the constraint from Super-Kamiokande~\cite{SK}. These constraints are
currently not sensitive enough} to cut
into the range of the $(\mneu1, \ssd)$ plane allowed in our global fit.} {We also show the neutrino
`floor' for \ssd, taken from~\cite{Ng}: wee that values of \ssd\ below this floor are quite possible in the sub-GUT model.}

\section{Impacts of the LHC 13-TeV and New Direct Detection Constraints}
\label{sec:impacts}

{We show in Fig.~\ref{fig:comparisons} some two-dimensional projections of the regions of sub-GUT MSSM
parameters favoured at the 68\% (red lines), 95\% (blue lines) and 99.7\% CL (green lines), comparing
the results of fits including the LHC 13-TeV data and recent direct searches for spin-independent dark
matter scattering (solid lines) and discarding them (dashed lines). The upper left panel shows the
$(m_{\tilde q_R}, \mgl)$ plane, the upper right plane shows the $(m_{\tilde q_R}, \mneu1)$ plane,
the lower left plane shows the $(\mgl, \mneu1)$ plane, and the lower right panel shows the $(\mneu1, \ssi)$ plane.
We see that in the upper panels that the new data restrict the favoured parameter space for $m_{\tilde q_R} \sim 2 \tev$,
the two left panels show a restriction for $\mgl \sim 1.3 \tev$, and the right and lower panels show that
the new data also restrict the range of $\mneu1$ {to $\gtrsim 800 \gev$}. However, the lower right panel does not show any
new restriction on the range of possible values of \ssi.}

%%%%%%%%%%%%%%%%%%%%%%%%% F I G U R E %%%%%%%%%%%%%%%%%%%%%%%%%%%%%%%%%%%%%%%%
\begin{figure*}[htbp!]
\centering
\includegraphics[width=0.45\textwidth]{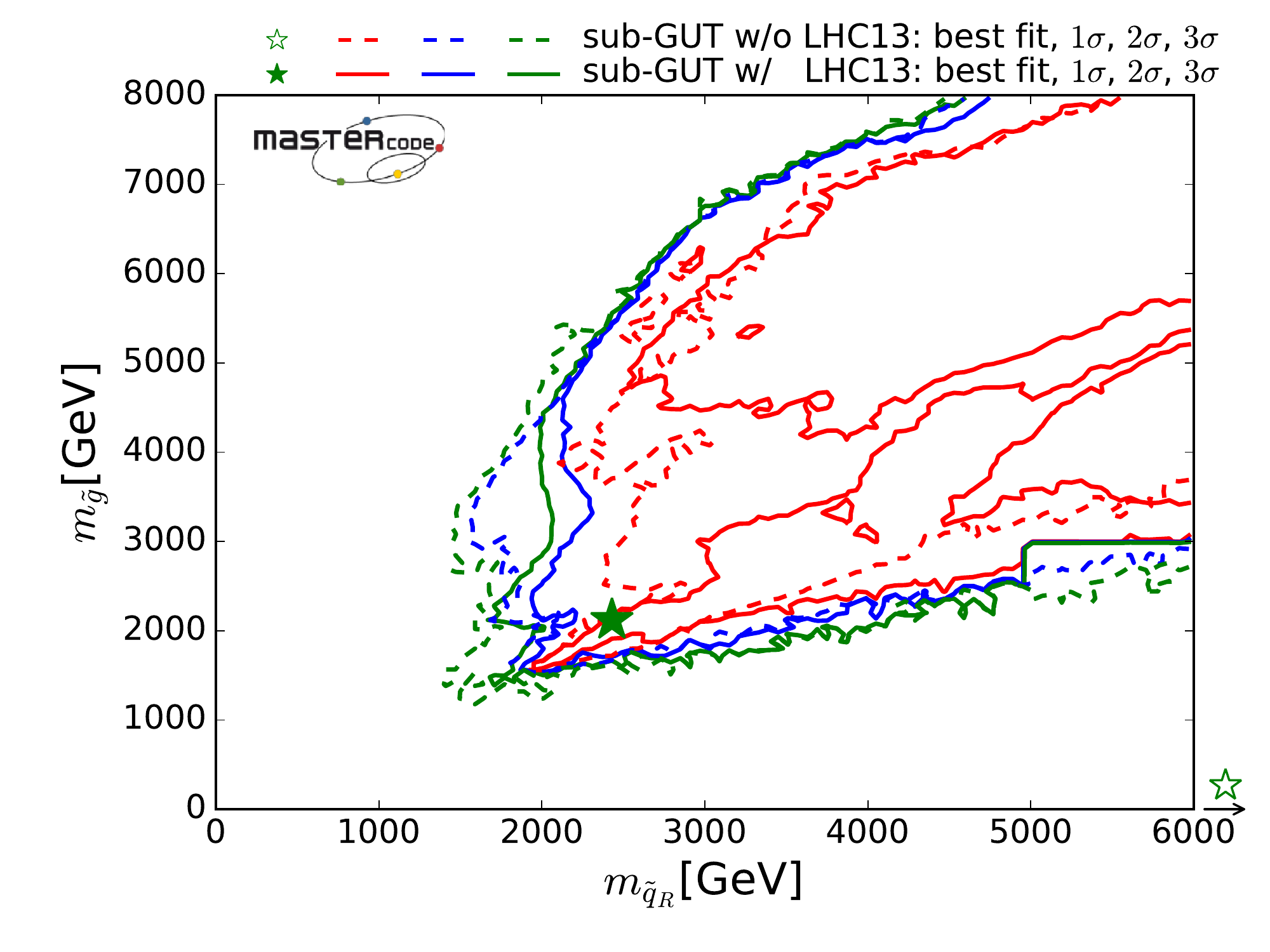}
\includegraphics[width=0.45\textwidth]{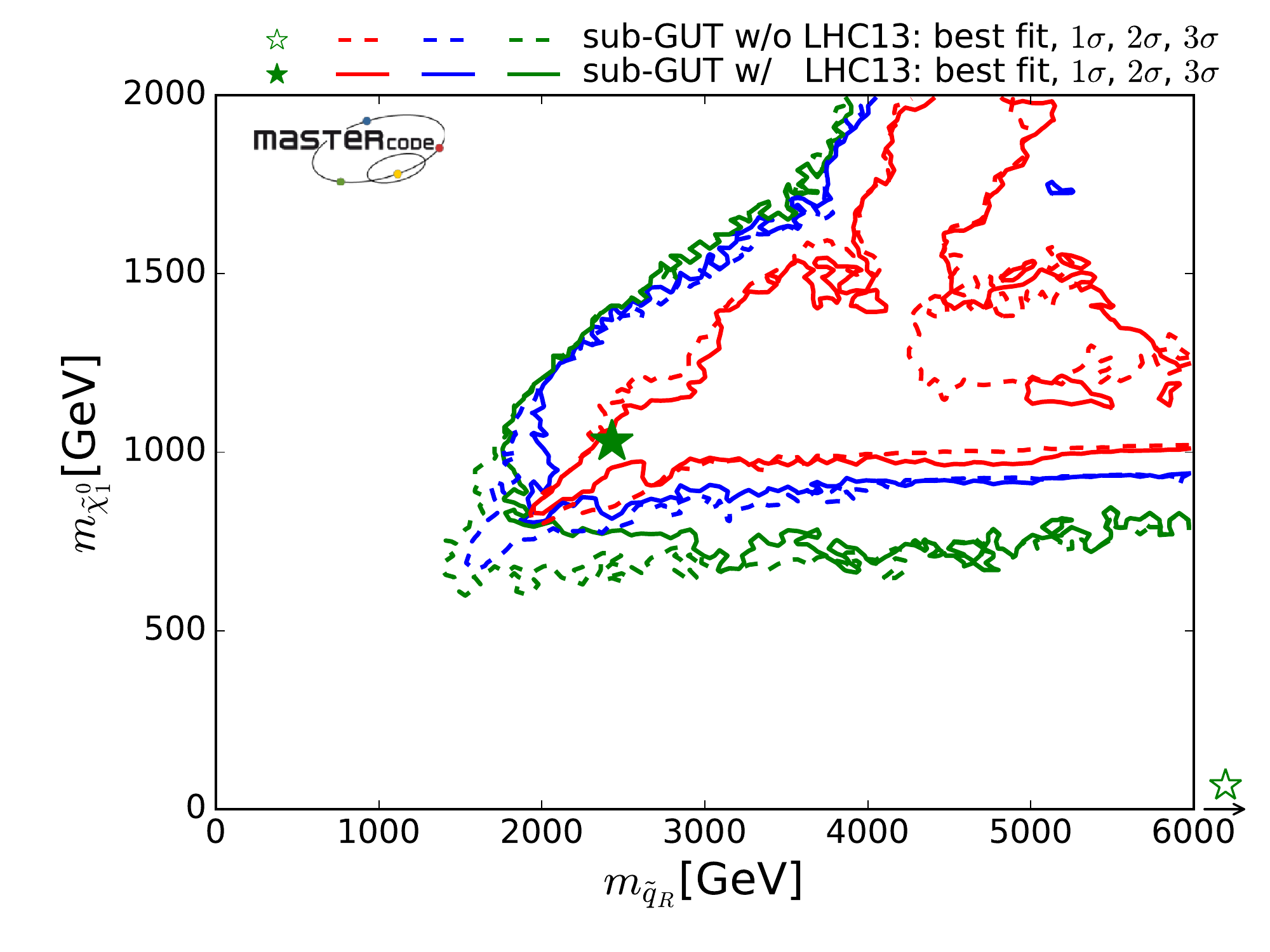} \\
\includegraphics[width=0.45\textwidth]{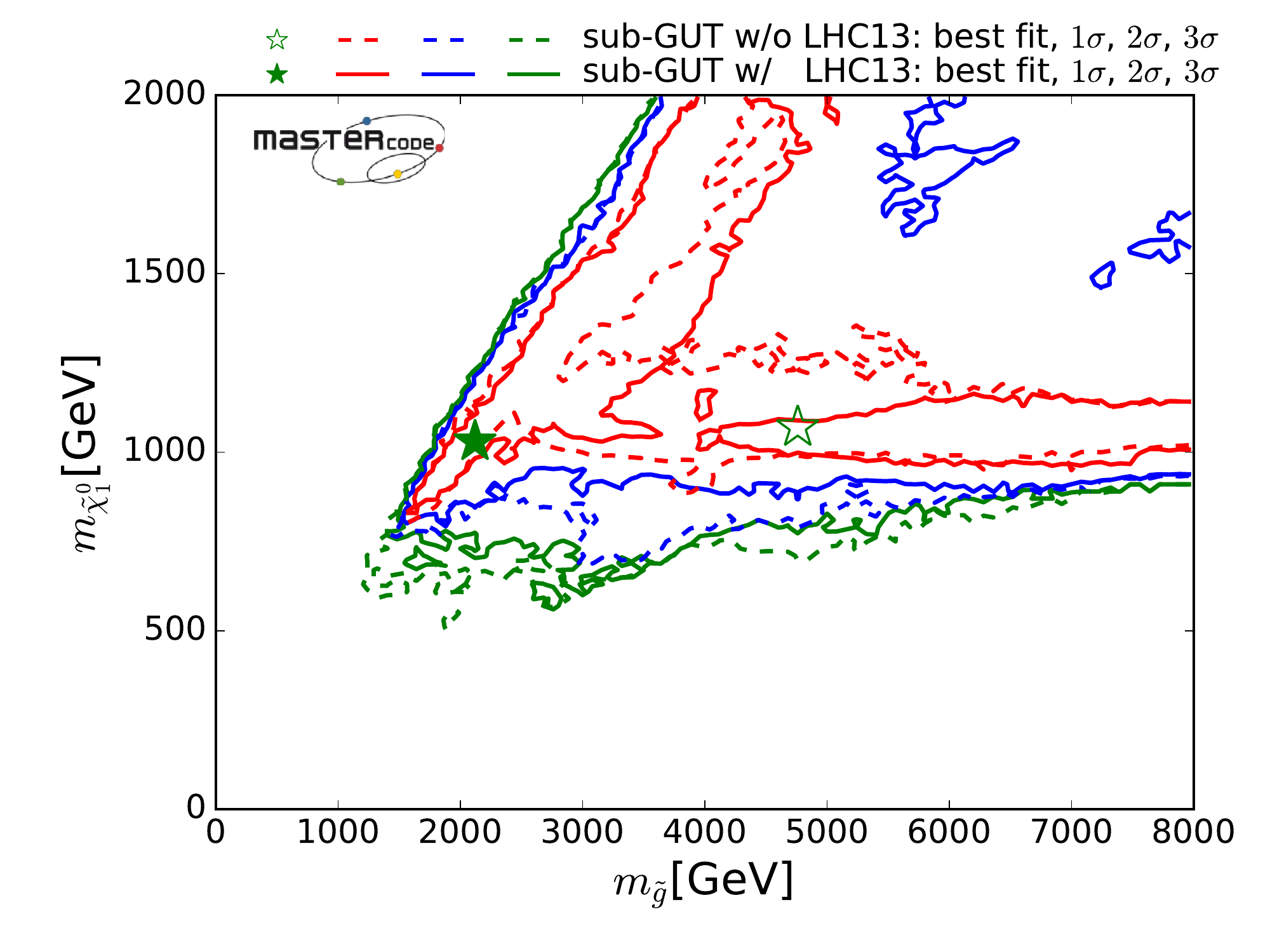}
\includegraphics[width=0.45\textwidth]{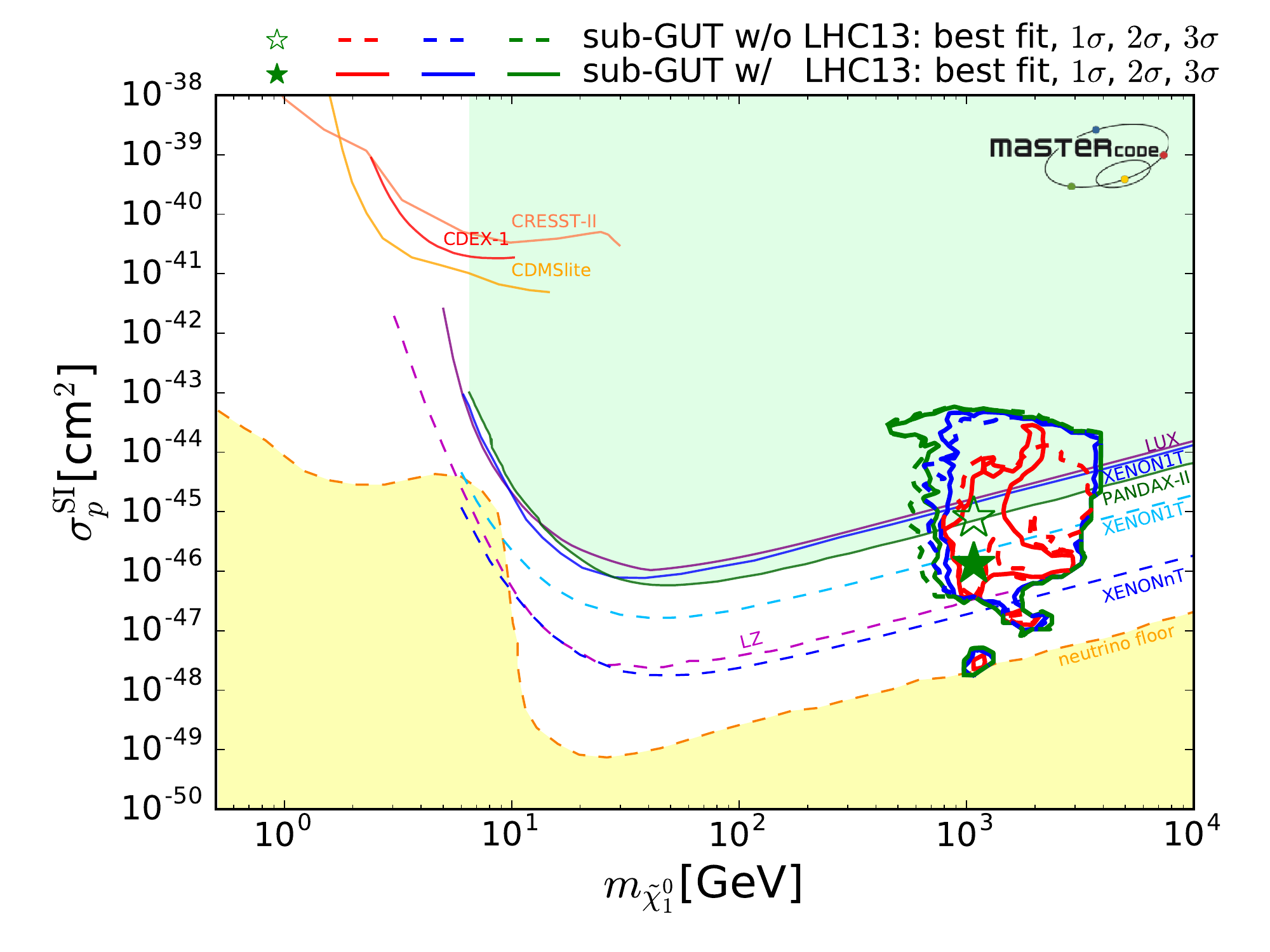} \\
\caption{\it {Two-dimensional projections of the global likelihood function for the sub-GUT MSSM in
the $(m_{\tilde q_R}, \mgl)$ plane (upper left panel), the $(m_{\tilde q_R}, \mneu1)$ plane (upper right panel),
the $(\mgl, \mneu1)$ plane (lower left panel), and the $(\mneu1, \ssi)$ plane (lower right panel).
In each panel we compare the projections of the sub-GUT parameter regions
favoured at the 68\% (red lines), 95\% (blue lines) and 99.7\% CL (green lines) in global fits with
the  LHC 13-TeV data and results from LUX, XENON1T, and PandaX-II~\protect\cite{lux16,XENON1T,pandax} (solid lines),
and without them (dashed lines).}}
\label{fig:comparisons}
\end{figure*}
%%%%%%%%%%%%%%%%%%%%%%%%% F I G U R E %%%%%%%%%%%%%%%%%%%%%%%%%%%%%%%%%%%%%%%%

\section{Best-Fit Points, Spectra and Decay Modes}
\label{sec:spectra}

{The values of the input parameters at the best-fit points with and without the
\gmt\ and LHC 13-TeV constraints have been shown in Table~\ref{tab:inputs}. {The best fits
have $\Min$ between $1.6 \times 10^5$ and $4.1 \times 10^8 \gev$, and we note that
the input parameters are rather insensitive to the inclusion of the 13-TeV
data when \gmt\ is dropped}. Table~\ref{tab:outputs} displays the mass spectra obtained as outputs at the best-fit
point including the 13-TeV data (quoted to 3 significant figures) and including (left column) or dropping (right column) the
\gmt\ constraint. As could be expected, the sparticle masses are generally heavier when
\gmt\ is dropped. However, the differences are small in the cases of the $\neu1, \neu2$ and $\cha1$,
being generally $< 10 \gev$. We also give in the next-to-last line of Table~\ref{tab:outputs} the values of
the global $\chi^2$ function at these best-fit points, dropping the {\tt HiggsSignals} contributions,
as was done previously~\cite{mc-su5,mc-amsb} to avoid biasing the analysis.}

\begin{table}[htb!]
\centering
\resizebox{0.475\textwidth}{!}{
		\begin{tabular}{ c c c }
			& With $(g-2)_{\mu}$ & Without $(g-2)_{\mu}$ \\
			\hline
			\hline
			$M_{H,A,H^+}$ & 2060 & 2220 \\
			$\tilde d_L, \tilde u_L, \tilde s_L, \tilde c_L$ & 2510 & 5050 \\
			$\tilde d_R, \tilde u_R, \tilde s_R, \tilde c_R$ & 2450 & 4835 \\
			$\tilde b_1$ & 1830 & 4100 \\
                        $\tilde b_2$ & 2190 & 4210 \\
			$\tilde t_1$ & 1130 & 3430 \\
			$\tilde t_2$ & 1850 & 4150 \\
			$\tilde e_L, \tilde \nu_{e_L}, \tilde \mu_L, \tilde \nu_{\mu_L}$ & 2040 & 3740 \\
			$\tilde e_R, \tilde \mu_R$ & 1980 & 3510 \\
			$\tilde \tau_1$ & 1380 & 2740 \\
                        $\tilde \tau_2$ & 1780 & 3390 \\
			$\tilde \nu_{\tau_L}$ & 1770 & 3390 \\
			$\tilde g$ & 2120 & 7240 \\
			$m_{\tilde \chi_1^0}$ & 1040 & 1060 \\
                        $m_{\tilde \chi_2^0}$ & 1270 & 1060 \\
                        $m_{\tilde \chi_3^0}$ & 1740 & 6010 \\
                        $m_{\tilde \chi_4^0}$ & 1740 & 6300 \\
                        $m_{\tilde \chi_1^\pm}$ & 1270 & 1060 \\
                        $m_{\tilde \chi_2^\pm}$ & 1740 & 6310 \\
			\hline \hline
			$\chi^2$ without & & \\
			{\tt HiggsSignals} & 28.86 & 18.02 \\
			\hline
			Number of d.o.f. & 24 & 23 \\
			\hline
			\hline
			p-value & 23\% & 76\% \\
			\hline
		\end{tabular}}
\caption{ \it {The spectra at the best-fit points including the LHC 13-TeV data
and including (left column) or dropping (right column) the \gmt\ constraint. The masses
are quoted in GeV. The three bottom lines give the values of the $\chi^2$ function dropping
{\tt HiggsSignals}, {the numbers of degrees of freedom (d.o.f.)} and the corresponding p-values.}}
\label{tab:outputs}
\end{table}

{The contributions of different observables to the global likelihood
function at the best-fit points including LHC13 data are shown in \reffi{fig:chi2break}.
We compare the contributions when \gmt\ is included (pink histograms) and
without \gmt\ (blue histograms). We note, in particular, that the contribution of \bsdmm\
is very small in both cases, which is a distinctive feature of sub-GUT models.

%%%%%%%%%%%%%%%%%%%%%%%%% F I G U R E %%%%%%%%%%%%%%%%%%%%%%%%%%%%%%%%%%%%%%%%
\begin{figure*}[htbp!]
\centering
\vspace{-4cm}
\hspace{-2cm}
	\begin{overpic}[scale=0.6]{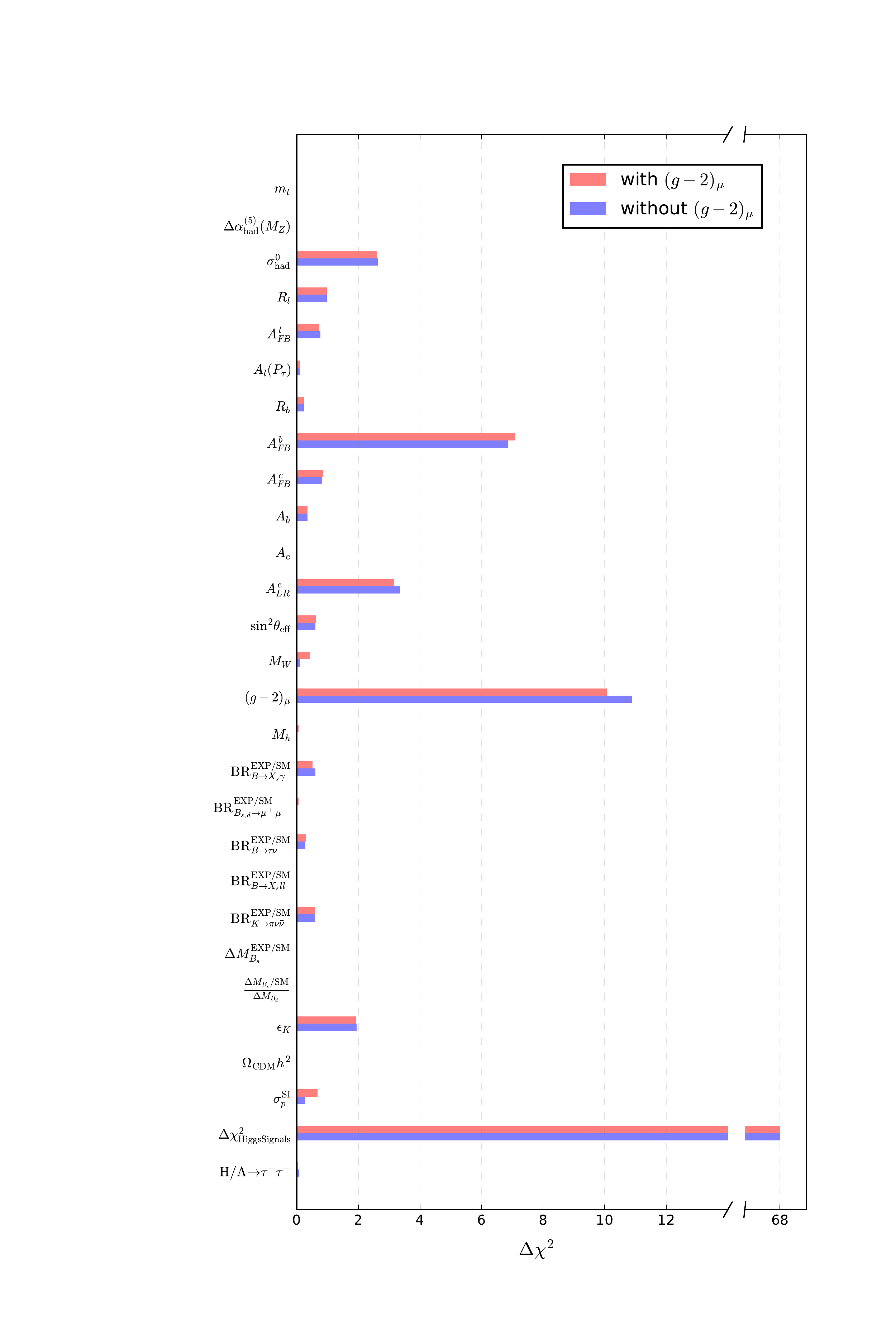}
		\put(25,85){\includegraphics[width=0.1\linewidth]{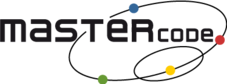}}
	\end{overpic}
\vspace{-2cm}
  \caption{\it {Contributions to the global $\chi^2$ function at the best-fit points found in our
  sub-GUT analysis {including LHC 13-TeV data,} in the cases with and without the \gmt\ constraint (pink and blue histograms, respectively).}}
  \label{fig:chi2break}
\end{figure*}
%%%%%%%%%%%%%%%%%%%%%%%%% F I G U R E %%%%%%%%%%%%%%%%%%%%%%%%%%%%%%%%%%%%%%%%

{The last line of Table~\ref{tab:outputs} shows the p-values for the best fits with and without \gmt,
which were calculated as follows. In the case with (without) \gmt, setting aside {\tt HiggsSignals}
so as to avoid biasing the analysis~\cite{mc-su5,mc-amsb}, the
number of constraints making non-zero contributions to the global $\chi^2$ function
(not including nuisance parameters) is 29 (28), and
the number of {free parameters}
is 5 in each case. Hence the numbers of degrees of freedom are 24 (23) in the two cases. The values of the total
$\chi^2$ function at the best-fit points, dropping the {\tt HiggsSignals} contribution, are 28.9 (18.0) and
the corresponding p-values are 23\% (76\%). The qualities of the global fits with and without \gmt\ are therefore both good.}
and the fit including \gmt\ is not poor enough to reject this fit hypothesis.}

{The spectra for the best fits are displayed graphically in \reffi{fig:spectrum},
including the \gmt\ constraint (upper panel) and dropping it (lower panel).
Also shown are the decay modes with branching ratios { $> 5\%$}, as dashed lines
whose intensities increase with the branching ratios. The heavy Higgs bosons decay predominantly
to SM final states, hence no dashed lines are shown.
We see that in both cases the squarks and gluino are probably too heavy to be discovered at the LHC,
and the sleptons are too heavy to be discovered at any planned $e^+e^-$ collider.
The best prospects for sparticle discovery may be for $\cha1$ and $\neu2$
production at CLIC running at $E_{\rm CM} \gtrsim 2 \tev$~\cite{CLIC}.}

\begin{figure*}[htbp!]
\begin{center}
	\begin{overpic}[scale=0.6]{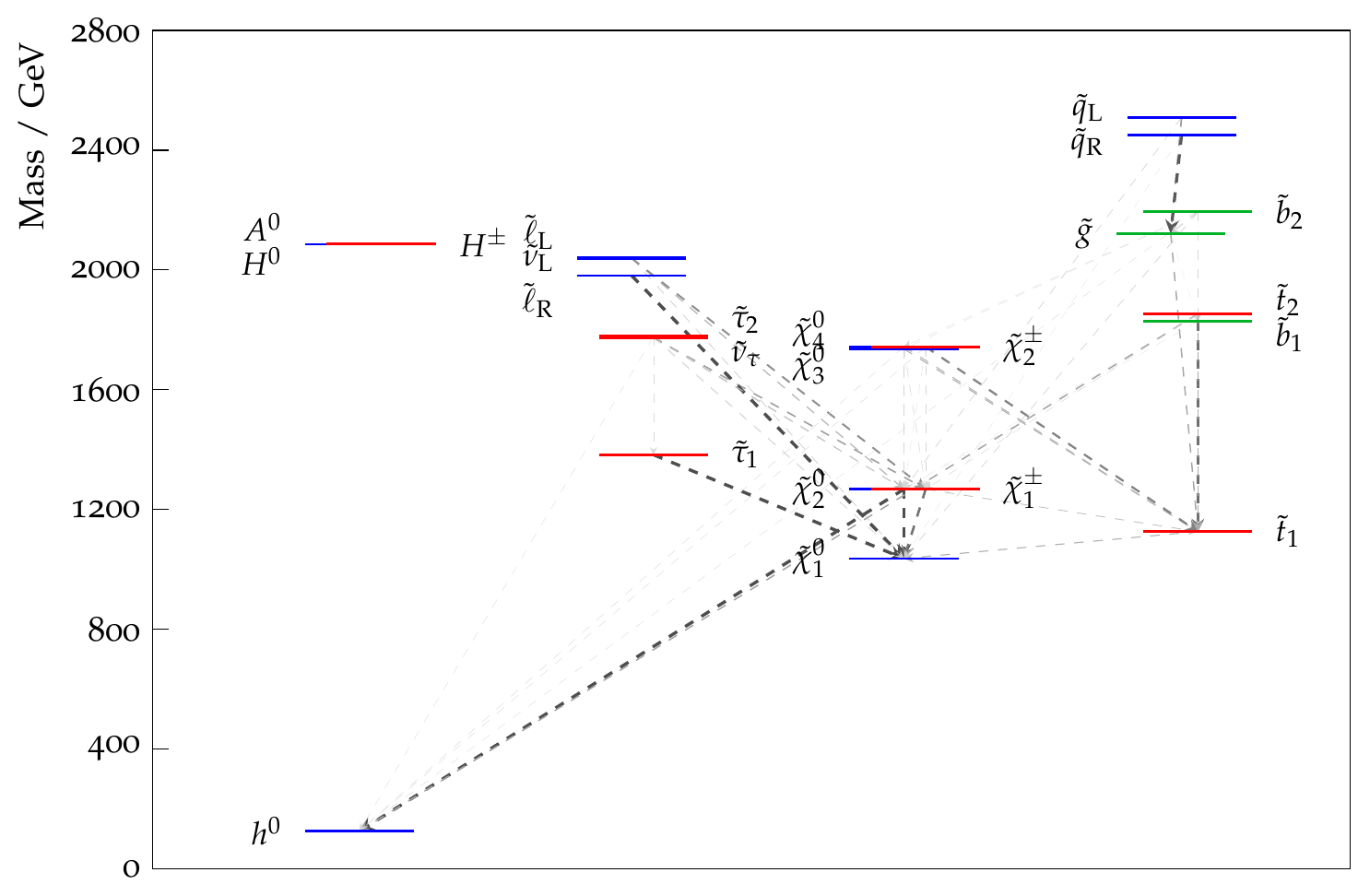}
		\put(15,55){\includegraphics[width=0.1\linewidth]{mastercode-logo.png}}
        \end{overpic}
        \vspace{0.5cm}
	\begin{overpic}[scale=0.6]{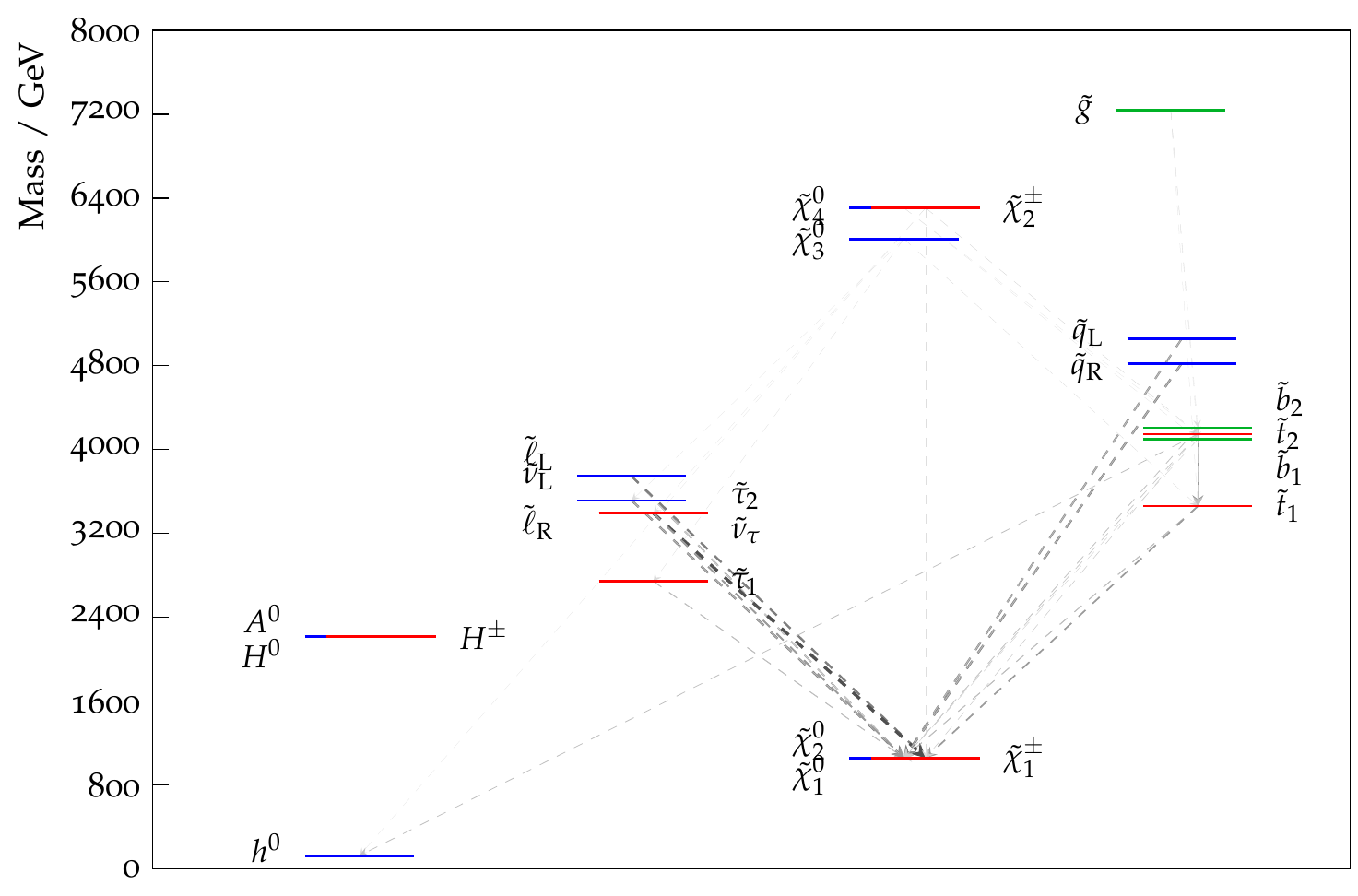} 
		\put(15,55){\includegraphics[width=0.1\linewidth]{mastercode-logo.png}}
	\end{overpic}
        \vspace{-0.5cm}
\end{center}
  \caption{\it {The spectra of Higgs bosons and sparticles at the best-fit points in the sub-GUT model {including LHC 13-TeV data,}
  including the \gmt\ constraint (upper panel) and dropping it (lower panel), with dashed lines indicating the
	decay modes with branching ratios  { $> 5\%$}. These plots were made using {\tt PySLHA}~\protect\cite{PySLHA}.}
}
  \label{fig:spectrum}
\end{figure*}

{The global likelihood function is quite flat at large sparticle masses, and very different
spectra are consistent with the data, within the current uncertainties. The 68 and 95\% CL
ranges of Higgs and sparticle masses are displayed in \reffi{fig:6895CLspectra} as orange
and yellow bands, respectively, with the best-fit values indicated by blue lines. The upper
panel is for a fit including the \gmt\ constraint, which is dropped in the lower panel. At the 68\%
CL there are possibilities for squark and gluino discovery at the LHC and the $\staue,
{\tilde \mu_R}$ and ${\tilde e_R}$ become potentially discoverable at CLIC if it
operates at $E_{\rm CM} = 3 \tev$~\cite{CLIC}.}

\begin{figure*}[htbp!]
	\begin{overpic}[width=\linewidth]{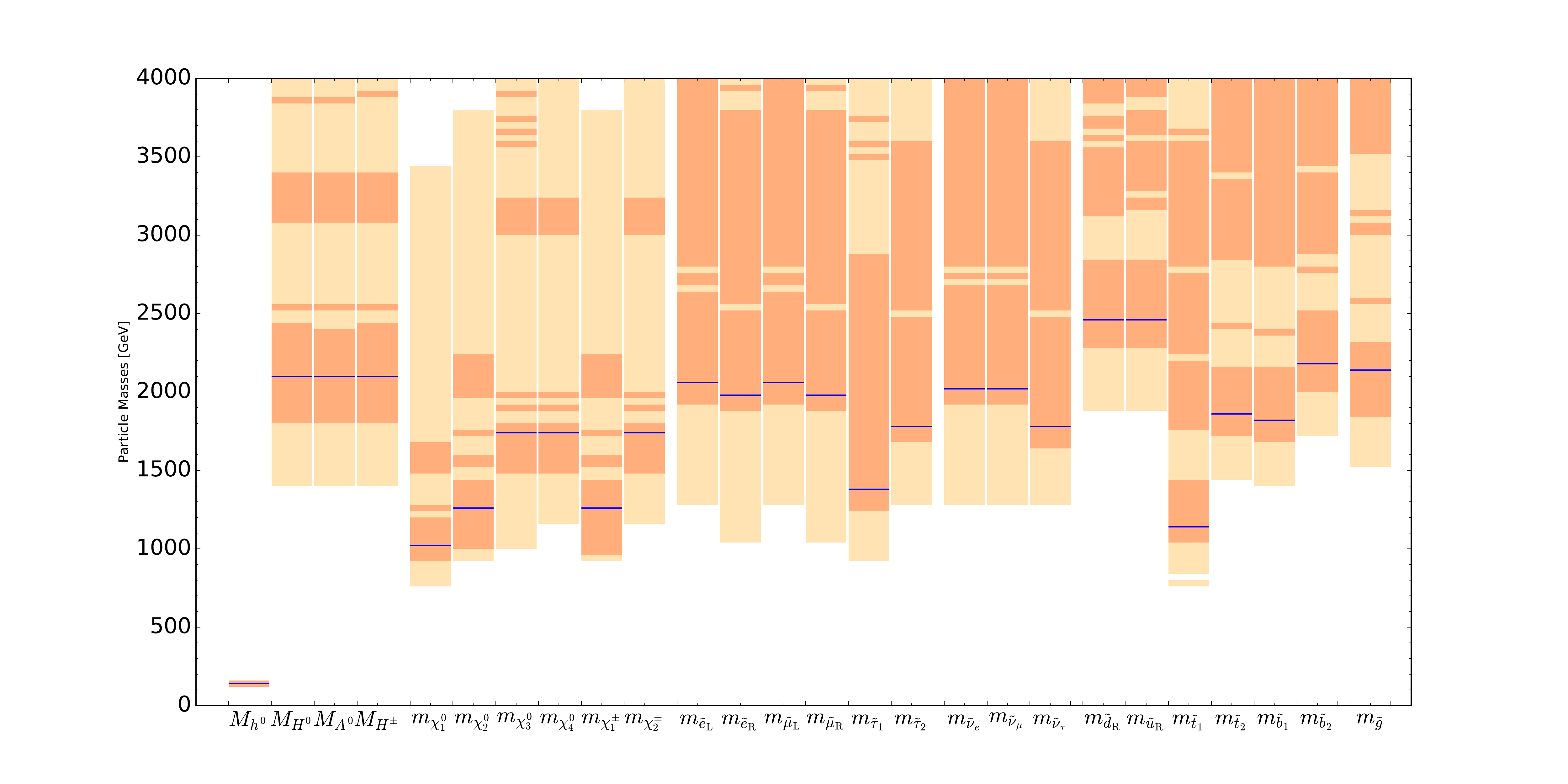}
		\put(77,7){\includegraphics[width=0.1\linewidth]{mastercode-logo.png}}
        \end{overpic}
	\begin{overpic}[width=\linewidth]{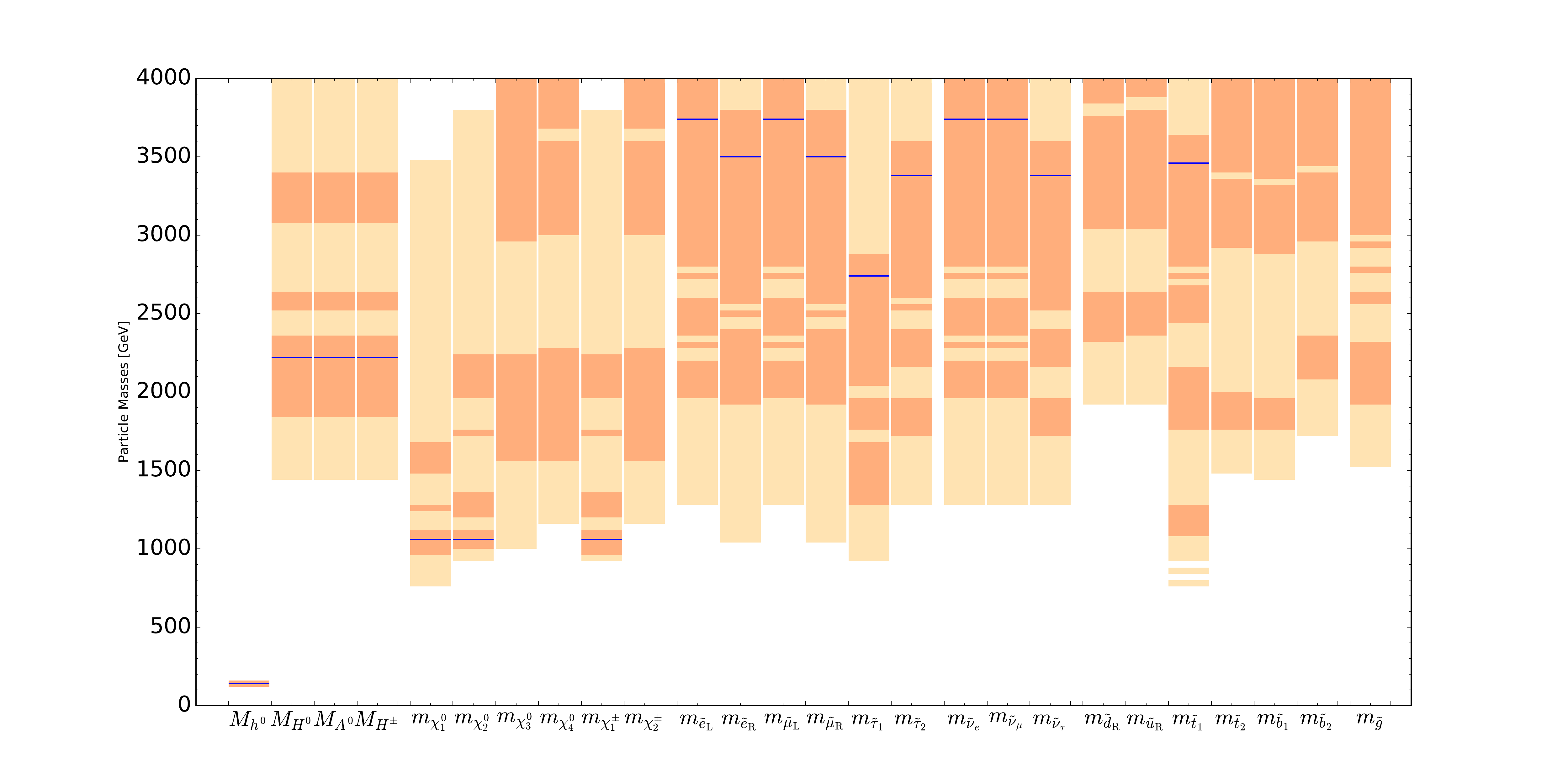}
		\put(77,7){\includegraphics[width=0.1\linewidth]{mastercode-logo.png}}
        \end{overpic}
	\caption{\it {The spectra in the sub-GUT model {including LHC 13-TeV data,} with (upper panel)
  and without (lower panel) the \gmt\ constraint,  displaying the best-fit values as blue lines,
  the 68\% CL ranges as orange bands, and the 95\% CL ranges as yellow bands.}}
  \label{fig:6895CLspectra}
\end{figure*}

\section{Summary and Perspectives}

{We have performed in this paper a frequentist analysis of sub-GUT models
in which soft supersymmetry-breaking parameters are assumed to be universal
at some input scale $\Min < \MGUT$. {The best-fit input parameters with
and without \gmt\ and the LHC 13-TeV data are shown in Table~\ref{tab:inputs}.
The physical sparticle masses including the LHC data, with
and without \gmt, are shown in Table~\ref{tab:outputs} and in Fig.~\ref{fig:spectrum},
where decay patterns are also indicated. As seen in the bottom line of Table~\ref{tab:outputs},
the p-values for the fits with and without \gmt\ are $\simeq 23$\% and $76$\%,
respectively.

Compared to the best fits with $\Min = \MGUT$,} we have found that the minimum value of
the global $\chi^2$ function may be reduced by $\Delta \chi^2 \sim 2$ in the sub-GUT model, with the
exact amount depending whether the \gmt\ constraint and/or LHC13 data are
included in the fit. Whether these observables are included, or not, the global $\chi^2$
minimum occurs for $\Min \sim 10^7 \gev$, and is due to the sub-GUT model's
ability to provide a better fit to the measured value of \bsdmm\ than in the CMSSM.
Although intriguing, this improvement in the fit quality is not very
significant, but it will be interesting
to monitor how the experimental measurement of \bsdmm\ evolves.}

{In all the scenarios studied (with/without \gmt\ and/or LHC13), the profile
likelihood function for $\mgl$ ($\msq$) varies by $\lesssim 1$ for $\mgl \gtrsim 1.9 \tev$
($\msq \gtrsim 2.2 \tev$). The corresponding slowly-varying ranges of $\chi^2$
for $m_{\tilde t_1}$ ($m_{\tilde b_1}$) start at $\sim 1 \tev$ ($\sim 1.6 \tev$),
respectively. On the other hand, we find a more marked preference for
$\mneu1 \sim 1 \tev$, with the $\cha1$ and $\neu2$ being slightly heavier and
large mass values being disfavoured at the $\Delta \chi^2 \sim 3$
level. The best-fit point is in a {region where rapid annihilation via $H/A$ poles is
hybridized with stop coannihilation, with} chargino coannihilation and stau coannihilation
also playing roles} in both the 68 and 95\% CL regions. Within the 95\% CL region,
the chargino lifetime may exceed $10^{-12}$~s, and the stau lifetime may
be as long as one second, motivating continued searches for long-lived sparticles
at the LHC.}

{Taking the LHC13 constraints into account, we find that the spin-independent
DM cross-section, \ssi, may be just below the present upper limits
from the LUX, XENON1T and PandaX-II experiments, and within the reaches of
the planned XENONnT and LZ experiments. On the other hand, the spin-dependent
DM cross-section, \ssd, may be between some 2 and 5 orders of magnitude
below the current upper limit from the PICO experiment.}

{Within the sub-GUT framework, therefore, we find interesting perspectives
for LHC searches for strongly-interacting sparticles via the conventional missing-energy signature.
{Future $\ETslash$ searches for electroweakly-interacting sparticles and for long-lived massive
charged particles may also have interesting prospects. The best-fit region of parameter space
accommodates the observed deviation of \bsdmm\ from its value in the SM,
and it will be interesting to see further improvement in the precision of this measurement.}
A future $e^+ e^-$ collider with centre-of-mass energy
above 2~TeV, such as CLIC~\cite{CLIC}, would have interesting perspectives for discovering
and measuring the properties of electroweakly-interacting sparticles. There are
also interesting perspectives for direct DM searches via spin-independent
scattering.}

\section*{Acknowledgements}

The work of E.B. and G.W. is supported in part by the Collaborative Research Center
SFB676 of the DFG, ``Particles, Strings and the early Universe''.
The work of M.B. and D.M.S. is supported by the European Research Council
via Grant BSMFLEET 639068. {The work of J.C.C. is supported by CNPq (Brazil).}
The work of M.J.D. is supported in part by the Australia Research Council.
The work of J.E. is supported in part by STFC (UK) via the research grant ST/L000326/1
and in part via the Estonian Research Council via a Mobilitas Pluss grant,
and the work of H.F. is also supported in part by STFC (UK). The work of S.H. is supported in
part by the MEINCOP Spain under contract FPA2016-78022-P, in part by the
Spanish Agencia Estatal de Investigaci{\' o}n (AEI) and the EU Fondo Europeo de
Desarrollo Regional (FEDER) through the project FPA2016-78645-P, in part by
the AEI through the grant IFT Centro de Excelencia Severo Ochoa SEV-2016-0597,
and by the Spanish MICINN Consolider-Ingenio 2010 Program under Grant
MultiDark CSD2009-00064. The work of M.L. and I.S.F. is supported by XuntaGal.
The work of K.A.O. is supported in part by DOE grant de-sc0011842 at the
University of Minnesota. K.S. thanks the TU Munich for hospitality during the final stages
of this work and has been partially supported by the DFG cluster of excellence
EXC 153 ``Origin and Structure of the UniverseÓ, by the Collaborative Research Center SFB1258.
The work of K.S. is also partially supported by the National Science Centre, Poland,
under research grants DEC-2014/15/B/ST2/02157, DEC-2015/18/M/ST2/00054 and
DEC-2015/19/D/ST2/03136. The work of G.W. is also supported in part by the European Commission
through the ``HiggsTools'' Initial Training Network PITN-GA-2012-316704.

\end{document}